\documentclass[a4paper,aps,prd,preprint,superscriptaddress,nofootinbib,showpacs,longbibliography]{revtex4-1}

\usepackage[utf8]{inputenc}
\usepackage{amsmath}
\usepackage{amsfonts}
\usepackage{amssymb}
\usepackage{amsthm}
\usepackage{graphicx}
\usepackage{bm}
\usepackage{bbm}
\usepackage{hyperref}
\usepackage{epstopdf}
\allowdisplaybreaks

\renewcommand{\k}{\bm{k}}
\newcommand{\q}{\bm{q}}

\renewcommand{\r}{\bm{r}}
\newcommand{\R}{\bm{R}}

\newcommand{\E}{\bm{E}}

\begin{document}

\preprint{TUM-EFT 67/15}

\title{Polyakov loop correlator in perturbation theory}

\author{Matthias~Berwein}
\affiliation{Department of Physics, Tohoku University, Sendai 980-8578, Japan}
\affiliation{Physik-Department, Technische Universit\"{a}t M\"{u}nchen, James-Franck-Stra\ss{}e 1, 85748 Garching, Germany}

\author{Nora~Brambilla}
\affiliation{Physik-Department, Technische Universit\"{a}t M\"{u}nchen, James-Franck-Stra\ss{}e 1, 85748 Garching, Germany}
\affiliation{Institute for Advanced Study, Technische Universit\"{a}t M\"{u}nchen, Lichtenbergstra\ss{}e 2a, 85748 Garching, Germany}

\author{P\'{e}ter~Petreczky}
\affiliation{Physics Department, Brookhaven National Laboratory, Upton, New York 11973, USA}

\author{Antonio~Vairo}
\affiliation{Physik-Department, Technische Universit\"{a}t M\"{u}nchen, James-Franck-Stra\ss{}e 1, 85748 Garching, Germany}


\begin{abstract}
  We study the Polyakov loop correlator in the weak coupling expansion and show how the perturbative series reexponentiates into singlet and adjoint contributions. We calculate the order $g^7$ correction to the Polyakov loop correlator in the short distance limit. We show how the singlet and adjoint free energies arising from the reexponentiation formula of the Polyakov loop correlator are related to the gauge invariant singlet and octet free energies that can be defined in pNRQCD, namely we find that the two definitions agree at leading order in the multipole expansion, but differ at first order in the quark-antiquark distance.
\end{abstract}


\maketitle

\section{Introduction}

The Polyakov loop correlator defines the free energy of a static quark-antiquark ($Q\bar{Q}$) pair and is an important quantity for the understanding of deconfinement and screening in the quark gluon plasma~\cite{McLerran:1981pb}. It has been extensively studied on the lattice both in pure SU(N) gauge theories~\cite{Kaczmarek:1999mm,Kaczmarek:2002mc,Digal:2003jc} as well as in QCD~\cite{Karsch:2000kv,Petreczky:2004pz,Borsanyi:2015yka}. However, the behavior of the Polyakov loop correlator in the weak coupling expansion is still poorly understood.

The leading order result has been known for several decades now~\cite{McLerran:1981pb}, both for small and large separation between the static quark and antiquark. The next-to-leading order (NLO) and next-to-next-to-leading order (NNLO) calculation of the Polyakov loop correlator in the short distance regime has been performed relatively recently~\cite{Brambilla:2010xn}. This calculation provided qualitatively new insight into the behavior of the Polyakov loop correlator, showing the exponentiation into singlet and adjoint contributions as well as showing how the free energy of the static $Q\bar{Q}$ pair goes over into the zero temperature static energy. The use of the potential nonrelativistic QCD (pNRQCD) at finite temperature approach was essential in obtaining this result.

At short distances, the calculation of the Polyakov loop correlator in perturbation theory is important if one wants to establish a connection to lattice QCD calculations. For distances of the order of the inverse Debye mass, the Polyakov loop correlator was calculated by Nadkarni~\cite{Nadkarni:1986cz}, while for distances much larger than the inverse Debye mass, the behavior of the Polyakov loop correlator was discussed by Braaten and Nieto~\cite{Braaten:1994qx} and by Laine and Veps\"{a}l\"{a}inen~\cite{Laine:2009dh}. These studies are based on dimensionally reduced effective field theories. 

Also the singlet free energy, defined in terms of the correlator of two Polyakov loop operators inside a single trace in Coulomb gauge, is a useful quantity for understanding color screening in the deconfined medium. This is due to the fact that it is more closely related to the static $Q\bar{Q}$ energy and, unlike Wilson loops, has only divergences associated with self-energy contributions~\cite{Burnier:2009bk,Berwein:2012mw}, which are identical to those in the vacuum energy of a static $Q\bar Q$ pair~\cite{Kaczmarek:2002mc}. Furthermore, the singlet free energy is used in modeling the in-medium properties of quarkonia (see, e.g., Ref.~\cite{Mocsy:2013syh} for a review). The singlet free energy was studied at NLO in Ref.~\cite{Burnier:2009bk}, where also a comparison with lattice QCD calculations was performed. However, no contact of the weak coupling calculations of the singlet free energy and pNRQCD has been made.

In this paper, we discuss the reexponentiation of the Polyakov loop correlator into singlet and adjoint contributions on general grounds using techniques developed for the reexponentiation of Wilson lines~\cite{Gardi:2010rn,Gardi:2013ita}. Then, we calculate the next-to-next-to-next-to-leading order (NNNLO) contribution to the Polyakov loop correlator at short distances. Furthermore, we analyze the short distance behavior of the singlet free energy in terms of pNRQCD and also calculate the corresponding NNLO contribution. We also give an NLO result for intermediate distances.

The rest of the paper is organized as follows. In Sec.~\ref{sec:general}, we describe the general framework used to calculate the Polyakov loop correlator. The actual calculation of the Polyakov loop correlator using Coulomb gauge is presented in Sec.~\ref{sec:polycorr}. In Sec.~\ref{sec:pnrqcd}, the relation of the singlet and adjoint contributions, which appear in the perturbative expression of the Polyakov loop correlator, to the gauge invariant definition of singlet and octet free energies in pNRQCD is discussed. Finally, in Sec.~\ref{sec:concl}, we present our conclusions. Technical details of the calculations are presented in the Appendices.

\section{Free energies}
\label{sec:general}

The free energies of static quarks are related to the Polyakov loop or correlators thereof in the following way. The Polyakov loop operator is defined in the imaginary time formalism as
\begin{equation}
 L(\bm{r})=\mathcal{P}\exp\left[ig\int_0^{1/T}d\tau\,A_0(\tau,\bm{r})\right]\,,
\end{equation}
where $\mathcal{P}$ denotes path ordering, $T$ is the temperature, $g$ is the coupling constant, and $A_0$ is the matrix valued temporal gauge field.

The thermal expectation value of the trace of a single Polyakov loop operator gives the free energy of a static quark, $F_Q$:
\begin{equation}
 \exp\left[-\frac{F_Q}{T}\right]=\frac{1}{N}\bigl\langle\mathrm{Tr}[L(\bm{r})]\bigr\rangle\,,
\end{equation}
where $N$ is the number of colors. This quantity is what we will usually understand in this paper by Polyakov loop, unless we explicitly refer to the operator. Because of translational invariance, it does not depend on the position $\bm{r}$.

The free energy of a static quark-antiquark pair, $F_{Q\bar{Q}}$, is correspondingly given by the Polyakov loop correlator:
\begin{equation}
 \exp\left[-\frac{F_{Q\bar{Q}}(r)}{T}\right]=\frac{1}{N^2}\left\langle\mathrm{Tr}\bigl[L(\bm{r})\bigr]\mathrm{Tr}\left[L^\dagger(\bm{0})\right]\right\rangle\,.
\end{equation}
The dagger on the second Polyakov loop, which corresponds to the antiquark contribution, turns the fundamental into the antifundamental representation. This quantity depends only on the absolute value of the relative distance $r$ as opposed to its direction because of rotational invariance. Translational invariance also excludes a dependence on the center of mass coordinate, so we have set it to $\bm{r}/2$ in the above expression for simplicity.

In both the single quark and the $Q\bar{Q}$ cases, the free energies are defined with respect to the medium, i.e., $F_Q$ is the difference between the free energy of the medium in the presence of one static quark and the free energy of the medium without static quarks, and analogously for $F_{Q\bar{Q}}$ (see discussions in Ref.~\cite{McLerran:1981pb}).

In the weak coupling regime (i.e., for large temperatures) these quantities can be calculated in perturbation theory. For the Polyakov loop there exists an exponentiation formula, which makes it possible to express the free energy directly through a set of Feynman diagrams (cf.~\cite{Berwein:2015ayt}). For the correlator, a similar expression has been obtained in~\cite{Berwein:2013xza}, however, while that calculation is correct, a more useful expression can be found by a slight modification of that approach.

The method we use is the replica trick for Wilson lines~\cite{Gardi:2010rn,Gardi:2013ita}, which we will outline here. First, consider the Polyakov loop correlator in terms of an amplitude with uncontracted indices, $\langle\mathcal{M}\rangle_{ij,\,kl}$:
\begin{equation}
 \exp\left[-\frac{F_{Q\bar{Q}}(r)}{T}\right]=\frac{\delta_{ij}\delta_{kl}}{N^2}\left\langle L_{ij}(\bm{r})L^*_{kl}(\bm{0})\right\rangle\equiv\frac{\delta_{ij}\delta_{kl}}{N^2}\langle\mathcal{M}\rangle_{ij,\,kl}\,,
\end{equation}
where $i$ and $k$ are the color indices of the Polyakov loop operator at imaginary time $\tau=1/T$, while $j$ and $l$ are at $\tau=0$. Since the uncontracted amplitude is gauge dependent, $\langle\mathcal{M}\rangle_{ij,\,kl}$ requires evaluation in a gauge fixed theory. Then, we define a multiplication of amplitudes $\mathcal{A}$ and $\mathcal{B}$ as $\mathcal{A}_{ij',\,kl'}\mathcal{B}_{j'j,\,l'l}$\footnote{Note that herein lies the difference to the approach in~\cite{Berwein:2013xza}, where the multiplication was defined as $\mathcal{A}_{ij',l'l}\mathcal{B}_{j'j,kl'}$. Since the Polyakov loop correlator itself does not depend on this multiplication, both definitions are valid and lead to different but equivalent exponentiations.}. Exponentiation is to be understood as a power series with respect to this multiplication. In order to find the exponentiated expression of the thermal average of the amplitude $\langle\mathcal{M}\rangle$, we have to determine an amplitude that can be interpreted as the logarithm of $\langle\mathcal{M}\rangle$.

Now consider the $n$th power of this amplitude and expand in $n$:
\begin{equation}
 \langle\mathcal{M}\rangle^n_{ij,\,kl}=\exp[n\ln\langle\mathcal{M}\rangle]_{ij,\,kl}=\delta_{ij}\delta_{kl}+n\ln\langle\mathcal{M}\rangle_{ij,\,kl}+\mathcal{O}\left(n^2\right)\,.
\end{equation}
In order to find the logarithm of $\langle\mathcal{M}\rangle$, we have to calculate the linear term in an expansion of $\mathcal{M}^n$ in $n$. There is an alternative way of doing this. We can define a theory that contains $n$ exact copies (or replicas) of the QCD fields, which interact like in QCD for each replica, but there is no interaction between different replica fields. In this theory, we can write the $n$th power of the thermal average of the amplitude as the thermal average of $n$ replicas of the amplitude:
\begin{equation}
 \langle\mathcal{M}\rangle^n_{ij,\,kl}=\left\langle\mathcal{M}^{(n)}_{ii',\,kk'}\mathcal{M}^{(n-1)}_{i'i'',\,k'k''}\cdots\mathcal{M}^{(1)}_{j'j,\,l'l}\right\rangle\,,
\end{equation}
where the upper indices label the different replicas.

The Feynman diagrams in this replica theory are almost the same as in QCD, except that now there is replica path ordering: all color matrices associated to gluons of a higher replica index are to be placed to the left of those associated to a lower index. Therefore, it makes sense to split the calculation of the Feynman diagrams $\mathcal{D}$ into a color and a kinematic part, where the color part $\mathcal{C}$ contains all color matrices and structure constants and the kinematic part $\mathcal{K}$ contains everything else:
\begin{equation}
 \mathcal{D}^{\{\rho\}}_{ij,\,kl}=\mathcal{C}^{\{\rho\}}_{ij,\,kl}(\mathcal{D})\mathcal{K}(\mathcal{D})\,,
\end{equation}
where ${\{\rho\}}$ denotes the set of all replica indices, while the absence of such an index denotes the corresponding expression in QCD without replicas. In this way, diagrams that differ only in the replica indices of the fields have the same kinematic part, which is the same as in QCD, so the sum over different replica indices and the expansion in $n$ can be performed exclusively in the color part. Consequently, the amplitude $\langle\mathcal{M}\rangle$ and its logarithm can be written as a sum over the same Feynman diagrams, but the color parts for each diagram have to be modified in the following way:
\begin{equation}
 \langle\mathcal{M}\rangle_{ij,\,kl}=\sum_\mathcal{D}\mathcal{C}_{ij,\,kl}(\mathcal{D})\mathcal{K}(\mathcal{D})=\exp\bigl[\ln\langle\mathcal{M}\rangle\bigr]_{ij,\,kl}=\exp\left[\sum_\mathcal{D}\widetilde{\mathcal{C}}(\mathcal{D})\mathcal{K}(\mathcal{D})\right]_{ij,\,kl}\,,
\end{equation}
where
\begin{equation}
 \sum_{\{\rho\}}\mathcal{C}^{\{\rho\}}_{ij,\,kl}(\mathcal{D})=n\,\widetilde{\mathcal{C}}_{ij,\,kl}(\mathcal{D})+\mathcal{O}\left(n^2\right)\,.
\end{equation}
We will present an explicit example of such a determination of the coefficients $\widetilde{\mathcal{C}}(\mathcal{D})$ in the following section and Appendix~\ref{app:color}.

Here, we will show how the exponential can be evaluated once the coefficients have been determined. In principle, this requires computing the exponential of an $N^2\times N^2$ matrix, however, in this case it will turn out to be much simpler. We may use the Fierz identity
\begin{equation}
 \delta_{ij}\delta_{kl}=\frac{1}{N}\delta_{ik}\delta_{lj}+2T^a_{ik}T^a_{lj}\equiv(P_S)_{ik}(P_S)^*_{jl}+(P_A)^a_{ik}(P_A)^{a\,*}_{jl}\,,
 \label{identity}
\end{equation}
where the first part can be understood as a projector on the color singlet space with ${(P_S)_{ik}=\delta_{ik}/\sqrt{N}}$ and the second part as a projector on the color adjoint space with ${(P_A)^a_{ik}=\sqrt{2}T^a_{ik}}$. As projectors they satisfy
\begin{equation}
 (P_R)^{a\,*}_{ik}(P_{R'})^b_{ik}=\delta_{RR'}\delta^{ab}\,,
\end{equation}
where the representation indices $R$ and $R'$ can stand for either singlet $S$ or adjoint $A$, and the color indices $a$ and $b$ are absent for the singlet or run from $1$ to $N^2-1$ for the adjoint projector.

With these projectors we can split any amplitude $\mathcal{A}$ like
\begin{equation}
 \mathcal{A}_{ij,\,kl}=(P_R)^a_{ik}(P_R)^{a\,*}_{i'k'}\mathcal{A}_{i'j',\,k'l'}(P_{R'})^b_{j'l'}(P_{R'})^{b\,*}_{jl}\equiv(P_R)^a_{ik}\mathcal{A}^{ab}_{RR'}(P_{R'})^{b\,*}_{jl}\,,
\end{equation}
and because of the orthogonality of the projectors the exponential of $\mathcal{A}$ can be expressed as
\begin{equation}
 \exp[\mathcal{A}]_{ij,\,kl}=\exp[P_R^a\mathcal{A}^{ab}_{RR'}P^{b\,*}_{R'}]_{ij,\,kl}=(P_R)^a_{ik}\exp[\mathcal{A}]^{ab}_{RR'}(P_{R'})^{b\,*}_{jl}\,.
\end{equation}
This amounts to a basis transformation for the amplitudes; the matrix exponential with the new indices $R$, $R'$ and $a$, $b$ still has $N^2\times N^2$ elements. But through the specific nature of the Feynman diagrams the exponential in this basis will be greatly simplified.

All color coefficients $\widetilde{\mathcal{C}}$ can be expressed as linear combinations of products of color matrices with all color indices contracted. We can use the Fierz identity~\eqref{identity} to show that any two fundamental color matrices with their color indices contracted can be expressed entirely through Kronecker deltas, hence we can write any color coefficient as
\begin{equation}
 \widetilde{\mathcal{C}}_{ij,\,kl}=c_1\delta_{ij}\delta_{kl}+c_2\delta_{ik}\delta_{jl}\,.
\end{equation}
With this and the other properties of the fundamental color matrices, $\mathrm{Tr}[T^a]=0$ and $\mathrm{Tr}[T^aT^b]=\delta^{ab}/2$,
it is straightforward to see that the projected color coefficients satisfy
\begin{equation}
 \widetilde{\mathcal{C}}_{RR'}^{ab}=(P_R)^{a\,*}_{ik}\widetilde{\mathcal{C}}_{ij,\,kl}(P_{R'})^b_{jl}=\widetilde{\mathcal{C}}_R\delta_{RR'}\delta^{ab}\,.
 \label{diagonal}
\end{equation}
Note that, contrary to our usual convention of summing over all repeated indices, in the last expression of this equation there is no summation over $R$ implied.

This means that $\ln\langle\mathcal{M}\rangle$ is diagonal in this projection and exponentiation is trivial:
\begin{align}
 \exp\left[-\frac{F_{Q\bar{Q}}}{T}\right]&=\frac{\delta_{ij}\delta_{kl}}{N^2}\exp\biggl[\sum_\mathcal{D}\widetilde{\mathcal{C}}(\mathcal{D})\mathcal{K}(\mathcal{D})\biggr]_{ij,\,kl}\notag\\
 &=\frac{\delta_{ij}\delta_{kl}}{N^2}(P_R)_{ik}^a\exp\biggl[\sum_\mathcal{D}\widetilde{\mathcal{C}}(\mathcal{D})\mathcal{K}(\mathcal{D})\biggr]^{ab}_{RR'}(P_{R'})^{b\,*}_{jl}\notag\\
 &=\frac{\delta_{ij}\delta_{kl}}{N^2}\biggl((P_S)_{ik}\exp\biggl[\sum_\mathcal{D}\widetilde{\mathcal{C}}_S(\mathcal{D})\mathcal{K}(\mathcal{D})\biggr](P_S)^*_{jl}\notag\\
 &\phantom{=\frac{\delta_{ij}\delta_{kl}}{N^2}}+(P_A)^a_{ik}\exp\biggl[\sum_\mathcal{D}\widetilde{\mathcal{C}}_A(\mathcal{D})\mathcal{K}(\mathcal{D})\biggr](P_A)^{a\,*}_{jl}\biggr)\notag\\
 &=\frac{1}{N^2}\exp\biggl[\sum_\mathcal{D}\widetilde{\mathcal{C}}_S(\mathcal{D})\mathcal{K}(\mathcal{D})\biggr]+\frac{N^2-1}{N^2}\exp\biggl[\sum_\mathcal{D}\widetilde{\mathcal{C}}_A(\mathcal{D})\mathcal{K}(\mathcal{D})\biggr]\notag\\
 &\equiv\frac{1}{N^2}\exp\left[-\frac{F_S}{T}\right]+\frac{N^2-1}{N^2}\exp\left[-\frac{F_A}{T}\right]\,.
\end{align}
In this way, we have split the free energy of a static quark-antiquark pair into a singlet and an adjoint free energy, which can also be defined directly as
\begin{align}
 \exp\left[-\frac{F_S}{T}\right]&=\frac{1}{N}\left\langle\mathrm{Tr}\left[L(\bm{r})L^\dagger(\bm{0})\right]\right\rangle\,,\\
 \exp\left[-\frac{F_A}{T}\right]&=\frac{2}{N^2-1}\left\langle\mathrm{Tr}\left[L(\bm{r})T^aL^\dagger(\bm{0})T^a\right]\right\rangle\,.
\end{align}

This procedure can be easily generalized to similar correlators of Polyakov loops in different representations or with more than two loops. For example, in a diquark Polyakov loop correlator (i.e., a correlator of two Polyakov loops without complex conjugation) one has antitriplet and sextet projectors [or rather $N(N-1)/2$ and $N(N+1)/2$ projectors for general $N$], which add up to a unit operator in a similar fashion as in Eq.~\eqref{identity}, and the projected color coefficients are still diagonal as in Eq.~\eqref{diagonal}. This gives an analogous definition of antitriplet and sextet free energies.

In the case of a baryonic Polyakov loop correlator (consisting of three Polyakov loops with $N=3$), one has one singlet, two octet, and one decuplet projector, but the projected color coefficients are no longer fully diagonal, for the two octet representations can mix. As a consequence, Eq.~\eqref{diagonal} has to be modified into
\begin{equation}
 \widetilde{\mathcal{C}}_{RR'}^{ab}=\widetilde{\mathcal{C}}_{RR'}\delta_{d(R)d(R')}\delta^{ab}\,,
\end{equation}
where $d(R)$ is the dimension of the representation $R$. The reason for this is that the baryonic projectors with mixed symmetries are only orthogonal to each other, if the indices are contracted in the right order: $\left(P_8\right)_{ikm}^{a\,*}\left(P_{8'}\right)_{ikm}^b=0$, but, e.g., $\left(P_8\right)_{ikm}^{a\,*}\left(P_{8'}\right)_{imk}^b\propto\delta^{ab}$. The exponentiated color factors contain terms that change the order in which the indices of the projectors are contracted, so they are no longer diagonal in the two octet channels. The singlet or decuplet projectors are fully (anti)symmetric in their indices, so a different order of the contracted indices does not matter and the projections are still diagonal. In fact, this generalization of Eq.~\eqref{diagonal} also applies to the diquark or quark-antiquark Polyakov loop correlators, therefore it may be true for any combination of representations and loops, although we will not attempt a proof in this paper.

In any case, this projection of the amplitudes in the baryonic Polyakov loop correlator then defines a singlet and a decuplet free energy through simple exponentials and two octet free energies through the trace of the exponential of a $2\times2$ matrix:
\begin{equation}
 \exp\left[-\frac{F_{3Q}}{T}\right] = 
\frac{1}{27}\exp\left[-\frac{F_1}{T}\right]+\frac{8}{27}\mathrm{Tr}\left\{\exp\left[-\frac{1}{T}\begin{pmatrix} F_{88} & F_{88'} \\ F_{8'8} & F_{8'8'} \end{pmatrix}\right]\right\}
+\frac{10}{27}\exp\left[-\frac{F_{10}}{T}\right]\,.
\end{equation}
The same structure, in particular the mixing of the two octet channels, has also been found in the context of a direct NLO calculation of the static potentials in a baryonic configuration in Ref.~\cite{Brambilla:2009cd}.

There are, however, two major problems related to the definition of singlet, adjoint, or other free energies such as these. First, the definition is gauge dependent, and second, each of these free energies contains ultraviolet divergences, which cancel in the full expression of the Polyakov loop correlator.

We will discuss the divergences in more detail (and return to the quark-antiquark case). There are two types of divergences, the first is a linear divergence proportional to the length of a Wilson line, in this case $1/T$, and can be understood as a mass correction to the (infinite) mass of the static quark. It factorizes (cf.~\cite{Berwein:2013xza}), which means that it affects singlet and adjoint free energies in the same way, and can be removed by multiplication with $\exp[-2\Lambda_F/T]$, where $\Lambda_F$ is a divergent constant and the index $F$ refers to the fundamental representation. In dimensional regularization such a divergence is absent.

The second kind of divergence is logarithmic and comes from gluons clustering around the endpoints of a Polyakov loop~\cite{Polyakov:1980ca,Dotsenko:1979wb,Brandt:1981kf}. All gluons contributing to this divergence have to be contained in an infinitesimal area around the endpoints, which means that the divergence does not depend on any characteristics of the Wilson line like length or curvature, except for when two or more endpoints coincide (i.e., at cusps or intersections), in which case the divergence also depends on the angles at this point. Such a divergent cluster can be added to any Feynman diagram and will factorize from the sum over all diagrams (before taking any traces), hence the divergence of the correlator is proportional to the correlator itself. Keeping in mind that the divergences at the endpoints of the two Polyakov loops are unrelated, we can write
\begin{equation}
 \mathrm{Div}\langle\mathcal{M}\rangle_{ij,\,kl}=\Delta_{ii',\,jj'}\langle\mathcal{M}\rangle_{i'j',\,kl}+\langle\mathcal{M}\rangle_{ij,\,k'l'}\Delta_{ll',\,kk'}-\Delta_{ii',\,jj'}\langle\mathcal{M}\rangle_{i'j',\,k'l'}\Delta_{ll',\,kk'}\,,
\end{equation}
where we have used the fact that both Polyakov loops have exactly the same configuration at their endpoints, since a Wilson line with final endpoint $k$ and initial endpoint $l$ in the antifundamental representation is equivalent to a Wilson line with final endpoint $l$ and initial endpoint $k$ in the fundamental representation. Accordingly the divergences $\Delta$ have to be identical. The last term is there to remove a double counting of terms with divergences at both Polyakov loops.

Then we define the renormalized correlator through the subtraction of the divergent part:
\begin{align}
 \langle\mathcal{M}\rangle^{(R)}_{ij,\,kl}&=
\langle\mathcal{M}\rangle_{ij,\,kl}-\mathrm{Div}\langle\mathcal{M}\rangle_{ij,\,kl}=(\delta_{ii'}\delta_{jj'}-\Delta_{ii',\,jj'})\langle\mathcal{M}\rangle_{i'j',\,k'l'}(\delta_{ll'}\delta_{kk'}-\Delta_{ll',\,kk'})\notag\\
 &\equiv Z_{ii',\,jj'}\langle\mathcal{M}\rangle_{i'j',\,k'l'}Z_{ll',\,kk'}\,.
 \label{renampl}
\end{align}
Again, we can use the Fierz identity~\eqref{identity} to argue that
\begin{equation}
 Z_{ii',\,jj'}=z_1\delta_{ii'}\delta_{jj'}+z_2\delta_{ij}\delta_{i'j'}\,.
 \label{renconst}
\end{equation}
Of course, we can multiply the renormalization tensors $Z_{ii',\,jj'}$ by some finite tensor, which corresponds to a different renormalization scheme. If we take the traces over the Polyakov loops, then the contour is smooth at their endpoints, which means that there are no logarithmic divergences~\cite{Polyakov:1980ca,Dotsenko:1979wb}. Therefore we can partially fix the renormalization scheme by requiring the renormalized Polyakov loop correlator to be identical to the unrenormalized one with respect to the logarithmic divergences: $\delta_{ij}Z_{ii',\,jj'}=\delta_{i'j'}$. From this it follows that
\begin{equation}
 z_1+Nz_2=1\,.
\end{equation}

If we now use the same projectors for the renormalized singlet and adjoint free energies as for the unrenormalized ones, then we have
\begin{align}
 \exp\left[-\frac{F^{(R)}_S}{T}\right]={}&(P_S)^*_{ik}\langle\mathcal{M}\rangle^{(R)}_{ij,\,kl}(P_S)_{jl}=(P_S)^*_{ik}Z_{ii',\,jj'}\langle\mathcal{M}\rangle_{i'j',\,k'l'}Z_{ll',\,kk'}(P_S)_{jl}\notag\\
 ={}&(P_S)^*_{ik}Z_{ii',\,jj'}\left((P_S)_{i'k'}\exp\left[-\frac{F_S}{T}\right](P_S)^*_{j'l'}\right.\notag\\
 &+\left.(P_A)^a_{i'k'}\exp\left[-\frac{F_A}{T}\right](P_A)^{a\,*}_{j'l'}\right)Z_{ll',\,kk'}(P_S)_{jl}\notag\\
 ={}&\frac{1+(N^2-1)z_1^2}{N^2}\exp\left[-\frac{F_S}{T}\right]+\frac{(N^2-1)(1-z_1^2)}{N^2}\exp\left[-\frac{F_A}{T}\right]\notag\\
 \equiv{}&(1-Z_S)\exp\left[-\frac{F_S}{T}\right]+Z_S\exp\left[-\frac{F_A}{T}\right]\,,\\
 \exp\left[-\frac{F^{(R)}_A}{T}\right]={}&\frac{1}{N^2-1}(P_A)^{a\,*}_{ik}\langle\mathcal{M}\rangle^{(R)}_{ij,\,kl}(P_A)^a_{jl}=\frac{1}{N^2-1}(P_A)^{a\,*}_{ik}Z_{ii',\,jj'}\langle\mathcal{M}\rangle_{i'j',\,k'l'}Z_{ll',\,kk'}(P_A)^a_{jl}\notag\\
 ={}&\frac{1}{N^2-1}(P_A)^{a\,*}_{ik}Z_{ii',\,jj'}\left((P_S)_{i'k'}\exp\left[-\frac{F_S}{T}\right](P_S)^*_{j'l'}\right.\notag\\
 &+\left.(P_A)^b_{i'k'}\exp\left[-\frac{F_A}{T}\right](P_A)^{b\,*}_{j'l'}\right)Z_{ll',\,kk'}(P_A)^a_{jl}\notag\\
 ={}&\frac{1-z_1^2}{N^2}\exp\left[-\frac{F_S}{T}\right]+\frac{N^2-1+z_1^2}{N^2}\exp\left[-\frac{F_A}{T}\right]\notag\\
 \equiv{}&Z_A\exp\left[-\frac{F_S}{T}\right]+(1-Z_A)\exp\left[-\frac{F_A}{T}\right]\,,
\end{align}
where we have introduced the renormalization constants
\begin{equation}
 Z_S=(N^2-1)Z_A=\frac{N^2-1}{N^2}(1-z_1^2)\,,
\end{equation}
such that $Z_S,\,Z_A=\mathcal{O}(\alpha_\mathrm{s})$. We see, therefore, that the singlet and adjoint free energies mix under renormalization. These relations can also be inverted as
\begin{align}
 \exp\left[-\frac{F_S}{T}\right]=\left(1-\widetilde{Z}_S\right)\exp\left[-\frac{F^{(R)}_S}{T}\right]+\widetilde{Z}_S\exp\left[-\frac{F^{(R)}_A}{T}\right]\,,\\
 \exp\left[-\frac{F_A}{T}\right]=\widetilde{Z}_A\exp\left[-\frac{F^{(R)}_S}{T}\right]+\left(1-\widetilde{Z}_A\right)\exp\left[-\frac{F^{(R)}_A}{T}\right]\,,
\end{align}
with
\begin{equation}
 \widetilde{Z}_S=(N^2-1)\widetilde{Z}_A=\frac{1-Z_S}{\dfrac{N^2}{N^2-1}(1-Z_S)-1}=\frac{N^2-1}{N^2}\frac{z_1^2-1}{z_1^2}\,.
\end{equation}
We also see that we can construct a multiplicatively renormalizable quantity through
\begin{align}
 \exp\left[-\frac{F^{(R)}_S}{T}\right]-\exp\left[-\frac{F^{(R)}_A}{T}\right]&=(1-Z_S-Z_A)\left(\exp\left[-\frac{F_S}{T}\right]-\exp\left[-\frac{F_A}{T}\right]\right)\notag\\
 &=z_1^2\left(\exp\left[-\frac{F_S}{T}\right]-\exp\left[-\frac{F_A}{T}\right]\right)\,.
\end{align}

\section{Calculation of the normalized Polyakov loop correlator}
\label{sec:polycorr}

The great advantage of exponentiated formulas, such as those that were derived in the previous section, is that they reduce the number of Feynman diagrams that one has to calculate at a given order in perturbation theory, since many of the color coefficients in the exponent are zero. We will show this explicitly for the two-gluon diagrams.

First, all diagrams where no gluons are exchanged between the two loops have color coefficients that are proportional to the identity $\delta_{ij}\delta_{kl}$, therefore they trivially factorize out of the exponentiation. They give a contribution that corresponds to the individual contributions of each Polyakov loop, i.e., $\exp[-2F_Q/T]$. Hence, it makes sense to divide the Polyakov loop correlator by these two Polyakov loops, which corresponds to calculating $F_{Q\bar{Q}}-2F_Q$ and can be interpreted as the interaction part of the correlator, because it contains only those diagrams where gluons are exchanged between the loops. We call this ratio the normalized Polyakov loop correlator.

\begin{figure}[t]
 \includegraphics[width=0.8\linewidth]{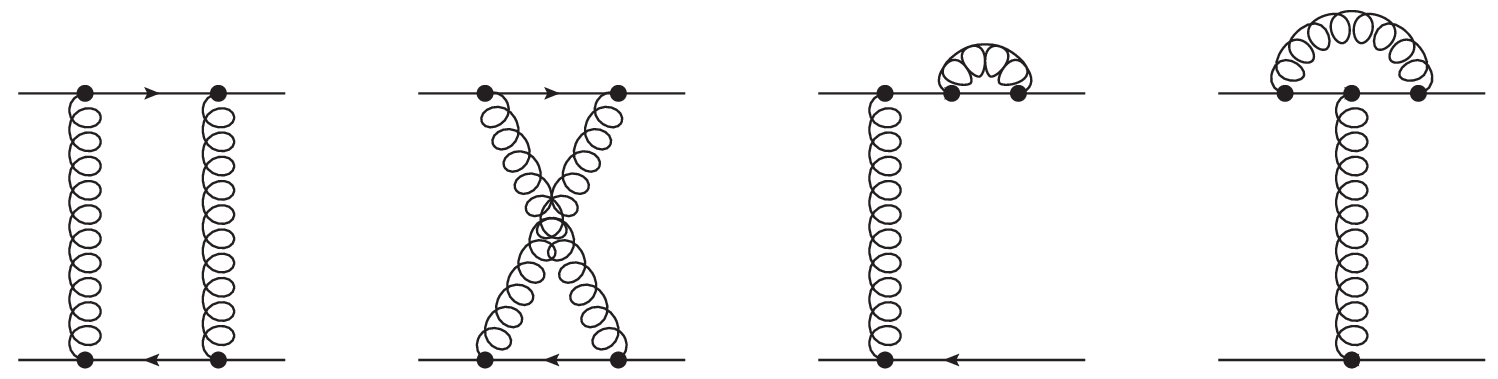}
 \caption{All unconnected two-gluon diagrams. Some diagrams can be flipped to give in total four other diagrams that are not explicitly displayed. The line with the right arrow (omitted in the last two diagrams) is the Polyakov loop contour for the quark, and the line with the left arrow corresponds to the antiquark.}
 \label{Fig1}
\end{figure}

For connected diagrams, i.e., diagrams where every gluon is connected to every other gluon through vertices or propagators, the color coefficient in the exponent is the same as the standard coefficient in QCD. The first diagrams for which the modification of the color coefficients obtained from the replica trick becomes relevant are the two-gluon diagrams shown in Fig.~\ref{Fig1}. For each diagram we have to sum over every possible assignment of replica indices and perform the corresponding replica path ordering:
\begin{align}
 \sum_{\{\rho\}}\mathcal{C}^{\{\rho\}}\left(\begin{minipage}{20pt}\includegraphics[width=\linewidth]{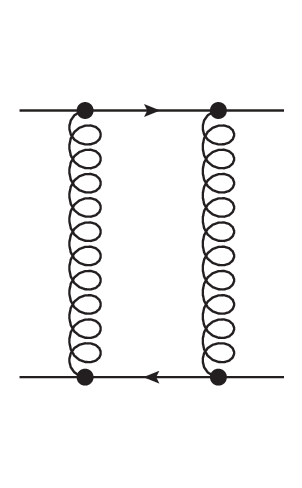}\end{minipage}\right)&=n(n-1)\,\mathcal{C}\left(\begin{minipage}{20pt}\includegraphics[width=\linewidth]{Fig1II.eps}\end{minipage}\right)+n\,\mathcal{C}\left(\begin{minipage}{20pt}\includegraphics[width=\linewidth]{Fig1II.eps}\end{minipage}\right)\,,\notag\\
 \sum_{\{\rho\}}\mathcal{C}^{\{\rho\}}\left(\begin{minipage}{20pt}\includegraphics[width=\linewidth]{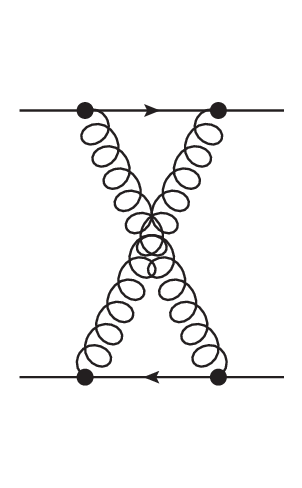}\end{minipage}\right)&=n(n-1)\,\mathcal{C}\left(\begin{minipage}{20pt}\includegraphics[width=\linewidth]{Fig1II.eps}\end{minipage}\right)+n\,\mathcal{C}\left(\begin{minipage}{20pt}\includegraphics[width=\linewidth]{Fig1X.eps}\end{minipage}\right)\,,\notag\\
 \sum_{\{\rho\}}\mathcal{C}^{\{\rho\}}\left(\begin{minipage}{20pt}\includegraphics[width=\linewidth]{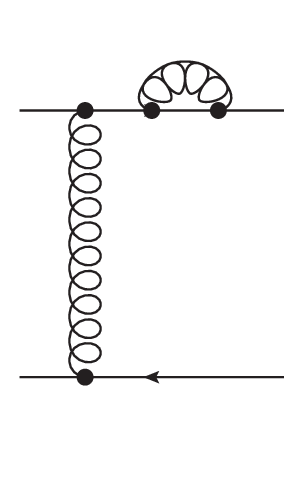}\end{minipage}\right)&=\frac{n(n-1)}{2}\,\mathcal{C}\left(\begin{minipage}{20pt}\includegraphics[width=\linewidth]{Fig1r.eps}\end{minipage}\right)+\frac{n(n-1)}{2}\,\mathcal{C}\left(\begin{minipage}{20pt}\includegraphics[width=\linewidth]{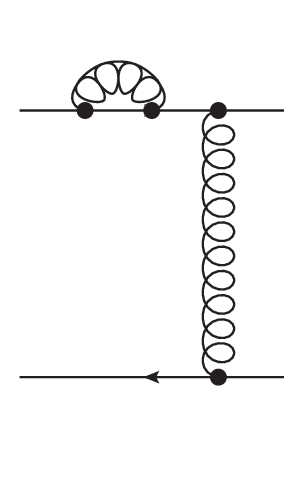}\end{minipage}\right)+n\,\mathcal{C}\left(\begin{minipage}{20pt}\includegraphics[width=\linewidth]{Fig1r.eps}\end{minipage}\right)\,,\notag\\
 \sum_{\{\rho\}}\mathcal{C}^{\{\rho\}}\left(\begin{minipage}{20pt}\includegraphics[width=\linewidth]{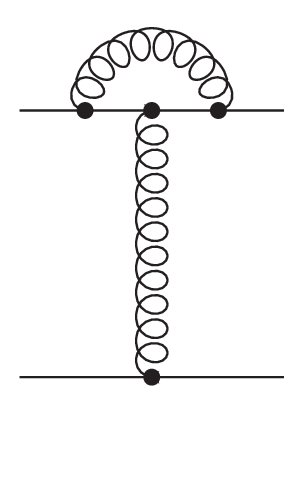}\end{minipage}\right)&=\frac{n(n-1)}{2}\,\mathcal{C}\left(\begin{minipage}{20pt}\includegraphics[width=\linewidth]{Fig1r.eps}\end{minipage}\right)+\frac{n(n-1)}{2}\,\mathcal{C}\left(\begin{minipage}{20pt}\includegraphics[width=\linewidth]{Fig1r2.eps}\end{minipage}\right)+n\,\mathcal{C}\left(\begin{minipage}{20pt}\includegraphics[width=\linewidth]{Fig1T.eps}\end{minipage}\right)\,,
\end{align}
where the first term counts the possibilities of having two gluons with different replica indices and the second term counts the possibilities of them having the same replica index. For the latter two diagrams, the first term is split into the possibilities of one gluon having a higher or a lower index than the other gluon, a distinction that is in fact unnecessary, because both orderings have the same standard color coefficient.

We see that for the first diagram the terms linear in $n$ cancel trivially, as a consequence this diagram does not contribute to the logarithm of the Polyakov loop correlator. For the third diagram it is straightforward to see that both standard color coefficients are equal, since the gluon attached only to the top Polyakov loop contributes with a unit matrix to the color coefficient, because $(T^aT^a)_{ij}=\delta_{ij}(N^2-1)/2N$, therefore also here the linear order of $n$ cancels. 

These two diagrams are the first examples of a more general statement: whenever one can draw a line cutting the upper and lower Polyakov loop such that there are gluons on both sides of it but no gluon crosses the line, then this diagram does not contribute to the logarithm of the correlator. This can be shown in the following way. Whenever it is possible to draw such a line, then the color coefficient $\mathcal{C}$ can be written as a product of two coefficients $\mathcal{A}$ and $\mathcal{B}$, one for the left and one for the right part. The statement that each color coefficient can be written through Kronecker deltas applies to both parts separately, so we can write
\begin{align}
 \mathcal{C}_{ij,\,kl}&=\mathcal{A}_{ij',\,kl'}\mathcal{B}_{j'j,\,l'l}=(a_1\delta_{ij'}\delta_{kl'}+a_2\delta_{ik}\delta_{j'l'})(b_1\delta_{j'j}\delta_{l'l}+b_2\delta_{j'l'}\delta_{jl})\notag\\
 &=a_1b_1\delta_{ij}\delta_{kl}+(a_1b_2+a_2b_1+a_2b_2N)\delta_{ik}\delta_{jl}=(b_1\delta_{ij'}\delta_{kl'}+b_2\delta_{ik}\delta_{j'l'})(a_1\delta_{j'j}\delta_{l'l}+a_2\delta_{j'l'}\delta_{jl})\notag\\
 &=\mathcal{B}_{ij',\,kl'}\mathcal{A}_{j'j,\,l'l}\,,
\end{align}
which means that the two parts (and in fact any two color coefficients) commute. But then the replica path ordering and counting of replica indices can be done for each part separately:
\begin{equation}
 \sum_{\{\rho\}}\mathcal{C}^{\{\rho\}}_{ij,\,kl}=\sum_{\{\rho_1\}}\mathcal{A}^{\{\rho_1\}}_{ij',\,kl'}\sum_{\{\rho_2\}}\mathcal{B}^{\{\rho_2\}}_{j'j,\,l'l}\,.
\end{equation}
Since each part is at least of order $n$, the sum over every replica index combination for the whole color coefficient will be at least of order $n^2$.

Using the replica method, it is also straightforward to calculate the projected color coefficients for each of the diagrams that contribute to the logarithm of the correlator. This calculation is presented in Appendix~\ref{app:color}. Putting together all the diagrams and the corresponding color factors for the singlet and adjoint contributions to the Polyakov loop correlator, we get
\begin{align}
 \frac{2F_Q-F_S}{T}=\mathcal{K}&\left\{\frac{N^2-1}{2N}\begin{minipage}{20pt}\includegraphics[width=\linewidth]{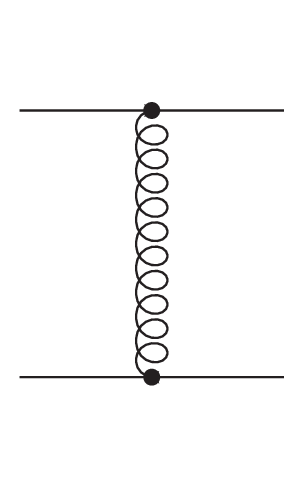}\end{minipage}-\frac{N^2-1}{4}\left(\begin{minipage}{20pt}\includegraphics[width=\linewidth]{Fig1X.eps}\end{minipage}+\begin{minipage}{20pt}\includegraphics[width=\linewidth]{Fig1T.eps}\end{minipage}+\begin{minipage}{20pt}\includegraphics[width=\linewidth]{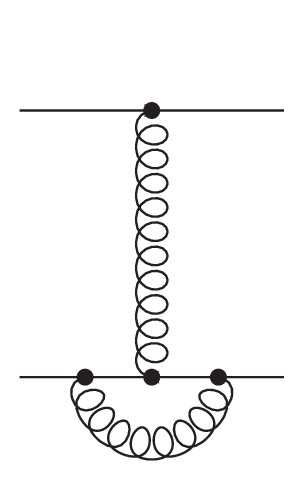}\end{minipage}\right)+\frac{N(N^2-1)}{8}\left(2\begin{minipage}{20pt}\includegraphics[width=\linewidth]{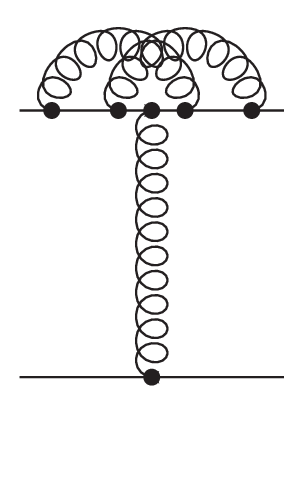}\end{minipage}+2\begin{minipage}{20pt}\includegraphics[width=\linewidth]{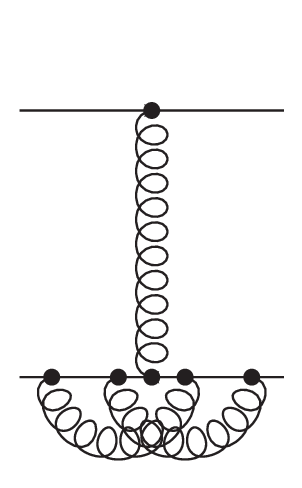}\end{minipage}\right.\right.\notag\\
 &+\begin{minipage}{20pt}\includegraphics[width=\linewidth]{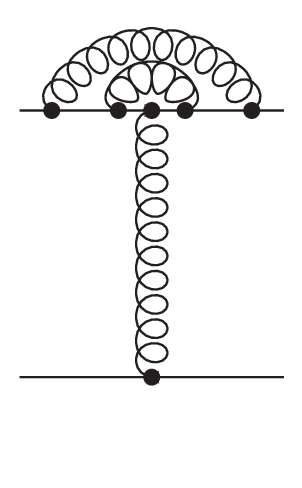}\end{minipage}+\begin{minipage}{20pt}\includegraphics[width=\linewidth]{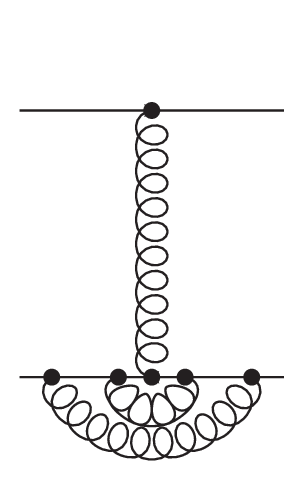}\end{minipage}+\begin{minipage}{20pt}\includegraphics[width=\linewidth]{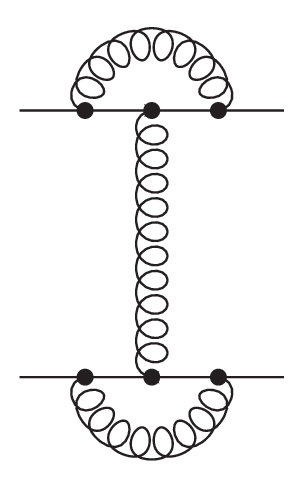}\end{minipage}+\begin{minipage}{20pt}\includegraphics[width=\linewidth]{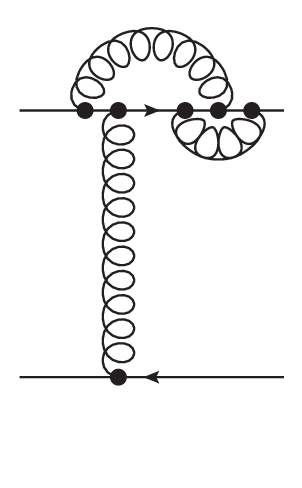}\end{minipage}+\begin{minipage}{20pt}\includegraphics[width=\linewidth]{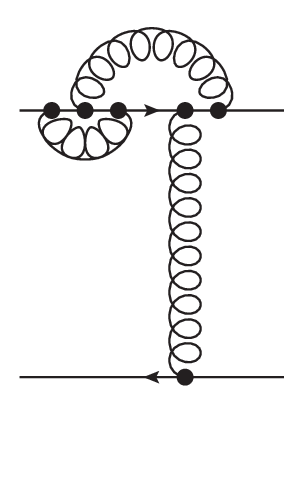}\end{minipage}+\begin{minipage}{20pt}\includegraphics[width=\linewidth]{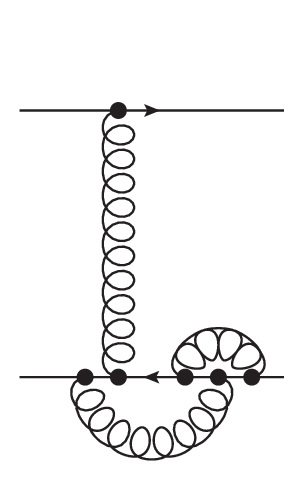}\end{minipage}+\begin{minipage}{20pt}\includegraphics[width=\linewidth]{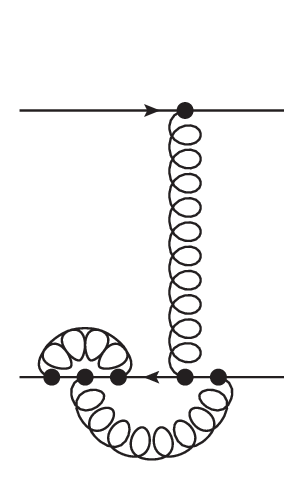}\end{minipage}+\begin{minipage}{20pt}\includegraphics[width=\linewidth]{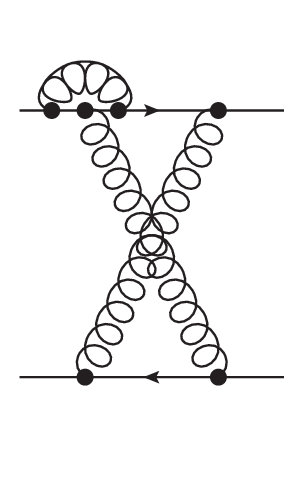}\end{minipage}+\begin{minipage}{20pt}\includegraphics[width=\linewidth]{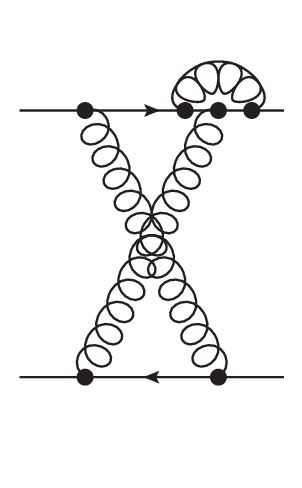}\end{minipage}+\begin{minipage}{20pt}\includegraphics[width=\linewidth]{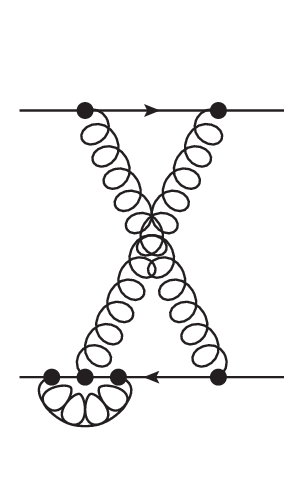}\end{minipage}+\begin{minipage}{20pt}\includegraphics[width=\linewidth]{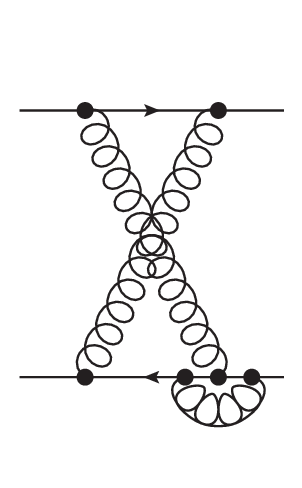}\end{minipage}\notag\\
 &+2\begin{minipage}{20pt}\includegraphics[width=\linewidth]{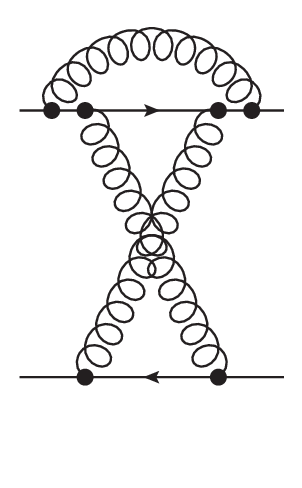}\end{minipage}+2\begin{minipage}{20pt}\includegraphics[width=\linewidth]{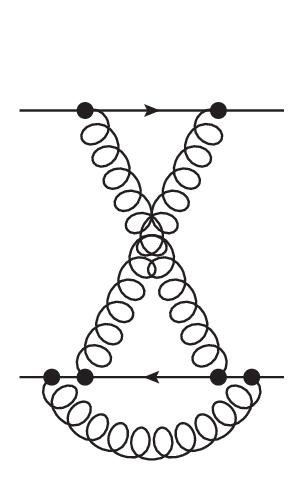}\end{minipage}+\begin{minipage}{20pt}\includegraphics[width=\linewidth]{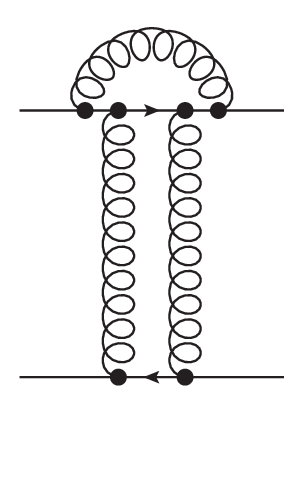}\end{minipage}+\begin{minipage}{20pt}\includegraphics[width=\linewidth]{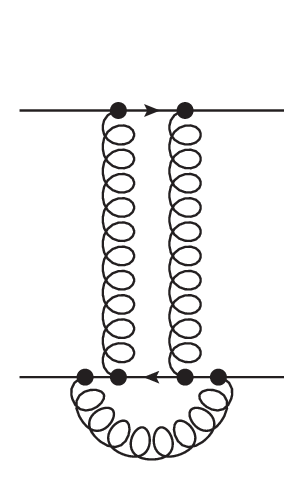}\end{minipage}+\begin{minipage}{20pt}\includegraphics[width=\linewidth]{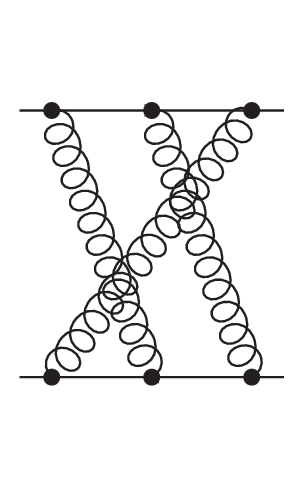}\end{minipage}+\begin{minipage}{20pt}\includegraphics[width=\linewidth]{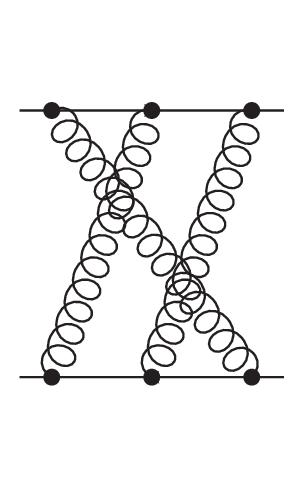}\end{minipage}+2\begin{minipage}{20pt}\includegraphics[width=\linewidth]{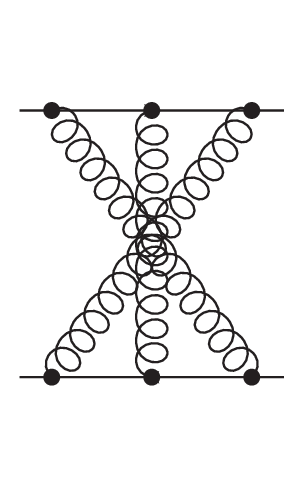}\end{minipage}-\begin{minipage}{20pt}\includegraphics[width=\linewidth]{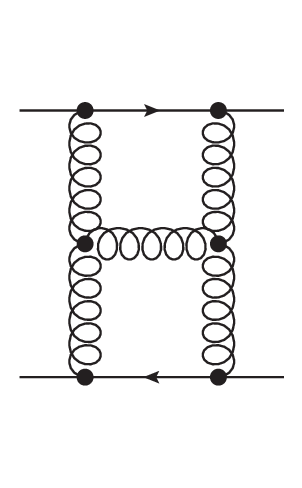}\end{minipage}-\begin{minipage}{20pt}\includegraphics[width=\linewidth]{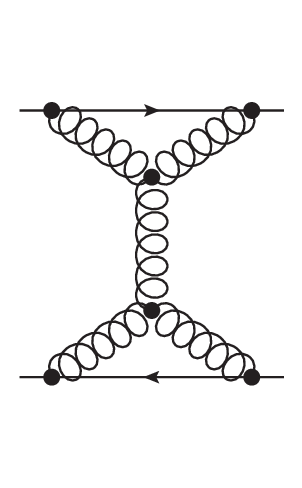}\end{minipage}\notag\\
 &-\left.\left.\begin{minipage}{20pt}\includegraphics[width=\linewidth]{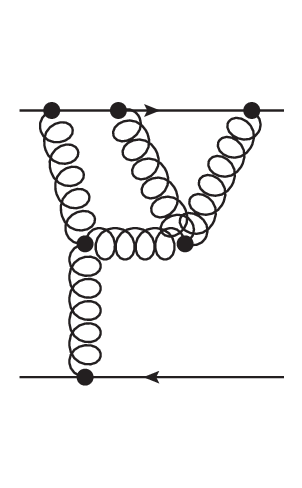}\end{minipage}-\begin{minipage}{20pt}\includegraphics[width=\linewidth]{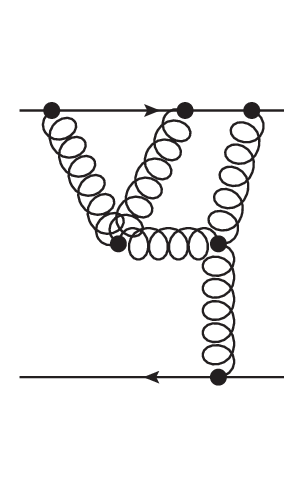}\end{minipage}-\begin{minipage}{20pt}\includegraphics[width=\linewidth]{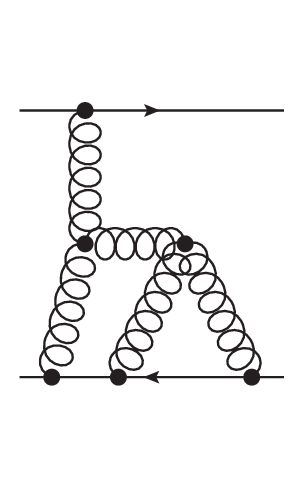}\end{minipage}-\begin{minipage}{20pt}\includegraphics[width=\linewidth]{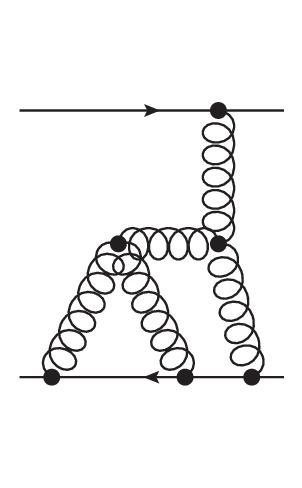}\end{minipage}\right)+\dots\right\}\,, \label{F1exp}\\
 \frac{2F_Q-F_A}{T}=\mathcal{K}&\left\{-\frac{1}{2N}\begin{minipage}{20pt}\includegraphics[width=\linewidth]{FigI.eps}\end{minipage}+\frac{1}{4}\left(\begin{minipage}{20pt}\includegraphics[width=\linewidth]{Fig1X.eps}\end{minipage}+\begin{minipage}{20pt}\includegraphics[width=\linewidth]{Fig1T.eps}\end{minipage}+\begin{minipage}{20pt}\includegraphics[width=\linewidth]{Fig1T2.eps}\end{minipage}\right)-\frac{N}{8}\left(2\begin{minipage}{20pt}\includegraphics[width=\linewidth]{Fig3gTB.eps}\end{minipage}+2\begin{minipage}{20pt}\includegraphics[width=\linewidth]{Fig3gTB2.eps}\end{minipage}+\begin{minipage}{20pt}\includegraphics[width=\linewidth]{Fig3gTT.eps}\end{minipage}+\begin{minipage}{20pt}\includegraphics[width=\linewidth]{Fig3gTT2.eps}\end{minipage}\right.\right.\notag\\
 &+\begin{minipage}{20pt}\includegraphics[width=\linewidth]{Fig3gTT3.eps}\end{minipage}+\begin{minipage}{20pt}\includegraphics[width=\linewidth]{Fig3gTS.eps}\end{minipage}+\begin{minipage}{20pt}\includegraphics[width=\linewidth]{Fig3gTS2.eps}\end{minipage}+\begin{minipage}{20pt}\includegraphics[width=\linewidth]{Fig3gTS3.eps}\end{minipage}+\begin{minipage}{20pt}\includegraphics[width=\linewidth]{Fig3gTS4.eps}\end{minipage}+\begin{minipage}{20pt}\includegraphics[width=\linewidth]{Fig3gXT.eps}\end{minipage}+\begin{minipage}{20pt}\includegraphics[width=\linewidth]{Fig3gXT2.eps}\end{minipage}+\begin{minipage}{20pt}\includegraphics[width=\linewidth]{Fig3gXT3.eps}\end{minipage}+\begin{minipage}{20pt}\includegraphics[width=\linewidth]{Fig3gXT4.eps}\end{minipage}-\begin{minipage}{20pt}\includegraphics[width=\linewidth]{Fig3gIT.eps}\end{minipage}-\begin{minipage}{20pt}\includegraphics[width=\linewidth]{Fig3gIT2.eps}\end{minipage}\notag\\
 &-\left.\left.\begin{minipage}{20pt}\includegraphics[width=\linewidth]{Fig3gIX.eps}\end{minipage}-\begin{minipage}{20pt}\includegraphics[width=\linewidth]{Fig3gIX2.eps}\end{minipage}+\begin{minipage}{20pt}\includegraphics[width=\linewidth]{Fig3gH.eps}\end{minipage}+2\begin{minipage}{20pt}\includegraphics[width=\linewidth]{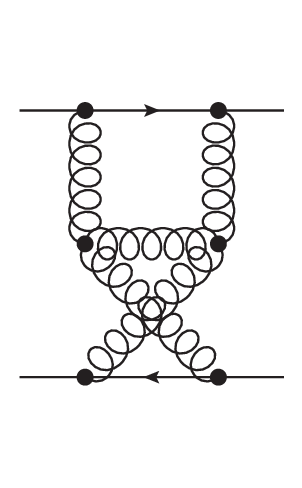}\end{minipage}-\begin{minipage}{20pt}\includegraphics[width=\linewidth]{Fig3gH3.eps}\end{minipage}-\begin{minipage}{20pt}\includegraphics[width=\linewidth]{Fig3gYB.eps}\end{minipage}-\begin{minipage}{20pt}\includegraphics[width=\linewidth]{Fig3gYB2.eps}\end{minipage}-\begin{minipage}{20pt}\includegraphics[width=\linewidth]{Fig3gYB3.eps}\end{minipage}-\begin{minipage}{20pt}\includegraphics[width=\linewidth]{Fig3gYB4.eps}\end{minipage}\right)+\dots\right\}\,,\label{FAexp}
\end{align}
where the dots include four-gluon diagrams and higher, and $\mathcal{K}$ denotes that all diagrams contribute with their kinematic part only, since the color factors are already written explicitly.

We have not drawn explicitly diagrams that differ from those shown above only by gluon self-energy insertions. Nevertheless, they are understood and contribute to the free energies. Their contributions will be computed by simply adding to the gluons in Eqs.~\eqref{F1exp} and~\eqref{FAexp} self energies whenever necessary to reach the desired accuracy: a first example is in~Sec.~\ref{secDI}. 

We have also neglected the several diagrams that vanish trivially in gauges where the gluon propagator is diagonal with respect to temporal and spatial components, such as Coulomb gauge, static gauge, or Feynman gauge. At the present order, there are 22 of such diagrams that vanish because a three-gluon vertex with three temporal indices gives zero, and 3 of such diagrams that vanish because a four-gluon vertex with four temporal indices gives zero.

The reexponentiation of the singlet contribution is analogous to the reexponentiation of the Wilson loop, while the reexponentiation of the adjoint contribution is a new result. From Eqs.~\eqref{F1exp} and~\eqref{FAexp}, we see that Casimir scaling for the singlet and adjoint free energies, i.e., the relation 
\begin{equation}
 \frac{F_S-2 F_Q}{F_A-2 F_Q}=-(N^2-1)\,,
 \label{Casimir}
\end{equation}
is broken at the order $\alpha_\mathrm{s}^3$.

We are interested in calculating the Polyakov loop correlator in the regime $\alpha_\mathrm{s}/(rT)\ll1$. In this regime, the exponentials of the singlet and adjoint contributions can be expanded and one finds that the contributions of many diagrams cancel out. As the result for the normalized Polyakov loop correlator, we get
\begin{align}
 &\exp\left[\frac{2F_Q-F_{Q\bar{Q}}}{T}\right]=\frac{1}{N^2}\exp\left[\frac{2F_Q-F_S}{T}\right]+\frac{N^2-1}{N^2}\exp\left[\frac{2F_Q-F_A}{T}\right]\notag\\
 &=1+\frac{N^2-1}{8N^2}\mathcal{K}^2\left(\begin{minipage}{20pt}\includegraphics[width=\linewidth]{FigI.eps}\end{minipage}\right)+\frac{\left(N^2-1\right)\left(N^2-2\right)}{48N^3}\mathcal{K}^3\left(\begin{minipage}{20pt}\includegraphics[width=\linewidth]{FigI.eps}\end{minipage}\right)\notag\\
 &\phantom{=}+\frac{N^2-1}{4N}\mathcal{K}\left(\begin{minipage}{20pt}\includegraphics[width=\linewidth]{Fig3gXT5.eps}\end{minipage}+\begin{minipage}{20pt}\includegraphics[width=\linewidth]{Fig3gXT6.eps}\end{minipage}+\begin{minipage}{20pt}\includegraphics[width=\linewidth]{Fig3gIT.eps}\end{minipage}+\begin{minipage}{20pt}\includegraphics[width=\linewidth]{Fig3gIT2.eps}\end{minipage}+\begin{minipage}{20pt}\includegraphics[width=\linewidth]{Fig3gIX.eps}\end{minipage}+\begin{minipage}{20pt}\includegraphics[width=\linewidth]{Fig3gIX2.eps}\end{minipage}+\begin{minipage}{20pt}\includegraphics[width=\linewidth]{Fig3gXX.eps}\end{minipage}-\begin{minipage}{20pt}\includegraphics[width=\linewidth]{Fig3gH.eps}\end{minipage}-\begin{minipage}{20pt}\includegraphics[width=\linewidth]{Fig3gH2.eps}\end{minipage}\right)\notag\\
 &\phantom{=}-\frac{N^2-1}{8N}\mathcal{K}\left(\begin{minipage}{20pt}\includegraphics[width=\linewidth]{FigI.eps}\end{minipage}\right)\mathcal{K}\left(\begin{minipage}{20pt}\includegraphics[width=\linewidth]{Fig1X.eps}\end{minipage}+\begin{minipage}{20pt}\includegraphics[width=\linewidth]{Fig1T.eps}\end{minipage}+\begin{minipage}{20pt}\includegraphics[width=\linewidth]{Fig1T2.eps}\end{minipage}\right)+\mathcal{O}\left(\alpha_\mathrm{s}^4\right)\,.
 \label{PLCDiag}
\end{align}
In order to obtain the weak coupling expansion of the Polyakov loop correlator, we need to evaluate the kinematic part of the diagrams entering the above equation. As we will see below, the evaluation of the kinematic parts becomes particularly simple in Coulomb gauge.

We will perform the calculations assuming two different scale hierarchies:
\begin{equation}
 \frac{1}{r}\gg\pi T\gg m_D\gg \frac{\alpha_\mathrm{s}}{r}\,,
 \label{hierarchy1}
\end{equation}
or
\begin{equation}
 \frac{1}{r}\sim m_D\,,
 \label{hierarchy2}
\end{equation}
where
\begin{equation}
 m_D(\mu)=\sqrt{\frac{N}{3}+\frac{n_f}{6}}g(\mu)T
\end{equation}
is the leading order Debye mass. We consider QCD with $n_f$ massless quarks.

We start the discussion with the case $1/r\gg\pi T\gg m_D\gg\alpha_\mathrm{s}/r$. Here, the sum of the unconnected two-gluon diagrams in the last line of Eq.~\eqref{PLCDiag}, which we denote as $D_X+2 D_T$, as well as the sum of all unconnected gluon diagrams appearing in the previous line of Eq.~\eqref{PLCDiag}, which we denote by $D_{3g}$, vanish in Coulomb gauge if the gluon propagators are taken without self-energy insertions. This is discussed in Appendix~\ref{app:unconn}. Therefore, in order to calculate the Polyakov loop correlator, we have to calculate the one-gluon exchange diagram $D_I$ and the last two H-shaped diagrams in the third line of Eq.~\eqref{PLCDiag}, which we denote by $D_H$.

The tree level result for $D_I$ is of order $g^2$, so the first nontrivial contribution (i.e., different from 1) to the Polyakov loop correlator is of order $g^4$, which is what we will call the leading order (LO). Since the Debye mass introduces odd powers of $g$ in the perturbative expansion, the NLO and NNLO contributions are of orders $g^5$ and $g^6$ respectively. Accordingly, the order $g^7$, which we calculate here for the first time, will be counted as NNNLO.

The kinematic parts of the diagrams will be determined through the method of integration by regions. This means that the integration over each gluon momentum is split into regions where the momentum scales as one of the relevant physical scales of the system. In this case, we have the inverse distance $1/r$ between the two Polyakov loops, the temperature scale $\pi T$, and the Debye mass scale $m_D$. In each region, the integrand is expanded according to the hierarchy~\eqref{hierarchy1}. Depending on the scale of the gluon momentum, the propagator can either be free or resummed. In the following subsection, we will discuss the evaluation of the diagrams $D_I$ and $D_H$ using this method.

The magnetic mass scale $m_M\sim g^2T$ is also present, but does not enter the calculation at this order. It has been shown in the context of the effective field theories (EFTs) EQCD and MQCD, which systematically incorporate the scale separation between $\pi T$, $m_D$, and $m_M$, that the magnetic scale enters the Polyakov loop only at order $g^7$~\cite{Braaten:1994qx}, even though it appears already at orders $g^5$ and $g^6$ in individual diagrams but gives canceling contributions~\cite{Berwein:2015ayt}. Since the dynamics of the magnetic scale take place at length scales much larger than those associated with the energy scales in Eq.~\eqref{hierarchy1}, we can expect a similar EFT argument to apply for the singlet and adjoint correlators, excluding the magnetic scale from entering the free energies until order $g^7$. For the Polyakov loop correlator itself, we expect the magnetic scale to be absent until order $g^9$, as it should enter through $D_I$, which contributes only quadratically and therefore raises the nonperturbative order by $g^2$. We have checked this explicitly in Appendix~\ref{scalemM}, showing that all magnetic scale contributions cancel up to order $g^8$ indeed in both hierarchies.

\subsection{\texorpdfstring{Calculation of $\bm{D_I}$ for $\bm{1/r\gg\pi T\gg m_D}$}{Calculation of D\_I for 1/r>>piT>>m\_D}}
\label{secDI}

We will start with the calculation of $D_I$:
\begin{equation}
 D_I=(ig)^2\int_0^{1/T}d\tau_1\int_{1/T}^0d\tau_2\sum_K\hspace{-16pt}\int\,e^{ik_0(\tau_1-\tau_2)+i\bm{k}\cdot\bm{r}}D_{00}(k_0,\bm{k})=\frac{g^2}{T}\int_ke^{i\bm{k}\cdot\bm{r}}D_{00}(0,\bm{k})\,.
\end{equation}
Splitting the integration into the different momentum regions, we have for $k\sim1/r$:
\begin{align}
 D_{I,1/r}&=\frac{g^2}{T}\int_{k\sim1/r}\frac{e^{i\bm{k}\cdot\bm{r}}}{k^2}\left(1-\frac{\Pi(0,k\gg\pi T)}{k^2}+\mathcal{O}\left(g^4\right)\right)\notag\\
 &=\frac{g^2}{T}\int_{k\sim1/r}\frac{e^{i\bm{k}\cdot\bm{r}}}{k^2}\left(1+\frac{g^2}{(4\pi)^2}\left[\frac{31}{9}N-\frac{10}{9}n_f+2\beta_0\ln\frac{\mu}{k}\right]\right.\notag\\
 &\phantom{=\frac{g^2}{T}\int_{k\sim1/r}\frac{e^{i\bm{k}\cdot\bm{r}}}{k^2}}\left.+\frac{Ng^2}{18}\frac{T^2}{k^2}-\left(\frac{44}{225}N+\frac{7}{45}n_f\right)g^2\pi^2\frac{T^4}{k^4}+\mathcal{O}\left(g^2(T/k)^6,g^4\right)\right)\notag\\
 &=\frac{\alpha_\mathrm{s}}{rT}\left(1+\frac{\alpha_\mathrm{s}}{4\pi}\left[\frac{31}{9}N-\frac{10}{9}n_f+2\beta_0(\gamma_E+\ln\mu r)\right]\right)\notag\\
 &\phantom{=}+\alpha_\mathrm{s}^2\left[-\frac{N}{9}r\pi T-\left(\frac{22}{675}N+\frac{7}{270}n_f\right)(r\pi T)^3\right]+\mathcal{O}\left(\alpha_\mathrm{s}^2 (r\pi T)^5,\alpha_\mathrm{s}^3\right)\,.
\end{align}
Here we have used the (charge-renormalized) temporal gluon self-energy in Coulomb gauge, expanded for momenta much larger than the temperature scale; $\beta_0=(11N-2n_f)/3$ is the first coefficient of the beta function. The second line corresponds to the vacuum part, while the third line corresponds to the matter part. Accordingly, the first part of the result gives the static potential in the vacuum (without the color factor) and the second part gives thermal corrections as a series in $r\pi T$.

The next contribution comes from the region $k\sim\pi T$, where we have to expand the numerator $\exp[i\bm{k}\cdot\bm{r}]$ for small $r$:
\begin{align}
 D_{I,\pi T}&=\frac{g^2}{T}\int_{k\sim\pi T}\frac{1-\frac{1}{2}(\bm{k}\cdot\bm{r})^2+\dots}{k^2}\left(1-\frac{\Pi(0,k\sim\pi T)}{k^2}+\mathcal{O}\left(g^4\right)\right)\notag\\*
 &=\alpha_\mathrm{s}^2\left[N\left(-\frac{1}{2\varepsilon}-1+\gamma_E+\ln\frac{T^2}{\pi\mu^2}\right)+n_f\ln2+\left(\frac{4}{3}N+n_f\right)\zeta(3)(rT)^2\right]+\mathcal{O}\left(\alpha_\mathrm{s}^3\right)\,.
\end{align}
The first term in the expansion of the numerator does not depend on $r$ and is exactly the same as $-2$ times the scale $\pi T$ contribution to a single Polyakov loop (without the color factor), which can be found in~\cite{Brambilla:2010xn}. The second order term in this expansion can be calculated by the same methods. The integrals without the self-energy are all scaleless and vanish in dimensional regularization. Furthermore, we have checked that higher powers in $r$ all vanish in the integral with the one-loop self-energy (cf.\ Appendix~\ref{Details}), so there are no $(r\pi T)^4$ or higher thermal corrections at order $\alpha_\mathrm{s}^2$. 

The last contribution comes from the region $k\sim m_D$, where again the numerator is expanded, but now the expansion of the denominator in terms of the self-energy is different:
\begin{align}
 D_{I,m_D}={}&\frac{g^2}{T}\int_{k\sim m_D}\hspace{-5pt}\frac{1-\frac{1}{2}(\bm{k}\cdot\bm{r})^2+\dots}{k^2+m_D^2}\notag\\
 &\times\left(1-\frac{\Pi(0,k\sim m_D)-m_D^2}{k^2+m_D^2}+\frac{\left(\Pi(0,k\sim m_D)-m_D^2\right)^2}{\left(k^2+m_D^2\right)^2}-\dots\right)\notag\\
 ={}&-\frac{\alpha_\mathrm{s}m_D}{T}+N\alpha_\mathrm{s}^2\left[\frac{1}{2\varepsilon}+\frac{1}{2}-\gamma_E+\ln\frac{\pi\mu^2}{m_D^2}\right]+\frac{(N^2-1)n_f}{4N}\frac{\alpha_\mathrm{s}^3T}{m_D}\notag\\
 &-\frac{3\alpha_\mathrm{s}^2m_D}{8\pi T}\left[3N+\frac{2}{3}n_f(1-4\ln2)+2\beta_0\left(\gamma_E+\ln\frac{\mu}{4\pi T}\right)\right]\notag\\
 &+\frac{N^2\alpha_\mathrm{s}^3T}{m_D}\left[\frac{89}{24}+\frac{\pi^2}{6}-\frac{11}{6}\ln2\right]-\frac{\alpha_\mathrm{s}m_D^3}{6T^3}(rT)^2+\mathcal{O}\left(\alpha_\mathrm{s}^3\right)\,.
\end{align}
Again, the terms coming from the zeroth order expansion of the numerator are equal to $-2$ times the scale $m_D$ contribution to a single Polyakov loop. The second order term in this expansion is a standard integral in dimensional regularization. Higher terms in $r$ also come with higher powers of $m_D$ by dimensional analysis, and therefore they are suppressed by additional powers of $g$.

Combining the contributions from the different scales, we have
\begin{align}
 D_I={}&\frac{\alpha_\mathrm{s}}{rT}\left(1+\frac{\alpha_\mathrm{s}}{4\pi}\left[\frac{31}{9}N-\frac{10}{9}n_f+2\beta_0(\gamma_E+\ln\mu r)\right]\right)\notag\\*
 &-\frac{\alpha_\mathrm{s}m_D}{T}+\alpha_\mathrm{s}^2\left[N\left(-\frac{1}{2}+\ln\frac{T^2}{m_D^2}\right)+n_f\ln2\right]\notag\\
 &-\frac{3\alpha_\mathrm{s}^2m_D}{8\pi T}\left[3N+\frac{2}{3}n_f(1-4\ln2)+2\beta_0\left(\gamma_E+\ln\frac{\mu}{4\pi T}\right)\right]\notag\\
 &+\frac{N^2\alpha_\mathrm{s}^3T}{m_D}\left[\frac{89}{24}+\frac{\pi^2}{6}-\frac{11}{6}\ln2\right]+\frac{(N^2-1)n_f}{4N}\frac{\alpha_\mathrm{s}^3T}{m_D}\notag\\
 &-\frac{N\alpha_\mathrm{s}^2}{9}r\pi T+\alpha_\mathrm{s}^2\left(\frac{4}{3}N+n_f\right)\zeta(3)(rT)^2-\frac{\alpha_\mathrm{s}m_D^3}{6T^3}(rT)^2\notag\\
 &-\alpha_\mathrm{s}^2\left(\frac{22}{675}N+\frac{7}{270}n_f\right)(r\pi T)^3+\mathcal{O}\left(\alpha_\mathrm{s}^2 (r\pi T)^5,\alpha_\mathrm{s}^3\right)\,,
\label{DIfull}
\end{align}
where the terms are ordered with increasing power of $r$ and $g$. The scale of $\alpha_\mathrm{s}$ is $\mu$ everywhere. The logarithms of $\mu$ can be absorbed in $g$ if evaluated at two different scales, which leads to an expression identical to the previous one up to terms of higher order:
\begin{align}
 D_I={}&\frac{\alpha_\mathrm{s}(1/r)}{rT}+\frac{\alpha_\mathrm{s}^2}{4\pi rT}\left[\frac{31}{9}N-\frac{10}{9}n_f+2\beta_0\gamma_E\right]\notag\\
 &-\frac{\alpha_\mathrm{s}(4\pi T)m_D(4\pi T)}{T}+\alpha_\mathrm{s}^2\left[N\left(-\frac{1}{2}+\ln\frac{T^2}{m_D^2}\right)+n_f\ln2\right]\notag\\
 &-\frac{3\alpha_\mathrm{s}^2m_D}{8\pi T}\left[3N+\frac{2}{3}n_f(1-4\ln2)+2\beta_0\gamma_E\right]\notag\\
 &+\frac{N^2\alpha_\mathrm{s}^3T}{m_D}\left[\frac{89}{24}+\frac{\pi^2}{6}-\frac{11}{6}\ln2\right]+\frac{(N^2-1)n_f}{4N}\frac{\alpha_\mathrm{s}^3T}{m_D}\notag\\
 &-\frac{N\alpha_\mathrm{s}^2}{9}r\pi T+\alpha_\mathrm{s}^2\left(\frac{4}{3}N+n_f\right)\zeta(3)(rT)^2-\frac{\alpha_\mathrm{s}m_D^3}{6T^3}(rT)^2\notag\\
 &-\alpha_\mathrm{s}^2\left(\frac{22}{675}N+\frac{7}{270}n_f\right)(r\pi T)^3+\mathcal{O}\left(\alpha_\mathrm{s}^2 (r\pi T)^5,\alpha_\mathrm{s}^3\right)\,.
\label{DI}
\end{align}
The choice of the scales is somewhat arbitrary, since, e.g., also the $\beta_0\gamma_E$ terms could be included by an extra factor $\exp[-\gamma_E]$ in the scale of $g$, but this ambiguity is a higher order effect.

Note that the $r$-independent part of the above expression is equal to twice the free energy of a single static quark, $F_Q$, calculated to NNLO~\cite{Berwein:2015ayt} [up to the factor $C_F=(N^2-1)/(2 N)$]:
\begin{align}
 \frac{F_Q}{T}={}&-\frac{(N^2-1)\alpha_\mathrm{s}(4\pi T)m_D(4\pi T)}{4NT}+\frac{(N^2-1)\alpha_\mathrm{s}^2}{4N}\left[N\left(-\frac{1}{2}+\ln \frac{T^2}{m_D^2}\right)+n_f\ln 2\right]\notag\\
 &-\frac{3(N^2-1)\alpha_\mathrm{s}^2m_D}{32N\pi T}\left[3N+\frac{2}{3}n_f(1-4\ln2)+2\beta_0\gamma_E\right]\notag\\
 &+\frac{N(N^2-1)\alpha_\mathrm{s}^3T}{m_D}\left[\frac{89}{96}+\frac{\pi^2}{24}-\frac{11}{24}\ln2\right]+\frac{(N^2-1)^2n_f}{16N^2}\frac{\alpha_\mathrm{s}^3T}{m_D}+\mathcal{O}\left(\alpha_\mathrm{s}^3\right).
 \label{FQexpr}
\end{align}

For the Polyakov loop correlator we need the square and cubic powers of this expression up to $\mathcal{O}\left(g^7\right)$:
\begin{align}
 D_I^2={}&\frac{\alpha_\mathrm{s}^2(1/r)}{r^2T^2}+\frac{\alpha_\mathrm{s}(1/r)\alpha_\mathrm{s}^2}{2\pi r^2T^2}\left[\frac{31}{9}N-\frac{10}{9}n_f+2\beta_0\gamma_E\right]\notag\\
 &-\frac{2\alpha_\mathrm{s}(1/r)\alpha_\mathrm{s}(4\pi T)m_D(4\pi T)}{rT^2}+\frac{\alpha_\mathrm{s}^2\alpha_\mathrm{s}(4\pi T)m_D(4\pi T)}{2\pi rT^2}\left[\frac{31}{9}N-\frac{10}{9}n_f+2\beta_0\gamma_E\right]\notag\\
 &+\frac{2\alpha_\mathrm{s}(1/r)\alpha_\mathrm{s}^2}{rT}\left[N\left(-\frac{1}{2}+\ln\frac{T^2}{m_D^2}\right)+n_f\ln2\right]+\frac{(N^2-1)n_f}{2N}\frac{\alpha_\mathrm{s}(1/r)\alpha_\mathrm{s}^3}{rm_D}\notag\\
 &-\frac{3\alpha_\mathrm{s}(1/r)\alpha_\mathrm{s}^2m_D}{4\pi rT^2}\left[3N+\frac{2}{3}n_f(1-4\ln2)+2\beta_0\gamma_E\right]\notag\\
 &+\frac{2N^2\alpha_\mathrm{s}(1/r)\alpha_\mathrm{s}^3}{rm_D}\left[\frac{89}{24}+\frac{\pi^2}{6}-\frac{11}{6}\ln2\right]-\frac{2\pi N\alpha_\mathrm{s}(1/r)\alpha_\mathrm{s}^2}{9}+\frac{\alpha_\mathrm{s}^2(4\pi T)m_D^2(4\pi T)}{T^2}\notag\\
 &-\frac{2\alpha_\mathrm{s}^2\alpha_\mathrm{s}(4\pi T)m_D(4\pi T)}{T}\left[N\left(-\frac{1}{2}+\ln\frac{T^2}{m_D^2}\right)+n_f\ln2\right]\notag\\
 &+2\alpha_\mathrm{s}(1/r)\alpha_\mathrm{s}^2\left(\frac{4}{3}N+n_f\right)\zeta(3)rT-\frac{\alpha_\mathrm{s}(1/r)\alpha_\mathrm{s}m_D^3}{3T^3}rT+\frac{2\pi N\alpha_\mathrm{s}^2\alpha_\mathrm{s}(4\pi T)m_D(4\pi T)}{9T}rT\notag\\
 &-2\pi\alpha_\mathrm{s}(1/r)\alpha_\mathrm{s}^2\left(\frac{22}{675}N+\frac{7}{270}n_f\right)(r\pi T)^2\notag\\
 &-\frac{2\alpha_\mathrm{s}^2\alpha_\mathrm{s}(4\pi T)m_D(4\pi T)}{T}\left(\frac{4}{3}N+n_f\right)\zeta(3)(rT)^2\notag\\
 &+\frac{2\alpha_\mathrm{s}^2\alpha_\mathrm{s}(4\pi T)m_D(4\pi T)}{T}\left(\frac{22}{675}N+\frac{7}{270}n_f\right)(r\pi T)^3+\mathcal{O}\left(\alpha_\mathrm{s}^3 (r\pi T)^4,\alpha_\mathrm{s}^4\right)\,,\\
 D_I^3={}&\frac{\alpha_\mathrm{s}^3(1/r)}{r^3T^3}-\frac{3\alpha_\mathrm{s}^2(1/r)\alpha_\mathrm{s}(4\pi T)m_D(4\pi T)}{r^2T^3}+\mathcal{O}\left(\alpha_\mathrm{s}^4\right)\,.
\end{align}
We have explicitly kept the same scale dependence of $\alpha_\mathrm{s}$ as in $D_I$.

\subsection{\texorpdfstring{Calculation of $\bm{D_H}$ for $\bm{1/r\gg\pi T\gg m_D}$}{Calculation of D\_H for 1/r>>piT>>m\_D}}
\label{secDHmain}

Now we discuss the contribution of the H-shaped diagrams to the Polyakov loop, i.e., the one that comes from the last two diagrams in the next-to-last line of Eq.~\eqref{PLCDiag}. The sum of those, which we call $D_H$, is much simpler to calculate than the individual diagrams, because in this case the contour integrations can be combined in such a way that they yield the condition that all Matsubara frequencies in the gluon propagators have to be zero. It turns out that $D_H$ is given by $g^4/2T$ times the spatial momentum integrals for the gluon propagators and vertices, which we will call $D_H'$.

This can be shown in the following way. We label the gluon momenta in the two diagrams in the same way, so that they are easier to combine. In the H-shaped diagram proper, the momentum $k$ flows from the antiquark loop to the quark loop along the temporal gluons on the left side, the momentum $p$ flows from the antiquark loop to the quark loop along the temporal gluons on the right side, and the momentum $q$ flows through the spatial gluon connecting the two temporal gluon legs from left to right, starting and ending on the quark loop. If we use labels $\tau_1$ to $\tau_4$ for the imaginary time coordinates in counterclockwise order starting from the antiquark loop, then $\tau_1$ connects to a propagator with momentum $k$, $\tau_2$ to $p$, $\tau_3$ to $p+q$, and $\tau_4$ to $k-q$. In the case of the second diagram, the lower two temporal gluon legs are crossed, so $\tau_1$ and $\tau_2$ change their roles.

Denoting the integral over the momenta by $D_H'(k_0,p_0,q_0)$, we get as a result of the contour integrations:
\begin{align}
 D_H={}&(ig)^4\int_{1/T}^0d\tau_1\int_{1/T}^{\tau_1}d\tau_2\int_0^{1/T}d\tau_3\int_0^{\tau_3}d\tau_4\notag\\
 &\times\sum_{k_0,p_0,q_0}\left(e^{-ik_0\tau_1}e^{-ip_0\tau_2}+e^{-ip_0\tau_1}e^{-ik_0\tau_2}\right)e^{i(p_0+q_0)\tau_3}e^{i(k_0-q_0)\tau_4}D_H'(k_0,p_0,q_0)\notag\\
 ={}&g^4\left(\int_{1/T}^0d\tau_1\int_{1/T}^{\tau_1}d\tau_2+\int_{1/T}^0d\tau_1\int_{\tau_1}^0d\tau_2\right)\int_0^{1/T}d\tau_3\int_0^{\tau_3}d\tau_4\notag\\
 &\times\sum_{k_0,p_0,q_0}e^{-ik_0\tau_1}e^{-ip_0\tau_2}e^{i(p_0+q_0)\tau_3}e^{i(k_0-q_0)\tau_4}D_H'(k_0,p_0,q_0)\notag\\
 ={}&g^4\int_0^{1/T}d\tau_3\int_0^{\tau_3}d\tau_4\,\sum_{k_0,p_0,q_0}\frac{\delta_{k_0}}{T}\frac{\delta_{p_0}}{T}e^{i(p_0+q_0)\tau_3}e^{i(k_0-q_0)\tau_4}D_H'(k_0,p_0,q_0)\notag\\
 ={}&g^4\int_0^{1/T}d\tau_3\int_0^{\tau_3}d\tau_4\,\sum_{q_0}e^{iq_0(\tau_3-\tau_4)}D_H'(0,0,q_0)\notag\\
 ={}&g^4\sum_{q_0}\left(\frac{\delta_{q_0}}{2T^2}+\frac{1-\delta_{q_0}}{iq_0T}\right)D_H'(0,0,q_0)\notag\\
 ={}&\frac{g^4}{2T}D_H'(0,0,0)\,,
 \label{eqDHprime}
\end{align}
where in the second step we interchanged the integration variables $\tau_1$ with $\tau_2$ and rewrote the boundaries of the integrations, $\delta_{k_0}$ means a Kronecker delta that selects the zero mode (so $\delta_{k_0}=\delta_{0n_k}$ for $k_0=2\pi Tn_k$), and the second term in the next-to-last line vanishes because it is odd in $q_0$ while $D_H'$ is even. Up to this point, the calculation does not depend on the chosen gauge. For Coulomb gauge we have
\begin{equation}
 D_H'=(ig)^2 \int_k\int_p\int_q\frac{-4\left((\bm{k}\cdot\bm{p})\bm{q}^2-(\bm{k}\cdot\bm{q})(\bm{p}\cdot\bm{q})\right)e^{i\bm{k}\cdot\bm{r}}e^{i\bm{p}\cdot\bm{r}}}{\bm{k}^2(\bm{k}-\bm{q})^2\left(\bm{q}^2\right)^2(\bm{p}+\bm{q})^2\bm{p}^2}\,.
\end{equation}
The calculation of $D_H'$ using the integration by region method is presented in Appendix~\ref{app:Hdiag} and the result reads
\begin{equation}
 D_H'(0,0,0)=-\frac{g^2}{(4\pi)^3r}\left(3-\frac{\pi^2}{4}\right)+\mathcal{O}\left(g^4\right)\,.
\end{equation}
This result can also be obtained by comparison to the $\mathcal{O}\left(\alpha_\mathrm{s}^3\right)$ result for the Polyakov loop correlator from~\cite{Brambilla:2010xn} (where static gauge was used instead of Coulomb gauge); we will see when we collect the different contributions to the final result for the Polyakov loop correlator that with this value for $D_H'$ the two calculations agree.

\subsection{NNNLO result for the Polyakov loop correlator at short distances}

We can now put all the different contributions together to get the final perturbative result for the Polyakov loop correlator in the case $1/r\gg\pi T\gg m_D\gg\alpha_\mathrm{s}/r$. We will first collect all terms up to $\mathcal{O}\left(\alpha_\mathrm{s}^3\right)$ and compare with the result from~\cite{Brambilla:2010xn}:
\begin{align}
 \exp&\left[\frac{2F_Q-F_{Q\bar{Q}}}{T}\right]_{{\rm up~to~}g^6}=\left.1+\frac{N^2-1}{8N^2}D_I^2+\frac{(N^2-1)(N^2-2)}{48N^3}D_I^3-\frac{N^2-1}{4N}D_H\right|_{{\rm up~to~}g^6} \notag\\
 ={}&1+\frac{N^2-1}{8N^2}\left\{\frac{\alpha_\mathrm{s}^2(1/r)}{r^2T^2}+\frac{\alpha_\mathrm{s}(1/r)\alpha_\mathrm{s}^2}{2\pi r^2T^2}\left(\frac{31}{9}N-\frac{10}{9}n_f+2\beta_0\gamma_E\right)\right.\notag\\
 &-\frac{2\alpha_\mathrm{s}(1/r)\alpha_\mathrm{s}(4\pi T)m_D(4\pi T)}{rT^2}+\frac{2\alpha_\mathrm{s}(1/r)\alpha_\mathrm{s}^2}{rT}\left[N\left(-\frac{1}{2}+\ln\frac{T^2}{m_D^2}\right)+n_f\ln2\right]\notag\\
 &-\frac{2\pi N\alpha_\mathrm{s}(1/r)\alpha_\mathrm{s}^2}{9}+\frac{\alpha_\mathrm{s}^2(4\pi T)m_D^2(4\pi T)}{T^2}+2\alpha_\mathrm{s}(1/r)\alpha_\mathrm{s}^2\left(\frac{4}{3}N+n_f\right)\zeta(3)rT\notag\\
 &-2\pi\alpha_\mathrm{s}(1/r)\alpha_\mathrm{s}^2\left(\frac{22}{675}N+\frac{7}{270}n_f\right)(r\pi T)^2+\frac{N^2-2}{6N}\frac{\alpha_\mathrm{s}^3(1/r)}{r^3T^3}\notag\\
 &+\left.\frac{N\alpha_\mathrm{s}^3}{rT}\left(3-\frac{\pi^2}{4}\right)\right\}+\mathcal{O}\left(g^6(r\pi T)^4\right)\notag\\
 ={}&1+\frac{N^2-1}{8N^2}\left\{\frac{\alpha_\mathrm{s}^2(1/r)}{r^2T^2} - \frac{2\alpha_\mathrm{s}(1/r)\alpha_\mathrm{s}(4\pi T)m_D(4\pi T)}{rT^2}\right.\notag\\
& + \frac{N^2-2}{6N}\frac{\alpha_\mathrm{s}^3(1/r)}{r^3T^3}  + \frac{\alpha_\mathrm{s}(1/r)\alpha_\mathrm{s}^2}{2\pi r^2T^2}\left(\frac{31}{9}N-\frac{10}{9}n_f+2\beta_0\gamma_E\right)\notag\\
 & +\frac{2\alpha_\mathrm{s}(1/r)\alpha_\mathrm{s}^2}{rT}\left[N\left(1-\frac{\pi^2}{8}+\ln\frac{T^2}{m_D^2}\right)+n_f\ln2\right]\notag\\
 &-\frac{2\pi N\alpha_\mathrm{s}(1/r)\alpha_\mathrm{s}^2}{9}+\frac{\alpha_\mathrm{s}^2(4\pi T)m_D^2(4\pi T)}{T^2}+2\alpha_\mathrm{s}(1/r)\alpha_\mathrm{s}^2\left(\frac{4}{3}N+n_f\right)\zeta(3)rT\notag\\
 &-\left.2\pi\alpha_\mathrm{s}(1/r)\alpha_\mathrm{s}^2\left(\frac{22}{675}N+\frac{7}{270}n_f\right)(r\pi T)^2\right\}+\mathcal{O}\left(g^6(r\pi T)^4\right)\,.
 \label{Og6result}
\end{align}
The result agrees with the one in~\cite{Brambilla:2010xn}, except that we have added a few more powers of $r\pi T$, and that we could also fix the scale of $\alpha_\mathrm{s}$ in some more terms through the relation to the one-gluon exchange diagram.

The next order is then
\begin{align}
 \exp\left[\frac{2F_Q-F_{Q\bar{Q}}}{T}\right]_{g^7}={}&\frac{N^2-1}{8N^2}\left\{-\frac{N^2-2}{2N}\frac{\alpha_\mathrm{s}^2(1/r)\alpha_\mathrm{s}(4\pi T)m_D(4\pi T)}{r^2T^3}\right.\notag\\
 &-\frac{2\alpha_\mathrm{s}^2\alpha_\mathrm{s}(4\pi T)m_D(4\pi T)}{4\pi rT^2}\left(\frac{31}{9}N-\frac{10}{9}n_f+2\beta_0\gamma_E\right)\notag\\
 &-\frac{3\alpha_\mathrm{s}(1/r)\alpha_\mathrm{s}^2m_D}{4\pi rT^2}\left[3N+\frac{2}{3}n_f(1-4\ln2)+2\beta_0\gamma_E\right]\notag\\
 &+\frac{(N^2-1)n_f}{2N}\frac{\alpha_\mathrm{s}(1/r)\alpha_\mathrm{s}^3}{rm_D}+\frac{2N^2\alpha_\mathrm{s}(1/r)\alpha_\mathrm{s}^3}{rm_D}\left[\frac{89}{24}+\frac{\pi^2}{6}-\frac{11}{6}\ln2\right]\notag\\
 &-\frac{2\alpha_\mathrm{s}^2\alpha_\mathrm{s}(4\pi T)m_D(4\pi T)}{T}\left[N\left(-\frac{1}{2}+\ln\frac{T^2}{m_D^2}\right)+n_f\ln2\right]\notag\\
 &-\frac{\alpha_\mathrm{s}(1/r)\alpha_\mathrm{s}m_D^3}{3T^3}rT+\frac{2\pi N\alpha_\mathrm{s}^2\alpha_\mathrm{s}(4\pi T)m_D(4\pi T)}{9T}rT\notag\\
 &-\frac{2\alpha_\mathrm{s}^2\alpha_\mathrm{s}(4\pi T)m_D(4\pi T)}{T}\left(\frac{4}{3}N+n_f\right)\zeta(3)(rT)^2\notag\\
 &+\left.\frac{2\alpha_\mathrm{s}^2\alpha_\mathrm{s}(4\pi T)m_D(4\pi T)}{T}\left(\frac{22}{675}N+\frac{7}{270}n_f\right)(r\pi T)^3\right\}\notag\\
 &+\mathcal{O}\left(g^7(r\pi T)^4\right)\,.
\label{Og7result}
\end{align}

\subsection{The singlet and adjoint free energies in Coulomb gauge at short distances}

Using the above analysis, it is straightforward to obtain the order $g^5$ result for the singlet and adjoint free energies for $r\pi T\ll1$. As discussed before, the sum of unconnected diagrams appearing at order $g^4$ vanishes apart from higher order loop corrections: $D_X+2 D_T=\mathcal{O}\left(g^6\right)$. Therefore we write
\begin{align}
 \frac{F_S}{T}&=-\frac{N^2-1}{2 N}D_I+2\frac{F_Q}{T}\,, & \frac{F_A}{T}&=\frac{1}{2N}D_I+2\frac{F_Q}{T}\,.
\end{align}
Using Eq.~\eqref{DIfull} for $D_I$ and Eq.~\eqref{FQexpr} for $F_Q/T$, we obtain
\begin{align}
 \frac{F_S}{T}={}&-\frac{N^2-1}{2N}\frac{\alpha_\mathrm{s}(1/r)}{rT}\left[1+\frac{\alpha_\mathrm{s}}{4\pi}\left(\frac{31}{9}N-\frac{10}{9}n_f+2\beta_0\gamma_E\right)\right]+\frac{1}{18}\left(N^2-1\right)\alpha_\mathrm{s}^2r\pi T\notag\\*
 &-\frac{N^2-1}{2N}\left(\frac{4}{3}N+n_f\right)\zeta(3)\alpha_\mathrm{s}^2r^2T^2+\frac{N^2-1}{12N}\frac{\alpha_\mathrm{s}m_D^3}{T^3}r^2T^2\notag\\
 &+\frac{N^2-1}{2 N}\left(\frac{22}{675}N+\frac{7}{270}n_f\right)\alpha_\mathrm{s}^2(r\pi T)^3+\mathcal{O}\left(\alpha_\mathrm{s}^2(r\pi T)^5,\alpha_\mathrm{s}^3\right)\,,
 \label{FS}\\
 \frac{F_A}{T}={}&-\frac{1}{N^2-1}\frac{F_S}{T}-\frac{N\alpha_\mathrm{s}(4\pi T)m_D(4\pi T)}{2T}+\frac{N\alpha_\mathrm{s}^2}{2}\left[N\left(-\frac{1}{2}+\ln\frac{T^2}{m_D^2}\right)+n_f\ln2\right]\notag\\
 &-\frac{3N\alpha_\mathrm{s}^2m_D}{16\pi T}\left[3N+\frac{2}{3}n_f(1-4\ln2)+2\beta_0\gamma_E\right]+\frac{N^3\alpha_\mathrm{s}^3T}{m_D}\left[\frac{89}{48}+\frac{\pi^2}{12}-\frac{11}{12}\ln2\right]\notag\\
 &+\frac{(N^2-1)n_f}{8}\frac{\alpha_\mathrm{s}^3T}{m_D}+\mathcal{O}\left(\alpha_\mathrm{s}^3\right)\,.
 \label{FA}
\end{align}

The result for $F_S$ agrees with the calculations by Laine and Burnier~\cite{Burnier:2009bk} up to order $g^4$, if the latter is Taylor expanded in $r\pi T$. This is shown in Appendix~\ref{Details}. The $g^5$ term in $F_S$ as well as the expression for $F_A$ are new results. It is interesting to note that the $g^5$ term in Eq.~\eqref{FS} can be guessed from the leading order result derived for $1/r\sim m_D$:
\begin{equation}
 F_S-2F_Q\Bigr|_\mathrm{LO}=-\frac{N^2-1}{2N}\frac{\alpha_\mathrm{s}}{r}e^{-rm_D},
\end{equation}
by expanding the exponent and keeping the term proportional to $m_D^3$. In the next subsection, we will discuss $F_S$ for $1/r\sim m_D$ in more details\footnote{It is also interesting to notice that the expression of $F_S$ given in the first two lines of~\eqref{FS} is identical with the real part of the real-time static potential computed in\cite{Brambilla:2008cx}. We thank J.\ Ghiglieri for communications on this point.}.

\subsection{The free energies in the screening regime}
\label{secfreescreening}

Let us now consider the singlet free energy for $1/r\sim m_D$. In this regime, there are only two separate scales larger than $\alpha_\mathrm{s}/r$: $\pi T$ and $m_D$. The exponentials in the propagators are no longer expanded for momenta of the order of the Debye mass, since their argument is now of order 1. In contrast, for momenta of the order of the temperature, the propagators of gluons exchanged between the two Polyakov loops are exponentially suppressed in coordinate space and do not contribute to the expansion.

The power counting of the different contributions also changes, since now each power of $1/r$ adds a power of $g$. Accordingly, the leading order contribution no longer counts as $g^2$, but as $g^3$. We will give the free energies up to order $g^4$, which is given only by $D_I$. One can show with simple power counting arguments that the two-gluon diagrams only start to contribute at higher orders: $D_X\sim D_I^2\sim g^4\exp(-2rm_D)/(rT)^2\sim g^6$ and $D_T\sim D_Ig^2m_D/T\sim g^6$ within this hierarchy. The calculations for $D_I$ at one-loop level are presented at the end of Appendix~\ref{massive_int}; we get
\begin{align}
 \frac{F_S-2F_Q}{T}={}&-\frac{N^2-1}{2N}\frac{\alpha_\mathrm{s}}{rT}e^{-rm_D}\notag\\
 &-\frac{N^2-1}{2}\alpha_\mathrm{s}^2e^{-rm_D}\left[2-\ln\left(2rm_D\right)-\gamma_E+e^{2rm_D}E_1\left(2rm_D\right)\right]+\mathcal{O}\left(g^5\right)\,.
 \label{FSmdr}
\end{align}
This result agrees with that of Ref.~\cite{Burnier:2009bk} up to terms $\mathcal{O}\left(g^5\right)$ [cf.\ Eq.~(3.22) of Ref.~\cite{Burnier:2009bk}]. Note that in our power counting scheme, the first term is of order $g^3$ and the second one of order~$g^4$.

In Eq.~\eqref{FSmdr}, there is no term that fixes the scale of $g$, not even in the leading contribution of order $g^3$. Such a term will appear at order $g^5$. However, in order to get the full result at order $g^5$, we would also need the calculation of $D_I$ with the two-loop self-energy at the scale $m_D$, which is not available at present. Nevertheless, all other contributions have been computed in Appendix~\ref{massive_int}. Since they include all contributions of order $g^5$ proportional to the number of light quarks and to the logarithm of the temperature, they are enough to fix the scale of $g$ at least in the leading order contribution of Eq.~\eqref{FSmdr}. They read
\begin{equation}
 \frac{\delta F_S}{T}=-\frac{N^2-1}{2N}\frac{\alpha_\mathrm{s}}{rT}e^{-rm_D}\left(1-\frac{rm_D}{2}\right)\delta Z_1\,,
\end{equation}
where
\begin{equation}
 \delta Z_1=\frac{\alpha_\mathrm{s}}{4\pi}\left[\frac{11}{3}N+\frac{2}{3}(1-4\ln2)n_f+2\beta_0\left(\gamma_E+\ln\frac{\mu}{4\pi T}\right)\right]\,.
\end{equation}
The logarithm in this term is proportional to the first coefficient of the beta function and determines the scale of $g$ in the leading order term of $F_S$ to be $4\pi T$, both in $\alpha_\mathrm{s}$ and in $m_D$ in the exponent. (Remember that $m_D\sim\sqrt{\alpha_\mathrm{s}}$, which explains the factor $1/2$ in the $m_D$ term of $\delta F_S$.) The expression of $\delta Z_1$ agrees with an analogous finding in~\cite{Burnier:2009bk} [cf.\ Eq.~(3.19) of Ref.~\cite{Burnier:2009bk}].

At this order, we have already seen in Eq.~\eqref{Casimir} that Casimir scaling still holds, hence the adjoint free energy is given by
\begin{equation}
\frac{F_A-2F_Q}{T}=-\frac{F_S-2F_Q}{\left(N^2-1\right)T}\,.
\end{equation}

\subsection{\texorpdfstring{The free energy of a $\bm{Q\bar{Q}}$ pair in the screening regime}{The free energy of a QQbar pair in the screening regime}}
\label{secfreescreeningresult}

We can also calculate the Polyakov loop correlator for $1/r\sim m_D$. In Coulomb gauge, it is easy to see that the unconnected two-gluon and three-gluon diagrams appearing in the last two lines of Eq.~\eqref{PLCDiag} give rise to contributions that are of order $g^9$. Therefore, we need to consider only the contributions from $D_I$ and $D_H$ for $1/r\sim m_D$. These calculations are discussed in Appendix~\ref{massive_int}. Using the results of these calculations, we obtain
\begin{align}
 \frac{2F_Q-F_{Q\bar Q}}{T}={}&\ln\left[1+\frac{N^2-1}{8N^2}D_I^2-\frac{N^2-1}{4N}D_H+\mathcal{O}\left(g^8\right)\right]\notag\\
 ={}&\frac{N^2-1}{8N^2}\left(\frac{\alpha_\mathrm{s}(4\pi T)e^{-rm_D(4\pi T)}}{rT}\right)^2+\frac{N^2-1}{8N}\frac{\alpha_\mathrm{s}^3e^{-2rm_D}}{rT}\biggl[3-\ln4rm_D-\gamma_E\notag\\
 &-e^{4rm_D}E_1(4rm_D)+\frac{2}{rm_D}\left(e^{2rm_D}E_1(2rm_D)+\gamma_E+\ln2rm_D\right)\notag\\
 &-\int_0^\infty dx\frac{e^{-2rm_Dx}}{x+1}\ln\frac{x+2}{x}\biggr]+\mathcal{O}\left(g^8\right)\,,
 \label{FQQresultscreening}
\end{align}
where we have fixed the scale in the leading term in the same way as for $F_S$. This result agrees with the one obtained by Nadkarni~\cite{Nadkarni:1986cz}, except for the fixing of the scale, which is new. The leading order term now scales as $g^6$, while the first correction is of order $g^7$.

\section{Free energies in pNRQCD}
\label{sec:pnrqcd}

The Polyakov loop correlator can be written as the correlator of static color sources $\psi$ and $\chi$ located at a distance $r$ and at imaginary times $0$ and $1/T$~\cite{Brambilla:2010xn}:
\begin{equation}
 \exp\left[-\frac{F_{Q\bar{Q}}}{T}\right]=\frac{1}{N^2\delta^6(0)}\mathrm{Tr}\left\langle\left(\psi(1/T,\bm{r})\chi^\dagger(1/T,\bm{0})\right)\left(\chi(0,\bm{0})\psi^\dagger(0,\bm{r})\right)\right\rangle\,.
\label{NRQCDcorr}
\end{equation}
The delta functions in the denominator are necessary for a correct normalization. Due to the equal-time anticommutators of the static sources: $\left\{\psi_i(\tau,\bm{x}),\psi_j^\dagger(\tau,\bm{y})\right\}=\delta_{ij}\delta^{(3)}(\bm{x}-\bm{y})$ and $\left\{\chi_i^\dagger(\tau,\bm{x}),\chi_j(\tau,\bm{y})\right\}=\delta_{ij}\delta^{(3)}(\bm{x}-\bm{y})$, the operators in the correlator~\eqref{NRQCDcorr}, which have the same spatial arguments, would lead to diverging delta functions. Exactly those are canceled through the normalization. The contraction of the indices of the Kronecker deltas requires the normalization factor $1/N^2$.

Accordingly, the singlet and adjoint free energies are given by
\begin{align}
 \exp\left[-\frac{F_S}{T}\right]&=\frac{1}{N\delta^{6}(0)}\left\langle\mathrm{Tr}\left[\psi(1/T,\bm{r})\chi^\dagger(1/T,\bm{0})\right]\mathrm{Tr}\left[\chi(0,\bm{0})\psi^\dagger(0,\bm{r})\right]\right\rangle\,,\\*
 \exp\left[-\frac{F_A}{T}\right]&=\frac{2}{(N^2-1)\delta^{6}(0)}\left\langle\mathrm{Tr}\left[T^a\psi(1/T,\bm{r})\chi^\dagger(1/T,\bm{0})\right]\mathrm{Tr}\left[T^a\chi(0,\bm{0})\psi^\dagger(0,\bm{r})\right]\right\rangle\,.
\end{align}
The dynamics of the static sources are described by the Lagrangian density:
\begin{equation}
 \mathcal{L} = \psi^\dagger D_0\psi+\chi^\dagger D_0\chi+\frac{1}{4}F_{\mu\nu}^aF_{\mu\nu}^a+\sum_{l=1}^{n_f}\bar{q}_lD_\mu\gamma_\mu q_l \,.
\end{equation}
This is the QCD Lagrangian for a static quark field $\psi$, a static antiquark field $\chi$, and $n_f$ massless quark fields $q_l$. 

If we assume the hierarchy $1/r \gg \pi T \gg m_D\gg\alpha_\mathrm{s}/r$, we may use an EFT where the expansion for small $r$ is systematically incorporated: pNRQCD~\cite{Pineda:1997bj,Brambilla:1999xf,Brambilla:2004jw} (pNRQCD at finite temperature has been discussed in~\cite{Brambilla:2008cx} in real time and in~\cite{Brambilla:2010xn} in imaginary time). In this EFT, the effective degrees of freedom are quark-antiquark fields in color singlet or octet configurations: $S$ and $O^a$. Up until this point we have always kept the number of colors $N$ general, however, pNRQCD is usually defined for $N=3$, hence the name octet for the adjoint field. But since the generalization to arbitrary $N$ is straightforward, we will keep $N$ general while still calling the adjoint field ``octet'' out of convention.

In Euclidean space-time, the pNRQCD Lagrangian density for static fields up to linear order in $r$ is given by~\cite{Brambilla:2010xn}
\begin{align}
 \mathcal{L}_\mathrm{pNRQCD}=&\int d^3r\left[S^\dagger(\partial_0+V_s)S+O^{\dagger\,a}\left(D_0^{ab}+V_o\delta^{ab}\right)O^b\right]+\frac{1}{4}F_{\mu\nu}^aF_{\mu\nu}^a+\sum_{l=1}^{n_f}\bar{q}_lD_\mu\gamma_\mu q_l\notag\\
 &-\int d^3r\left[\frac{V_A}{\sqrt{2N}}\left(S^\dagger(\r\cdot ig\E^a)O^a+O^{\dagger\,a}(\r\cdot ig\E^a)S\right)+\frac{V_B}{2}d^{abc}O^{\dagger\,a}(\r\cdot ig\E^b)O^c\right]\,,
\end{align}
where the singlet and octet fields $S$ and $O^a$ depend on both the relative coordinate $\r$ and the center of mass coordinate $\R$, while gluons and light quarks depend only on $\R$. The Wilson coefficients at next-to-leading order are given by
\begin{align}
 V_s(r)&=-(N^2-1)V_o(r)=-\frac{N^2-1}{2N}\frac{\alpha_\mathrm{s}(1/r)}{r}\left[1+\frac{\alpha_\mathrm{s}}{4\pi}\left(\frac{31N}{9}-\frac{10n_f}{9}+2\beta_0\gamma_E\right)\right]\,,\notag\\
 V_A(r)&=V_B(r)=1\,.
\end{align}

In pNRQCD one can also define singlet and octet (adjoint) free energies in a gauge invariant way:
\begin{align}
 \frac{f_s}{T}\equiv{}&-\ln\frac{1}{\delta^{6}(0)}\left\langle S(1/T,\bm{R},\bm{r})S^\dagger(0,\bm{R},\bm{r})\right\rangle\notag\\
 ={}&-\frac{N^2-1}{2N}\frac{\alpha_\mathrm{s}(1/r)}{rT}\left[1+\frac{\alpha_\mathrm{s}}{4\pi}\left(\frac{31N}{9}-\frac{10n_f}{9}+2\beta_0\gamma_E\right)\right]+\frac{1}{9}\left(N^2-1\right)\alpha_\mathrm{s}^2r\pi T\notag\\
 &-\frac{N^2-1}{2N}\left(\frac{4N}{3}+n_f\right)\zeta(3)\alpha_\mathrm{s}^2r^2T^2+\frac{(N^2-1)\alpha_\mathrm{s}}{12N}\frac{m_D^3}{T^3}r^2T^2+\mathcal{O}\left(\alpha_\mathrm{s}^2(r\pi T)^3,\alpha_\mathrm{s}^3\right)\,,\\
 \frac{f_o}{T}\equiv{}&-\ln\frac{1}{(N^2-1)\delta^{6}(0)}\left\langle O^a(1/T,\bm{R},\bm{r})O^{a\,\dagger}(0,\bm{R},\bm{r})\right\rangle\notag\\
 ={}&-\frac{1}{N^2-1}\frac{f_s}{T}-\frac{N\alpha_\mathrm{s}(4\pi T)m_D(4\pi T)}{2T}+\frac{N\alpha_\mathrm{s}^2}{2}\left[N\left(-\frac{1}{2}+\ln\frac{T^2}{m_D^2}\right)+n_f\ln 2\right]\notag\\
 &-\frac{3N\alpha_\mathrm{s}^2m_D}{16\pi T}\left[3N+\frac{2}{3}n_f(1-4\ln2)+2\beta_0\gamma_E\right]+\frac{N^3\alpha_\mathrm{s}^3T}{m_D}\left[\frac{89}{48}+\frac{\pi^2}{12}-\frac{11}{12}\ln2\right]\notag\\
 &+\frac{(N^2-1)n_f}{8}\frac{\alpha_\mathrm{s}^3T}{m_D}+\mathcal{O}\left(\alpha_\mathrm{s}^2(r\pi T)^3,\alpha_\mathrm{s}^3\right)\,.
\end{align}
We have taken these results from~\cite{Brambilla:2010xn} and added the information from~\cite{Berwein:2015ayt} about the $\mathcal{O}\left(g^5\right)$ Polyakov loop in the adjoint representation. The value of the center of mass coordinate is irrelevant because of translational invariance, however, for comparison with the expressions in the QCD correlator~\eqref{NRQCDcorr} we set it to $\bm{R}=\bm{r}/2$. We can also express the Polyakov loop correlator with these free energies~\cite{Brambilla:2010xn}:
\begin{equation}
 \exp\left[-\frac{F_{Q\bar{Q}}}{T}\right]=\frac{1}{N^2}\exp\left[-\frac{f_s}{T}\right]+\frac{N^2-1}{N^2}\exp\left[-\frac{f_o}{T}\right]+\mathcal{O}\left(\alpha_\mathrm{s}^3(r\pi T)^4\right)\,.
\end{equation}

If we compare $f_s$ and $f_o$ with the singlet and adjoint free energies, $F_S$ and $F_A$, given in Coulomb gauge by Eqs.~\eqref{FS} and~\eqref{FA} we see that they almost agree, but there is a difference of a factor $2$ in the linear term in $r\pi T$. This is not surprising since $f_s$, $f_o$ and $F_S$, $F_A$ do not describe exactly the same quantities: $F_S$ and $F_A$ depend on the choice of gauge while $f_s$ and $f_o$ do not. In addition, $f_s$ and $f_o$ give the Polyakov loop correlator up to corrections of order $\alpha_\mathrm{s}^3(r\pi T)^4$. Still we can quantify the difference by a proper matching calculation.

More specifically, we will match the operator $\psi(\bm{r})\chi^\dagger(\bm{0})$. It transforms as $N_{\bm{r}}\times\overline{N}_{\bm{0}}$ under gauge transformations (here $N$ and $\overline{N}$ refer to fundamental and anti-fundamental representations, transforming locally at the points $\bm{r}$ and $\bm{0}$ respectively). Hence, also the matching pNRQCD operators have to transform in the same way. This requires that they are of the form $\phi(\bm{r},\bm{r}/2)(\dots)\phi^\dagger(\bm{0},\bm{r}/2)$, where the dots stand for the most general expression made of gauge covariant pNRQCD operators located at the center of mass coordinate that are consistent with the discrete symmetries, $P$, $C$ and $T$, of the QCD operator\footnote{Note that in imaginary time $\tau=it\xrightarrow{T}(-i)(-t)=\tau$, and thus $A_0\xrightarrow{T}-A_0$, $\bm{A}\xrightarrow{T}-\bm{A}$, and $\bm{E}\xrightarrow{T}-\bm{E}$. This means that the imaginary time version of the $T$ symmetry involves replacing the gauge fields by their negative, while keeping the static quark fields invariant and complex conjugating the coefficients.}. These operators will be made in general by combinations of chromoelectric or chromomagnetic fields with one color singlet field $S$ or one color octet field $O^a$, this last requirement following from the heavy quark number conservation. The operator $\phi(\bm{x}_1,\bm{x}_2)$ stands for the spatial Wilson line connecting the points $\bm{x}_1$ and $\bm{x}_2$:
\begin{align}
 \phi(\bm{x}_1,\bm{x}_2)\equiv\mathcal{P}\exp\left[ig\int_0^1ds(\bm{x}_1-\bm{x}_2)\cdot\bm{A}(s\bm{x}_1+(1-s)\bm{x}_2)\right]\,,
\end{align}
where we suppressed the imaginary time argument. The Wilson lines guarantee that the matching pNRQCD operators transform as the QCD operator also under gauge transformations.

At $\mathcal{O}\left(r^2\right)$ in the multipole expansion, the matching condition therefore reads
\begin{align}
 \psi(\bm{r})\chi^\dagger(\bm{0}) \rightarrow {}&\phi(\bm{r},\bm{r}/2)\left[\frac{Z_s}{\sqrt{N}}S\mathbbm{1}+\sqrt{2}Z_oO^aT^a+\sqrt{2}Z_{Es}\,r\,\left(\bm{r}\cdot ig\bm{E}^a\right)ST^a\right.\notag\\
 &+\left.\frac{Z_{Eo}\,r}{\sqrt{N}}\left(\bm{r}\cdot ig\bm{E}^a\right)O^a\mathbbm{1}+\sqrt{2}Z'_{Eo}\,d^{abc}\,r\left(\bm{r}\cdot ig\bm{E}^a\right)O^bT^c+\mathcal{O}\left(r^3\right)\right]\phi^\dagger(\bm{0},\bm{r}/2)\,.
\label{matchingcondition}
\end{align}
All the fields inside the square brackets are located at the center of mass coordinate $\bm{R}=\bm{r}/2$. The factors $Z$ are the matching coefficients. They have been chosen such that $Z_s$ and $Z_o$ are~$1$ at leading order.

The Wilson lines in the right-hand side of Eq.~\eqref{matchingcondition} can be multipole expanded. In particular, if the Wilson lines go from $\bm{R}$ to $\bm{R}\pm\bm{r}/2$ their expansion is 
\begin{align}
 \phi(\bm{R}\pm\bm{r}/2,\bm{R})={}&\mathbbm{1}\pm \frac{1}{2}\int_0^1ds\,\bm{r}\cdot ig\bm{A}(\bm{R}\pm s\bm{r}/2)\notag\\
 &+\frac{1}{4}\int_0^1ds_1\int_0^{s_1}ds_2\,(\bm{r}\cdot ig\bm{A}(\bm{R}\pm s_1\bm{r}/2))(\bm{r}\cdot ig\bm{A}(\bm{R}\pm s_2\bm{r}/2))+\dots\notag\\
 ={}&\mathbbm{1}\pm\frac{1}{2}\bm{r}\cdot ig\bm{A}(\bm{R})+\frac{1}{8}(\bm{r}\cdot\bm{\nabla}_R)(\bm{r}\cdot ig\bm{A}(\bm{R}))+\frac{1}{8}(\bm{r}\cdot ig\bm{A}(\bm{R}))^2+\dots\,,
\label{multipoleWilsonlines}
\end{align}
where $\mathbbm{1}$ is the unit matrix in color space, and the dots contain cubic terms and higher in the multipole expansion. 

Different projections of the matching condition~\eqref{matchingcondition} are required to generate $F_S$ and $F_A$:
\begin{align}
 \frac{1}{\sqrt{N}}\mathrm{Tr}\left[\psi\chi^\dagger\right] \rightarrow {}&Z_sS+\frac{Z_o}{\sqrt{2N}}\left(\bm{r}\cdot ig\bm{A}^a\right)O^a+\frac{Z_s}{4N}\left(\bm{r}\cdot ig\bm{A}^a\right)\left(\bm{r}\cdot ig\bm{A}^a\right)S\notag\\*
 &+\frac{Z_o}{4\sqrt{2N}}d^{abc}\left(\bm{r}\cdot ig\bm{A}^a\right)\left(\bm{r}\cdot ig\bm{A}^b\right)O^c+Z_{Eo}\,r\left(\bm{r}\cdot ig\bm{E}^a\right)O^a+\mathcal{O}\left(r^3\right),\label{matchsing}\\
 \sqrt{2}\mathrm{Tr}\left[T^a\psi\chi^\dagger\right] \rightarrow {}&Z_oO^a+\frac{Z_s}{\sqrt{2N}}\left(\bm{r}\cdot ig\bm{A}^a\right)S+\frac{Z_o}{2}d^{abc}\left(\bm{r}\cdot ig\bm{A}^b\right)O^c\notag\\
 &+\frac{Z_s}{4\sqrt{2N}}d^{abc}\left(\bm{r}\cdot ig\bm{A}^b\right)\left(\bm{r}\cdot ig\bm{A}^c\right)S+\frac{Z_o}{4N}\left(\bm{r}\cdot ig\bm{A}^a\right)\left(\bm{r}\cdot ig\bm{A}^b\right)O^b\notag\\
 &+\frac{Z_o}{8}d^{abe}d^{ecd}\left(\bm{r}\cdot ig\bm{A}^b\right)\left(\bm{r}\cdot ig\bm{A}^c\right)O^d+\frac{Z_o}{8}if^{abc}\left[(\bm{r}\cdot\bm{\nabla}),\left(\bm{r}\cdot ig\bm{A}^b\right)\right]O^c\notag\\
 &+Z_{Es}\,r\left(\bm{r}\cdot ig\bm{E}^a\right)S+Z'_{Eo}\,d^{abc}\,r\left(\bm{r}\cdot ig\bm{E}^b\right)O^c+\mathcal{O}\left(r^3\right),\label{matchoct}
\end{align}
where we have multipole expanded the Wilson lines according to Eq.~\eqref{multipoleWilsonlines}. As it will turn out, the matching conditions~\eqref{matchsing} and~\eqref{matchoct} are sufficient to match the free energies. One reason is that the singlet and adjoint free energies $F_S$ and $F_A$ in Coulomb gauge are finite and therefore do not mix under renormalization. We recall that this is a specific feature of the Coulomb gauge, for in general $F_S$ and $F_A$ do mix as discussed at the end of Sec.~\ref{sec:general}.

We can now compute $F_S$ and $F_A$ in pNRQCD by inserting the matching conditions~\eqref{matchsing} and~\eqref{matchoct} into the respective correlators:
\begin{align}
 \exp\left[-\frac{F_S}{T}\right]={}&\frac{1}{N\delta^{6}(0)}\left\langle\mathrm{Tr}\left[\psi(1/T)\chi^\dagger(1/T)\right]\mathrm{Tr}\left[\chi(0)\psi^\dagger(0)\right]\right\rangle\notag\\
 ={}&\frac{|Z_s|^2}{\delta^6(0)}\left\langle S(1/T)S^\dagger(0)\right\rangle+\frac{Z_s^*Z_o}{\sqrt{2N}\delta^6(0)}\left\langle\left(\bm{r}\cdot ig\bm{A}^a\right)O^a(1/T)S^\dagger(0)\right\rangle\notag\\
 &-\frac{Z_o^*Z_s}{\sqrt{2N}\delta^6(0)}\left\langle S(1/T)\left(\bm{r}\cdot ig\bm{A}^a\right)O^{a\,\dagger}(0)\right\rangle\notag\\
 &+\frac{|Z_s|^2}{2N\delta^6(0)}\left\langle\left(\bm{r}\cdot ig\bm{A}^a\right)\left(\bm{r}\cdot ig\bm{A}^a\right)S(1/T)S^\dagger(0)\right\rangle\notag\\
 &-\frac{|Z_o|^2}{2N\delta^6(0)}\left\langle\left(\bm{r}\cdot ig\bm{A}^a\right)\left(\bm{r}\cdot ig\bm{A}^b\right)O^a(1/T)O^{b\,\dagger}(0)\right\rangle+\mathcal{O}\left(\alpha_\mathrm{s}^2(r\pi T)^3\right),\\
 (N^2-1)\exp\left[-\frac{F_A}{T}\right]={}&\frac{2}{\delta^{6}(0)}\left\langle\mathrm{Tr}\left[T^a\psi(1/T)\chi^\dagger(1/T)\right]\mathrm{Tr}\left[T^a\chi(0)\psi^\dagger(0)\right]\right\rangle\notag\\
 ={}&\frac{|Z_o|^2}{\delta^6(0)}\left\langle O^a(1/T)O^a(0)\right\rangle-\frac{Z_s^*Z_o}{\sqrt{2N}\delta^6(0)}\left\langle\left(\bm{r}\cdot ig\bm{A}^a\right)O^a(1/T)S^\dagger(0)\right\rangle\notag\\
 &+\frac{Z_o^*Z_s}{\sqrt{2N}\delta^6(0)}\left\langle S(1/T)\left(\bm{r}\cdot ig\bm{A}^a\right)O^{a\,\dagger}(0)\right\rangle\notag\\
 &-\frac{|Z_s|^2}{2N\delta^6(0)}\left\langle\left(\bm{r}\cdot ig\bm{A}^a\right)\left(\bm{r}\cdot ig\bm{A}^a\right)S(1/T)S^\dagger(0)\right\rangle\notag\\
 &+\frac{|Z_o|^2}{2N\delta^6(0)}\left\langle\left(\bm{r}\cdot ig\bm{A}^a\right)\left(\bm{r}\cdot ig\bm{A}^b\right)O^a(1/T)O^{b\,\dagger}(0)\right\rangle+\mathcal{O}\left(\alpha_\mathrm{s}^2(r\pi T)^3\right).
\end{align}
We have suppressed the time arguments of the gauge fields: since they obey periodic boundary conditions, it does not matter if they are evaluated at imaginary time $0$ or $1/T$. Some terms have been neglected, because they do not contribute at this order in $r$, and several terms cancel. We see that the corrections to the pNRQCD free energies are gauge dependent, for they involve the gauge fields $\bm{A}$ instead of gauge invariant combinations of $\bm{E}$ and $\bm{B}$ fields.

The calculation of the correlators for the leading order corrections can be done in the following way. The quark-antiquark fields can be replaced by the leading order propagators:
\begin{align}
 \left\langle\left(\bm{r}\cdot ig\bm{A}^a\right)\left(\bm{r}\cdot ig\bm{A}^a\right)S(1/T)S^\dagger(0)\right\rangle&=\delta^6(0)e^{-V_s/T}\left\langle\left(\bm{r}\cdot ig\bm{A}^a\right)\left(\bm{r}\cdot ig\bm{A}^a\right)\right\rangle+\dots\,,\\
 \left\langle\left(\bm{r}\cdot ig\bm{A}^a\right)\left(\bm{r}\cdot ig\bm{A}^b\right)O^a(1/T)O^{b\,\dagger}(0)\right\rangle&=\delta^6(0)e^{-V_o/T}\left\langle\left(\bm{r}\cdot ig\bm{A}^a\right)\left(\bm{r}\cdot ig\bm{A}^a\right)\right\rangle+\dots\,,
\end{align}
where the dots contain additional vertex insertions or higher order expansion terms of the adjoint Polyakov loop appearing in the octet propagator. When both singlet and octet fields appear, then the insertion of a vertex is necessary:
\begin{align}
 &\left\langle\left(\bm{r}\cdot ig\bm{A}^a\right)O^a(1/T)S^\dagger(0)\right\rangle\notag\\
 &=\frac{V_A\delta^6(0)}{\sqrt{2N}}\int_0^{1/T}d\tau\,e^{-V_o(1/T-\tau)-V_s\tau}\left\langle\left(\bm{r}\cdot ig\bm{E}^a(\tau)\right)\left(\bm{r}\cdot ig\bm{A}^a\right)\right\rangle+\dots\,,\\
 &\left\langle\left(\bm{r}\cdot ig\bm{A}^a\right)S(1/T)O^{a\,\dagger}(0)\right\rangle\notag\\
 &=\frac{V_A\delta^6(0)}{\sqrt{2N}}\int_0^{1/T}d\tau\,e^{-V_s(1/T-\tau)-V_o\tau}\left\langle\left(\bm{r}\cdot ig\bm{E}^a(\tau)\right)\left(\bm{r}\cdot ig\bm{A}^a\right)\right\rangle+\dots\,.
\end{align}
The leading contribution from the electric fields comes from the $-\partial_\tau\bm{A}^a$ term, and we can use the imaginary time derivative to integrate by parts:
\begin{align}
 &\int_0^{1/T}d\tau\,e^{-V_o(1/T-\tau)-V_s\tau}\left\langle\left(\bm{r}\cdot ig\bm{E}^a(\tau)\right)\left(\bm{r}\cdot ig\bm{A}^a\right)\right\rangle\notag\\
 &\hspace{4cm} =\left(e^{-V_o/T}-e^{-V_s/T}\right)\left\langle\left(\bm{r}\cdot ig\bm{A}^a\right)\left(\bm{r}\cdot ig\bm{A}^a\right)\right\rangle+\mathcal{O}\left(\alpha_\mathrm{s}^3\right)\,,\\
 &\int_0^{1/T}d\tau\,e^{-V_s(1/T-\tau)-V_o\tau}\left\langle\left(\bm{r}\cdot ig\bm{E}^a(\tau)\right)\left(\bm{r}\cdot ig\bm{A}^a\right)\right\rangle\notag\\*
 &\hspace{4cm} =\left(e^{-V_s/T}-e^{-V_o/T}\right)\left\langle\left(\bm{r}\cdot ig\bm{A}^a\right)\left(\bm{r}\cdot ig\bm{A}^a\right)\right\rangle+\mathcal{O}\left(\alpha_\mathrm{s}^3\right)\,.
\end{align}

We may replace $V_A$ by 1, because higher order corrections to this coefficient, which start at order $\alpha_\mathrm{s}^2$~\cite{Brambilla:2006wp}, are beyond the accuracy of this calculation. For the same reason, we may also replace $Z_s$ and $Z_o$ by $1$ in subleading terms. The corrections to the free energies then simplify to:
\begin{align}
 \exp\left[-\frac{F_S}{T}\right]={}&|Z_s|^2\exp\left[-\frac{f_s}{T}\right]\notag\\
 &+\frac{e^{-V_o/T}-e^{-V_s/T}}{2N}\left\langle\left(\bm{r}\cdot ig\bm{A}^a\right)\left(\bm{r}\cdot ig\bm{A}^a\right)\right\rangle+\mathcal{O}\left(\alpha_\mathrm{s}^2(r\pi T)^3,\alpha_\mathrm{s}^3\right)\,,\\
 \exp\left[-\frac{F_A}{T}\right]={}&|Z_o|^2\exp\left[-\frac{f_o}{T}\right]\notag\\
 &-\frac{e^{-V_o/T}-e^{-V_s/T}}{2N(N^2-1)}\left\langle\left(\bm{r}\cdot ig\bm{A}^a\right)\left(\bm{r}\cdot ig\bm{A}^a\right)\right\rangle+\mathcal{O}\left(\alpha_\mathrm{s}^2(r\pi T)^3,\alpha_\mathrm{s}^3\right)\,.
\end{align}

For the calculation of the gauge field correlator at tree level, we need to use the same gauge as for $F_S$ and $F_A$, i.e., Coulomb gauge:
\begin{align}
 \left\langle\left(\bm{r}\cdot ig\bm{A}^a\right)\left(\bm{r}\cdot ig\bm{A}^a\right)\right\rangle={}&-g^2\sum_K\hspace{-16pt}\int\,\frac{\left(\bm{r}^2\bm{k}^2-(\bm{r}\cdot\bm{k})^2\right)\delta^{aa}}{\bm{k}^2\left(k_0^2+\bm{k}^2\right)}=-g^2(N^2-1)\frac{d-1}{d}\sum_K\hspace{-16pt}\int\,\frac{r^2}{k_0^2+\bm{k}^2}\notag\\
 ={}&-\frac{g^2(N^2-1)r^2T^{d-1}\mu^{3-d}}{2\pi^{2-\frac{d}{2}}}\frac{d-1}{d}\Gamma\left(1-\tfrac{d}{2}\right)\zeta(2-d)\notag\\
 \stackrel{d=3}{=}&-\frac{2\pi}{9}(N^2-1)\alpha_\mathrm{s}r^2T^2\,.
\end{align}
When we insert this into the expression above, then we also expand the exponentials of the potentials, since they are of $\mathcal{O}(\alpha_\mathrm{s})$. Comparing both sides, we see that the matching coefficients $Z_s$ and $Z_o$ have to be $1$ up to corrections of order $\alpha_\mathrm{s}^3$. Finally, the leading order corrections read
\begin{align}
 \exp\left[-\frac{F_S}{T}\right]={}&\exp\left[-\frac{f_s}{T}\right]+\frac{1}{18}(N^2-1)\alpha_\mathrm{s}^2r\pi T+\mathcal{O}\left(\alpha_\mathrm{s}^2(r\pi T)^3,\alpha_\mathrm{s}^3\right)\,, \label{rels} \\
 \exp\left[-\frac{F_A}{T}\right]={}&\exp\left[-\frac{f_o}{T}\right]-\frac{1}{18}\alpha_\mathrm{s}^2r\pi T+\mathcal{O}\left(\alpha_\mathrm{s}^2(r\pi T)^3,\alpha_\mathrm{s}^3\right)\,. \label{relo}
\end{align}
This exactly reproduces the difference between the free energies in QCD and pNRQCD at the given order.

\section{Conclusions}
\label{sec:concl}

In this paper, we have studied the Polyakov loop correlator in perturbation theory. We showed, based on general considerations, how the perturbative expansion of the Polyakov loop correlator reexponentiates into singlet and adjoint contributions. The definition of the singlet and adjoint contributions depends on the renormalization scheme and gauge, however. Using the reexponentiation formulas, the $\overline{\mathrm{MS}}$-scheme, and Coulomb gauge, we have calculated the Polyakov loop correlator up to order $g^7$ in the case $1/r\gg\pi T\gg m_D\gg\alpha_\mathrm{s}/r$, and reproduced the previous order $g^6$ result, which was obtained using static gauge~\cite{Brambilla:2010xn}. Using Coulomb gauge has the advantage that the contributions of many diagrams vanish and the calculation is reduced to only three diagrams. The order $g^7$ contribution to the Polyakov loop correlator is given in Eq.~\eqref{Og7result} and is the main result of this paper. As a byproduct of this calculation, we obtain the singlet free energy in Coulomb gauge at order $g^5$. Furthermore, we have considered the singlet free energy and the Polyakov loop correlator in the regime $\pi T\gg1/r\sim m_D$. We have discussed the power counting in this regime and reproduced an earlier result for the singlet free energy~\cite{Burnier:2009bk}. We have also reproduced the NLO result for the Polyakov loop correlator by Nadkarni~\cite{Nadkarni:1986cz}, and extended it with a partial NNLO calculation that fixes the scale of the running coupling in the leading order expression.

We have also investigated the relation of the singlet and adjoint free energies in Coulomb gauge with the gauge invariant definition of singlet and octet free energies in pNRQCD. We found that the two definitions agree at leading order in the multipole expansion, but disagree by a term proportional to $\alpha_\mathrm{s}^2r\pi T$, cf.\ Eqs.~\eqref{rels} and~\eqref{relo}. This may explain why the singlet correlator in Coulomb gauge and the cyclic Wilson loops calculated on the lattice agree quite well at short distances~\cite{Bazavov:2013zha}.

Finally, we mention that the reexponentiation of the Polyakov loop correlator and the singlet correlator was also discussed in Ref.~\cite{Pisarski:2015xea}. There, only the contribution of diagrams made of tree level propagators has been resummed, in SU(2) or in the large-$N$ limit. The authors of Ref.~\cite{Pisarski:2015xea} did not reproduce the leading order perturbative result for the singlet contribution contrary to our analysis. As shown in Appendix~\ref{app:pisarski}, this is due to the fact that the contributions of certain diagrams have been omitted. There we also show that, once the contributions of the missing diagrams are included, the correct result for the singlet correlator is reproduced.

The work presented in this paper can be extended in at least two ways. First, it will be interesting to compare the weak coupling results for the singlet free energy and Polyakov loop correlators to the lattice results in the high temperature region and see to which extent the two agree. This will clarify the question whether the onset of color screening can be understood in perturbation theory. Second, the reexponentiation formula~\eqref{F1exp} and the results obtained in this paper set the stage for a future calculation of the order $g^6$ expression of the singlet free energy, which appears to be in reach.

\acknowledgments

This work has been supported by the DFG grant BR 4058/1-2 ``Effective Field theories for heavy probes of hot plasma," and by the DFG cluster of excellence ``Origin and structure of the universe" (www.universe-cluster.de). M.B.\ acknowledges support by the Japanese Society for the Promotion of Science (JSPS). P.P.\ was supported by the U.S.\ Department of Energy under Contract No.\ \protect{DE-SC0012704}. We thank Robert Pisarski for discussions.

\appendix

\section{Calculation of the projected color factors}
\label{app:color}

In this appendix, we compute the color factors of the diagrams contributing to the Polyakov loop correlator in detail. First, we clarify the conventions related to the complex conjugation of the antiquark Polyakov loop. There is a minus sign from the $ig$ factor in the exponent, which we will use to revert the direction of the contour integration in the kinematic parts of the diagrams (indicated as an arrow to the left in Fig.~\ref{Fig1}), so for the calculation of the color coefficients, we will only use charge conjugated color matrices without this minus sign. Then we have
\begin{align}
 \widetilde{\mathcal{C}}_S\left(\begin{minipage}{20pt}\includegraphics[width=\linewidth]{FigI.eps}\end{minipage}\right)&=\mathcal{C}_S\left(\begin{minipage}{20pt}\includegraphics[width=\linewidth]{FigI.eps}\end{minipage}\right)=\frac{\delta_{ik}\delta_{jl}}{N}T^a_{ij}T^{a\,*}_{kl}\notag\\
 &=\frac{1}{N}\mathrm{Tr}\bigl[T^aT^a\bigr]=\frac{1}{2N}(N^2-1)\,,\\
 \widetilde{\mathcal{C}}_A\left(\begin{minipage}{20pt}\includegraphics[width=\linewidth]{FigI.eps}\end{minipage}\right)&=\mathcal{C}_A\left(\begin{minipage}{20pt}\includegraphics[width=\linewidth]{FigI.eps}\end{minipage}\right)=\frac{2T^{b\,*}_{ik}T^b_{jl}}{N^2-1}T^a_{ij}T^{a\,*}_{kl}\notag\\
 &=\frac{2}{N^2-1}\mathrm{Tr}\bigl[T^aT^bT^aT^b\bigr]=-\frac{1}{2N}\,,\\
 \widetilde{\mathcal{C}}_S\left(\begin{minipage}{20pt}\includegraphics[width=\linewidth]{Fig1X.eps}\end{minipage}\right)&=\mathcal{C}_S\left(\begin{minipage}{20pt}\includegraphics[width=\linewidth]{Fig1X.eps}\end{minipage}\right)-\mathcal{C}_S\left(\begin{minipage}{20pt}\includegraphics[width=\linewidth]{Fig1II.eps}\end{minipage}\right)\notag\\
 &=\frac{\delta_{ik}\delta_{jl}}{N}\bigl[(T^aT^b)_{ij}(T^{b\,*}T^{a\,*})_{kl}-(T^aT^b)_{ij}(T^{a\,*}T^{b\,*})_{kl}\bigr]\notag\\
 &=\frac{1}{N}\mathrm{Tr}\bigl[T^aT^bT^aT^b-T^aT^bT^bT^a\bigr]=-\frac{1}{4}(N^2-1)\,,\\
 \widetilde{\mathcal{C}}_A\left(\begin{minipage}{20pt}\includegraphics[width=\linewidth]{Fig1X.eps}\end{minipage}\right)&=\mathcal{C}_A\left(\begin{minipage}{20pt}\includegraphics[width=\linewidth]{Fig1X.eps}\end{minipage}\right)-\mathcal{C}_A\left(\begin{minipage}{20pt}\includegraphics[width=\linewidth]{Fig1II.eps}\end{minipage}\right)\notag\\
 &=\frac{2T^{c\,*}_{ik}T^c_{jl}}{N^2-1}\bigl[(T^aT^b)_{ij}(T^{b\,*}T^{a\,*})_{kl}-(T^aT^b)_{ij}(T^{a\,*}T^{b\,*})_{kl}\bigr]\notag\\
 &=\frac{2}{N^2-1}\mathrm{Tr}\bigl[T^aT^bT^cT^aT^bT^c-T^aT^bT^cT^bT^aT^c\bigr]=\frac{1}{4}\,,\\
 \widetilde{\mathcal{C}}_S\left(\begin{minipage}{20pt}\includegraphics[width=\linewidth]{Fig1T.eps}\end{minipage}\right)&=\mathcal{C}_S\left(\begin{minipage}{20pt}\includegraphics[width=\linewidth]{Fig1T.eps}\end{minipage}\right)-\mathcal{C}_S\left(\begin{minipage}{20pt}\includegraphics[width=\linewidth]{Fig1r.eps}\end{minipage}\right)\notag\\
 &=\frac{\delta_{ik}\delta_{jl}}{N}\bigl[(T^aT^bT^a)_{ij}T^{b\,*}_{kl}-(T^aT^bT^b)_{ij}T^{a\,*}_{kl}\bigr]\notag\\
 &=\frac{1}{N}\mathrm{Tr}\bigl[T^aT^bT^aT^b-T^aT^bT^bT^a\bigr]=-\frac{1}{4}(N^2-1)\,,\\
 \widetilde{\mathcal{C}}_A\left(\begin{minipage}{20pt}\includegraphics[width=\linewidth]{Fig1T.eps}\end{minipage}\right)&=\mathcal{C}_A\left(\begin{minipage}{20pt}\includegraphics[width=\linewidth]{Fig1T.eps}\end{minipage}\right)-\mathcal{C}_A\left(\begin{minipage}{20pt}\includegraphics[width=\linewidth]{Fig1r.eps}\end{minipage}\right)\notag\\
 &=\frac{2T^{c\,*}_{ik}T^c_{jl}}{N^2-1}\bigl[(T^aT^bT^a)_{ij}T^{b\,*}_{kl}-(T^aT^bT^b)_{ij}T^{a\,*}_{kl}\bigr]\notag\\
 &=\frac{2}{N^2-1}\mathrm{Tr}\bigl[T^aT^bT^aT^cT^bT^c-T^aT^bT^bT^cT^aT^c\bigr]=\frac{1}{4}\,.
\end{align}

\section{Calculation of unconnected diagrams}
\label{app:unconn}

In this appendix, we compute the unconnected diagrams in Eq.~\eqref{PLCDiag} at short distances. First, we note that the contributions from the scales $1/r$ and $\pi T$ vanish for unconnected diagrams without loop insertions. This can be seen by calculating the free propagator for the temporal gluons in position space:
\begin{equation}
 D_{00}(\tau,\bm{r})=\sum_K\hspace{-16pt}\int\,\frac{e^{ik_0\tau+i\bm{k}\cdot\bm{r}}}{\bm{k}^2}=\frac{\Gamma\left(\frac{d}{2}-1\right)}{4\pi^{\frac{d}{2}}r^{d-2}}\sum_{n\in\mathbb{Z}}\delta\left(\tau-\frac{n}{T}\right)\,.
 \label{Coulombprop}
\end{equation}
For all practical purposes, only the $\delta(\tau)$ term is relevant, since the argument of the propagator will always lie inside $(-1/T,1/T)$, and the boundaries do not contribute to the integral. This delta function requires the propagators to have the same imaginary time arguments at both ends, hence any two- or three-gluon diagram in Eq.~\eqref{PLCDiag} with crossed propagators vanishes when the free propagator is used, which happens for $k\sim1/r$ and $k\sim\pi T$. For $k\sim m_D$ one has to use a resummed propagator, which depends on $k_0$, and this relation cannot be used.

Next, we compute the contribution from the diagrams in the last line of Eq.~\eqref{PLCDiag}. This is a product of two diagrams. Since the first is at least of $\mathcal{O}(\alpha_\mathrm{s})$, the others need to be calculated at $\mathcal{O}\left(g^5\right)$, hence we do not have to consider higher order diagrams with loop insertions. All these diagrams have crossed gluons, hence in Coulomb gauge they all vanish except for the scale $m_D$ contribution. But since gluons with a momentum of order $m_D$ increase the order of the diagram by $g$, only one gluon is allowed to have such a momentum, otherwise the diagram would be $\mathcal{O}\left(\alpha_\mathrm{s}^3\right)$.

In the two $D_T$ diagrams, if the gluon connecting the two Polyakov loops carries a momentum either of order $m_D$ or of order $\pi T$, then we obtain a scaleless integral that vanishes in dimensional regularization. Therefore, the only contribution to $D_T$ comes when the gluon connecting the Polyakov loops carries a momentum of order $1/r$ and the other of order $m_D$. Also in $D_X$, one gluon momentum needs to be of order $1/r $ and the other of order $m_D$, but here there are two possible distributions of these momenta.

We will now show that at leading order the sum of $D_X$ and $2D_T$ vanishes. This can be seen in the following way: for the gluon with momentum of order $m_D$, the separation $r$ between the two Polyakov loops vanishes at leading order, and the time arguments of the other gluon are identical because of the delta function in the Coulomb gauge propagator. Hence the scale $m_D$ gluon in $D_X$ has the same contour integration as in $D_T$ (one endpoint to the left and one to the right of the other gluon), but there is a relative minus sign because of the opposite orientation of the two loops. In $D_X$ there is also a factor $2$ because of the different possibilities to distribute the momenta. In the multipole expansion of $D_X$ there are higher terms $m_D^2r^2$ etc., which are not canceled by $2D_T$, but those are suppressed by higher powers in $g$ and can be neglected.

We will now show this with an explicit calculation. We can use the Coulomb gauge propagator~\eqref{Coulombprop} with $d=3$ (there are no divergences at this point), and the $d$-dimensional integral of $\left(k^2+m_D^2\right)^{-1}$ gives $-m_D/4\pi$ for $d\to3$. Then we have
\begin{align}
 D_X&=(ig)^4\int_{1/T}^0d\tau_1\int_{1/T}^{\tau_1}d\tau_2\int_0^{1/T}d\tau_3\int_0^{\tau_3}d\tau_4\left(\frac{T\delta(\tau_1-\tau_3)}{4\pi r}+\frac{T\delta(\tau_2-\tau_4)}{4\pi r}\right)\int_k\frac{1+\dots}{k^2+m_D^2}\notag\\
 &=g^4\left(\int_{1/T}^0d\tau_1\int_{1/T}^{\tau_1}d\tau_2\int_0^{\tau_1}d\tau_4+\int_{1/T}^0d\tau_1\int_{1/T}^{\tau_1}d\tau_2\int_{\tau_2}^{1/T}d\tau_3\right)\left(-\frac{Tm_D}{(4\pi)^2r}+\mathcal{O}\left(g^3\right)\right)\notag\\
 &=g^4\int_{1/T}^0d\tau_1\left(-\tau_1\left(\frac{1}{T}-\tau_1\right)-\frac{1}{2}\left(\frac{1}{T}-\tau_1\right)^2\right)\left(-\frac{Tm_D}{(4\pi)^2r}+\mathcal{O}\left(g^3\right)\right)\notag\\
 &=-\frac{\alpha_\mathrm{s}^2m_D}{3rT^2}+\mathcal{O}\left(g^7\right)\,,\\
 D_T&=(ig)^4\int_{1/T}^0d\tau_1\int_0^{1/T}d\tau_2\int_0^{\tau_2}d\tau_3\int_0^{\tau_3}d\tau_4\,\frac{T\delta(\tau_1-\tau_3)}{4\pi r}\int_k\frac{1}{k^2+m_D^2}\notag\\
 &=-g^4\int_0^{1/T}d\tau_2\int_0^{\tau_2}d\tau_3\int_0^{\tau_3}d\tau_4\left(-\frac{Tm_D}{(4\pi)^2r}\right)=\frac{\alpha_\mathrm{s}^2m_D}{6rT^2}\,,
\end{align}
where we labeled the imaginary time coordinates in clockwise order starting from the antiquark loop. The combination $D_X+2D_T$ is $\mathcal{O}\left(g^7\right)$ and, therefore, the last line of Eq.~\eqref{PLCDiag} does not contribute to the Polyakov loop correlator until $\mathcal{O}\left(g^9\right)$.

A similar mechanism is at work for the unconnected diagrams of the next-to-last line of Eq.~\eqref{PLCDiag} (i.e., all except for the last two). We need to calculate these diagrams at $\mathcal{O}\left(g^7\right)$, so again no loop insertions are required. If all gluon momenta are larger than $m_D$ then each of these diagrams vanishes in Coulomb gauge because of the crossed propagators, but on the other hand only one gluon may carry a momentum of order $m_D$, because otherwise it would be $\mathcal{O}\left(g^8\right)$ or smaller. For the first two unconnected diagrams and the last one, it does not matter which gluon carries the scale $m_D$ momentum: any choice leaves two other gluons with higher scale momenta that are crossed and therefore the first two and the last unconnected three-gluon diagrams in Eq.~\eqref{PLCDiag} vanish in Coulomb gauge.

Thus, we are left with only four unconnected three-gluon diagrams, namely the third, fourth, fifth and sixth diagram in the next-to-last line of Eq.~\eqref{PLCDiag}. We denote the sum of these diagrams as $D_{3g}$. For each of the four diagrams, there is only one possibility to choose a gluon carrying a momentum of order $m_D$ in such a way that the other gluons are not crossed. Since the scale $\pi T$ does not appear, for the corresponding integrals are scaleless for unresummed propagators, the remaining gluons each carry a momentum of order $1/r$.

Now we show that $D_{3g}$ vanishes at leading order. The argument is analogous to the one in the previous case: the scale $m_D$ gluon does not distinguish between the two Polyakov loops, it starts in front of and ends behind the two parallel gluons connecting the two loops in each case, but for two of them the direction of the integration is the opposite of the other two. We also give the explicit calculation, where we use the fact, that a diagram turned upside down is identical to the original diagram for symmetry reasons:
\begin{align}
 D_{3g}={}&2(ig)^6\int_{1/T}^0d\tau_1\int_{1/T}^{\tau_1}d\tau_2\int_0^{1/T}d\tau_3\int_0^{\tau_3}d\tau_4\int_0^{\tau_4}d\tau_5\int_0^{\tau_5}d\tau_6\notag\\*
 &\times\frac{T\delta(\tau_1-\tau_5)\delta(\tau_2-\tau_4)}{(4\pi r)^2}\int_k\frac{1}{k^2+m_D^2}\notag\\
 &+2(ig)^6\int_{1/T}^0d\tau_1\int_{1/T}^{\tau_1}d\tau_2\int_{1/T}^{\tau_2}d\tau_3\int_0^{1/T}d\tau_4\int_0^{\tau_4}d\tau_5\int_0^{\tau_5}d\tau_6\notag\\
 &\times\frac{T\delta(\tau_2-\tau_6)\delta(\tau_3-\tau_5)}{(4\pi r)^2}\int_k\frac{1+\dots}{k^2+m_D^2}\notag\\
 ={}&-2g^6\int_0^{1/T}d\tau_3\int_0^{\tau_3}d\tau_4\int_0^{\tau_4}d\tau_5\int_0^{\tau_5}d\tau_6\left(-\frac{Tm_D}{(4\pi)^3r^2}\right)\notag\\
 &-2g^6\int_{1/T}^0d\tau_1\int_{1/T}^{\tau_1}d\tau_2\int_{1/T}^{\tau_2}d\tau_3\int_{\tau_3}^{1/T}d\tau_4\left(-\frac{Tm_D}{(4\pi)^3r^2}\right)+\mathcal{O}\left(g^9\right)\notag\\
 ={}&\frac{\alpha_\mathrm{s}^3m_D}{12r^2T^3}-\frac{\alpha_\mathrm{s}^3m_D}{12r^2T^3}+\mathcal{O}\left(g^9\right)=\mathcal{O}\left(g^9\right)\,.
\end{align}
In summary, we have shown that the contribution of all unconnected diagrams to the Polyakov loop correlator vanishes at order $g^7$.

\section{The H-shaped diagrams at short distances}
\label{app:Hdiag}

In this appendix, we discuss the calculation of the H-shaped diagrams at leading order. First, we will show that there are no contributions from scales $\pi T$ and $m_D$ to $D_H'$ (see Sec.~\ref{secDHmain} for the definition), where the absence of the latter ensures that corrections to $D_H'$ are of order $g^4$. The absence of scale $\pi T$ contributions is immediately apparent: we have already seen that the contour integrations combine such that all Matsubara frequencies are zero [see Eq.~\eqref{eqDHprime}], hence the scale $\pi T$ is in fact not present in the calculation.

For the scale $m_D$ contributions, we first consider the case when one of the four temporal gluon legs carries a momentum of order $m_D$, with all the others of order $1/r$. We will discuss the case $k\sim m_D$ and $p\sim q\sim 1/r$, all other cases are analogous (we use the same labels for the momenta as in the main section). The top-left propagator as well as the $\exp[i\bm{k}\cdot\bm{r}]$ factor need to be expanded in $k$; since all higher order terms in this expansion are beyond $\mathcal{O}\left(g^3\right)$, we may just insert $k=0$ in these terms. The left vertex factor is proportional to $2\bm{k}-\bm{q}$, but also here we may neglect the $k$-term at $\mathcal{O}\left(g^3\right)$. The $k$-integral is then only over a single scale $m_D$ propagator and gives the known result. But for the remaining integrals, the momentum from the vertex factor multiplies the spatial gluon, $q_iD_{ij}(0,\bm{q})$, which vanishes because of the transverse projector in the spatial Coulomb gauge propagator.

There is another option when two of the gluon momenta are of scale $m_D$. Again we will only discuss the case $k\sim q\sim m_D$ and $p\sim1/r$, all others are analogous. Now the top-right propagator has to be expanded in $q$, while the $\exp[i\bm{k}\cdot\bm{r}]$ factor needs to be expanded in $k$, but still only the leading order terms are relevant at $\mathcal{O}\left(g^3\right)$. Also in the term $-(2\bm{p}+\bm{q})$ in the numerator from the right vertex, only $\bm{p}$ needs to be kept. Consequently, the integrand of the $k$ and $q$ integrations is odd under the transformation $\bm{k}\to-\bm{k}$ and $\bm{q}\to-\bm{q}$ and vanishes. 

This leaves only the case when all gluon momenta scale like $m_D$ and both exponentials need to be expanded:
\begin{align}
 D_H'(0,0,0)\Bigr|_{g^3}&=(ig)^2\int_k\int_p\int_q\frac{-4\left((\bm{k}\cdot\bm{p})\bm{q}^2-(\bm{k}\cdot\bm{q})(\bm{p}\cdot\bm{q})\right)+\dots}
{\left(\bm{k}^2+m_D^2\right)\left((\bm{k}-\bm{q})^2+m_D^2\right)\left(\bm{q}^2\right)^2\left((\bm{p}+\bm{q})^2+m_D^2\right)\left(\bm{p}^2+m_D^2\right)}\,.
\end{align}
The $k$ and $p$ integrations both have a vector $\bm{k}$ or $\bm{p}$ in the numerator, and the only other momentum in their denominators is $\bm{q}$, so the results of both these integrals have to be proportional to $\bm{q}$ for symmetry reasons. When these are contracted with the transverse projector from the spatial gluon propagator, then they vanish. Therefore there are no contributions to $D_H'$ at all from the scale $m_D$ at $\mathcal{O}\left(g^3\right)$.

Now we will calculate the first contribution to $D_H'$ from the scale $1/r$. The integral itself is finite in 3 dimensions, however, some of the operations we are going to perform are only allowed in the framework of dimensional regularization, hence for the moment we will assume general $d$ dimensions. First, we shift the momenta $\bm{k}\to\bm{k}-\bm{p}$ and $\bm{q}\to\bm{q}-\bm{p}$. Then, the integral contains only one momentum in the exponential:
\begin{equation}
 D_H=\frac{2g^6}{T}\int_k\int_p\int_q\frac{(k-p)_i\left(\delta_{ij}(\bm{q}-\bm{p})^2-(q-p)_i(q-p)_j\right)p_je^{i\bm{k}\cdot\bm{r}}}{(\bm{k}-\bm{p})^2(\bm{k}-\bm{q})^2\left((\bm{q}-\bm{p})^2\right)^2\bm{q}^2\bm{p}^2}\,.
\end{equation}
The $p$ and $q$ integrations can be put into the form of general $k$-dependent integrals:
\begin{equation}
 I_k(n_1,n_2,n_3,n_4,n_5)\equiv\int_p\int_q\frac{1}{\left((\bm{k}-\bm{p})^2\right)^{n_1}\left((\bm{k}-\bm{q})^2\right)^{n_2}\left((\bm{p}-\bm{q})^2\right)^{n_3}\left(\bm{p}^2\right)^{n_4}\left(\bm{q}^2\right)^{n_5}}\,.
\end{equation}
Through redefinitions of the integration momenta, one can show the following identities:
\begin{equation}
 I_k(n_1,n_2,n_3,n_4,n_5)=I_k(n_2,n_1,n_3,n_5,n_4)=I_k(n_4,n_5,n_3,n_1,n_2)=I_k(n_5,n_4,n_3,n_2,n_1)\,.
\end{equation}
Reexpressing the numerator through terms that can be canceled against terms in the denominator and using these identities, we get
\begin{align}
 D_H=\frac{g^6}{T}\int_ke^{i\bm{k}\cdot\bm{r}}&\left[I_k(1,0,2,1,0)-I_k(1,0,2,0,1)+\frac{1}{2}I_k(1,1,0,1,1)\right.\notag\\*
 &\left.\phantom{\frac{1}{2}}-2I_k(1,0,1,1,1)+k^2I_k(1,1,1,1,1)\right]\,.
\end{align}

In the first integral of this expression, the $q$ integration is scaleless, so $I_k(1,0,2,1,0)=0$. The other integrals, except for the last one, can all be calculated with standard methods. The last integral can be simplified by using integration-by-parts relations. In order to obtain these, we insert $\bm{\nabla}_p\cdot\bm{p}$ or $\bm{\nabla}_p\cdot\bm{q}$ into the general expression for $I_k(n_1,n_2,n_3,n_4,n_5)$. Because it is an integral over a total derivative, each of these expressions vanishes, but if one calculates the derivatives explicitly, then one can also express it through other integrals of this type. Other relations may also be obtained, but in this case these two are sufficient.
\begin{align}
 0={}&\int_p\int_q \bm{\nabla}_p\cdot\bm{p} \frac{1}{\left((\bm{k}-\bm{p})^2\right)^{n_1}\left((\bm{k}-\bm{q})^2\right)^{n_2}\left((\bm{p}-\bm{q})^2\right)^{n_3}\left(\bm{p}^2\right)^{n_4}\left(\bm{q}^2\right)^{n_5}}\notag\\
 ={}&-n_1I_k(n_1+1,n_2,n_3,n_4-1,n_5)+n_1k^2I_k(n_1+1,n_2,n_3,n_4,n_5)\notag\\
 &-n_3I_k(n_1,n_2,n_3+1,n_4-1,n_5)+n_3I_k(n_1,n_2,n_3+1,n_4,n_5-1)\notag\\
 &+(d-n_1-n_3-2n_4)I_k(n_1,n_2,n_3,n_4,n_5)\,,\\
 0={}&\int_p\int_q \bm{\nabla}_p\cdot\bm{q} \frac{1}{\left((\bm{k}-\bm{p})^2\right)^{n_1}\left((\bm{k}-\bm{q})^2\right)^{n_2}\left((\bm{p}-\bm{q})^2\right)^{n_3}\left(\bm{p}^2\right)^{n_4}\left(\bm{q}^2\right)^{n_5}}\notag\\
 ={}&-n_1I_k(n_1+1,n_2-1,n_3,n_4,n_5)+n_1I_k(n_1+1,n_2,n_3-1,n_4,n_5)\notag\\
 &-n_1I_k(n_1+1,n_2,n_3,n_4-1,n_5)+n_1k^2I_k(n_1+1,n_2,n_3,n_4,n_5)\notag\\
 &-n_3I_k(n_1,n_2,n_3+1,n_4-1,n_5)+n_3I_k(n_1,n_2,n_3+1,n_4,n_5-1)\notag\\
 &+n_4I_k(n_1,n_2,n_3-1,n_4+1,n_5)-n_4I_k(n_1,n_2,n_3,n_4+1,n_5-1)\notag\\
 &+(n_3-n_4)I_k(n_1,n_2,n_3,n_4,n_5)\,.
\end{align}
Subtracting the second relation from the first, we obtain
\begin{align}
 0={}&(d-n_1-2n_3-n_4)I_k(n_1,n_2,n_3,n_4,n_5)\notag\\
 &+n_1I_k(n_1+1,n_2-1,n_3,n_4,n_5)-n_1I_k(n_1+1,n_2,n_3-1,n_4,n_5)\notag\\
 &-n_4I_k(n_1,n_2,n_3-1,n_4+1,n_5)+n_4I_k(n_1,n_2,n_3,n_4+1,n_5-1)\,.
\end{align}
This relation can be used repeatedly to lower either the index $n_2$, $n_3$, or $n_5$ to $0$, at which point the integral is straightforward to calculate. In the case of $I_k(1,1,1,1,1)$, one iteration is sufficient:
\begin{equation}
 I_k(1,1,1,1,1)=\frac{2}{4-d}I_k(2,0,1,1,1)-\frac{2}{4-d}I_k(1,2,0,1,1)\,,
\end{equation}
where we have used the symmetry relations again.

We now give the results of the integrals when one index is $0$. Because of the symmetry relations we only need to consider two cases:
\begin{align}
 I_k(n_1,0,n_3,n_4,n_5)={}&\int_p\frac{1}{\left((\bm{k}-\bm{p})^2\right)^{n_1}\left(\bm{p}^2\right)^{n_4}}\int_q\frac{1}{\left((\bm{p}-\bm{q})^2\right)^{n_3}\left(\bm{q}^2\right)^{n_5}}\notag\\
 ={}&\int_p\frac{1}{\left((\bm{k}-\bm{p})^2\right)^{n_1}\left(\bm{p}^2\right)^{n_4}}
\frac{\mu^{d-3}\Gamma\left(n_3+n_5-\frac{d}{2}\right)\Gamma\left(\frac{d}{2}-n_3\right)\Gamma\left(\frac{d}{2}-n_5\right)}{(4\pi)^{d/2}\left(\bm{p}^2\right)^{n_3+n_5-d/2}\Gamma(n_3)\Gamma(n_5)\Gamma(d-n_3-n_5)}\notag\\
 ={}&\frac{\left(\bm{k}^2\right)^{d-n_1-n_3-n_4-n_5}\mu^{6-2d}}{(4\pi)^d}\frac{\Gamma\left(\frac{d}{2}-n_1\right)\Gamma\left(\frac{d}{2}-n_3\right)\Gamma\left(\frac{d}{2}-n_5\right)}{\Gamma(n_1)\Gamma(n_3)\Gamma(n_5)}\notag\\
 &\times\frac{\Gamma\left(n_3+n_5-\frac{d}{2}\right)\Gamma(d-n_3-n_4-n_5)\Gamma(n_1+n_3+n_4+n_5-d)}{\Gamma(d-n_3-n_5)\Gamma\left(n_3+n_4+n_5-\frac{d}{2}\right)\Gamma\left(\frac{3d}{2}-n_1-n_3-n_4-n_5\right)}\,,\\
 I_k(n_1,n_2,0,n_4,n_5)={}&\int_p\frac{1}{\left((\bm{k}-\bm{p})^2\right)^{n_1}\left(\bm{p}^2\right)^{n_4}}\int_q\frac{1}{\left((\bm{k}-\bm{q})^2\right)^{n_2}\left(\bm{q}^2\right)^{n_5}}\notag\\
 ={}&\frac{\mu^{3-d}\Gamma\left(n_1+n_4-\frac{d}{2}\right)\Gamma\left(\frac{d}{2}-n_1\right)\Gamma\left(\frac{d}{2}-n_4\right)}{(4\pi)^{d/2}\left(\bm{k}^2\right)^{n_1+n_4-d/2}\Gamma(n_1)\Gamma(n_4)\Gamma(d-n_1-n_4)}\notag\\
 &\times\frac{\mu^{3-d}\Gamma\left(n_2+n_5-\frac{d}{2}\right)\Gamma\left(\frac{d}{2}-n_2\right)\Gamma\left(\frac{d}{2}-n_5\right)}{(4\pi)^{d/2}\left(\bm{k}^2\right)^{n_2+n_5-d/2}\Gamma(n_2)\Gamma(n_5)\Gamma(d-n_2-n_5)}\notag\\
 ={}&\frac{\left(\bm{k}^2\right)^{d-n_1-n_2-n_4-n_5}\mu^{6-d}}{(4\pi)^d}
\frac{\Gamma\left(\frac{d}{2}-n_1\right)\Gamma\left(\frac{d}{2}-n_2\right)\Gamma\left(\frac{d}{2}-n_4\right)\Gamma\left(\frac{d}{2}-n_5\right)}{\Gamma(n_1)\Gamma(n_2)\Gamma(n_4)\Gamma(n_5)}\notag\\
 &\times\frac{\Gamma\left(n_1+n_4-\frac{d}{2}\right)\Gamma\left(n_2+n_5-\frac{d}{2}\right)}{\Gamma(d-n_1-n_4)\Gamma(d-n_2-n_5)}\,.
\end{align}

Then we have
\begin{align}
 D_H&=\frac{g^6}{T}\int_ke^{i\bm{k}\cdot\bm{r}}\left[-I_k(1,0,2,0,1)+\frac{1}{2}I_k(1,1,0,1,1)-2I_k(1,0,1,1,1)\right.\notag\\
 &\hspace{73pt}+\left.\frac{2k^2}{4-d}I_k(2,0,1,1,1)-\frac{2k^2}{4-d}I_k(2,1,0,1,1)\right]\notag\\
 &=\frac{g^6}{T}\int_k\frac{e^{i\bm{k}\cdot\bm{r}}\mu^{6-2d}}{(4\pi)^d\left(\bm{k}^2\right)^{4-d}}\left[-\frac{\Gamma^2\left(\frac{d}{2}-1\right)\Gamma\left(\frac{d}{2}-2\right)\Gamma(4-d)}{\Gamma\left(\frac{3d}{2}-4\right)}+\frac{\Gamma^4\left(\frac{d}{2}-1\right)\Gamma^2\left(2-\frac{d}{2}\right)}{2\Gamma^2(d-2)}\right.\notag\\
 &\hspace{121pt}-\frac{2\Gamma^3\left(\frac{d}{2}-1\right)\Gamma\left(2-\frac{d}{2}\right)\Gamma(d-3)\Gamma(4-d)}{\Gamma(d-2)\Gamma\left(3-\frac{d}{2}\right)\Gamma\left(\frac{3d}{2}-4\right)}\notag\\
 &\hspace{121pt}+\frac{2\Gamma^2\left(\frac{d}{2}-1\right)\Gamma\left(\frac{d}{2}-2\right)\Gamma\left(2-\frac{d}{2}\right)\Gamma(d-3)\Gamma(5-d)}{(4-d)\Gamma(d-2)\Gamma\left(3-\frac{d}{2}\right)\Gamma\left(\frac{3d}{2}-5\right)}\notag\\
 &\hspace{121pt}-\left.\frac{2\Gamma^3\left(\frac{d}{2}-1\right)\Gamma\left(\frac{d}{2}-2\right)\Gamma\left(2-\frac{d}{2}\right)\Gamma\left(3-\frac{d}{2}\right)}{(4-d)\Gamma(d-2)\Gamma(d-3)}\right]\notag\\
 &=\frac{g^6(\mu r)^{9-3d}}{4^4\pi^{3d/2}rT}\frac{\Gamma\left(\frac{3d}{2}-4\right)}{\Gamma(4-d)}\left[-\frac{2d\Gamma^3\left(\frac{d}{2}-1\right)\Gamma(4-d)}{(4-d)^2\Gamma\left(\frac{3d}{2}-4\right)}+\frac{(3d-8)\Gamma^4\left(\frac{d}{2}-1\right)\Gamma^2\left(2-\frac{d}{2}\right)}{2(4-d)\Gamma^2(d-2)}\right]\notag\\*
 &=\frac{\alpha_\mathrm{s}^3(\mu r)^{9-3d}}{4\pi^{3d/2-3}rT}\left[-\frac{2d}{(4-d)^2}\Gamma^3\left(\tfrac{d}{2}-1\right)+\frac{\Gamma\left(\frac{3d}{2}-3\right)\Gamma^4\left(\frac{d}{2}-1\right)\Gamma^2\left(2-\frac{d}{2}\right)}{\Gamma(5-d)\Gamma^2(d-2)}\right]\,.
\end{align}
For $d=3$ this gives
\begin{equation}
 D_H=\frac{\alpha_\mathrm{s}^3}{rT}\left(-\frac{3}{2}+\frac{\pi^2}{8}\right)\,.
\label{DHvalue}
\end{equation}
The result is consistent with a similar finding in Ref.~\cite{Kummer:1996jz}.

\section{Calculations of massive diagrams}
\label{massive_int}

In this appendix, we consider the calculations of $D_H$ and $D_I$ for the case $\pi T\gg1/r\sim m_D$. The temperature scale does not contribute, because momenta of this scale lead to exponentially suppressed terms that do not appear in the expansions, or to scaleless integrals. Therefore all momenta are of the order of the Debye mass $m_D$, which means that all temporal gluons have massive propagators.

We begin with the H-shaped diagrams, which give a leading contribution of $\mathcal{O}\left(g^7\right)$ in this regime:
\begin{equation}
 D_H=\frac{2g^6}{T}\int_k\int_p\int_q\frac{(k-p)_i\left(\delta_{ij}(\bm{p}-\bm{q})^2-(p-q)_i(p-q)_j\right)p_je^{i\bm{k}\cdot\bm{r}}}{\left((\bm{k}-\bm{p})^2+m_D^2\right)\left((\bm{k}-\bm{q})^2+m_D^2\right)\left((\bm{p}-\bm{q})^2\right)^2\left(\bm{p}^2+m_D^2\right)\left(\bm{q}^2+m_D^2\right)}\,.
\end{equation}
We will proceed in the same fashion as in the previous appendix; large portions of the calculation remain the same, we just have to introduce mass terms in the $k$-dependent integrals:
\begin{align}
 &J_k(n_1,n_2,n_3,n_4,n_5)\notag\\
 &\equiv\int_p\int_q\frac{1}{\left((\bm{k}-\bm{p})^2+m_D^2\right)^{n_1}\left((\bm{k}-\bm{q})^2+m_D^2\right)^{n_2}\left((\bm{p}-\bm{q})^2\right)^{n_3}\left(\bm{p}^2+m_D^2\right)^{n_4}\left(\bm{q}^2+m_D^2\right)^{n_5}}\,.
\end{align}
The identities are still valid, since none of them exchange the indices of a massive with the ones of a massless denominator:
\begin{equation}
 J_k(n_1,n_2,n_3,n_4,n_5)=J_k(n_2,n_1,n_3,n_5,n_4)=J_k(n_4,n_5,n_3,n_1,n_2)=J_k(n_5,n_4,n_3,n_2,n_1)\,.
\end{equation}
Again, we cancel the terms in the denominator and simplify the resulting expression through the identities; the result is almost unchanged:
\begin{align}
 D_H=\frac{g^6}{T}\int_ke^{i\bm{k}\cdot\bm{r}}&\left[J_k(1,0,2,1,0)-J_k(1,0,2,0,1)-2J_k(1,0,1,1,1)\phantom{\frac{1}{2}}\right.\notag\\
 &+\left.\frac{1}{2}J_k(1,1,0,1,1)+\left(k^2+2m_D^2\right)J_k(1,1,1,1,1)\right]\,.
\end{align}

The integration-by-parts relations change as follows:
\begin{align}
 &\int_p\int_q \bm{\nabla}_p\cdot\bm{p} \frac{1}{\left((\bm{k}-\bm{p})^2+m_D^2\right)^{n_1}\left((\bm{k}-\bm{q})^2+m_D^2\right)^{n_2}\left((\bm{p}-\bm{q})^2\right)^{n_3}\left(\bm{p}^2+m_D^2\right)^{n_4}\left(\bm{q}^2+m_D^2\right)^{n_5}}\notag\\
 &=-n_1J_k(n_1+1,n_2,n_3,n_4-1,n_5)+n_1\left(k^2+2m_D^2\right)J_k(n_1+1,n_2,n_3,n_4,n_5)\notag\\
 &\phantom{=}-n_3J_k(n_1,n_2,n_3+1,n_4-1,n_5)+n_3J_k(n_1,n_2,n_3+1,n_4,n_5-1)\notag\\
 &\phantom{=}+2n_4m_D^2J_k(n_1,n_2,n_3,n_4+1,n_5)+(d-n_1-n_3-2n_4)J_k(n_1,n_2,n_3,n_4,n_5)=0\,,\\
 &\int_p\int_q \bm{\nabla}_p\cdot\bm{q} \frac{1}{\left((\bm{k}-\bm{p})^2+m_D^2\right)^{n_1}\left((\bm{k}-\bm{q})^2+m_D^2\right)^{n_2}\left((\bm{p}-\bm{q})^2\right)^{n_3}\left(\bm{p}^2+m_D^2\right)^{n_4}\left(\bm{q}^2+m_D^2\right)^{n_5}}\notag\\
 &=-n_1J_k(n_1+1,n_2-1,n_3,n_4,n_5)+n_1J_k(n_1+1,n_2,n_3-1,n_4,n_5)\notag\\
 &\phantom{=}-n_1J_k(n_1+1,n_2,n_3,n_4-1,n_5)+n_1\left(k^2+2m_D^2\right)J_k(n_1+1,n_2,n_3,n_4,n_5)\notag\\
 &\phantom{=}-n_3J_k(n_1,n_2,n_3+1,n_4-1,n_5)+n_3J_k(n_1,n_2,n_3+1,n_4,n_5-1)\notag\\
 &\phantom{=}+n_4J_k(n_1,n_2,n_3-1,n_4+1,n_5)-n_4J_k(n_1,n_2,n_3,n_4+1,n_5-1)\notag\\
 &\phantom{=}+2n_4m_D^2J_k(n_1,n_2,n_3,n_4+1,n_5)+(n_3-n_4)J_k(n_1,n_2,n_3,n_4,n_5)=0\,.
\end{align}

Now there are a few more terms due to the Debye mass appearing in the numerators when terms are canceled with the denominators; however, when we take the difference between both expressions, those terms cancel again and the relation is identical to the massless case:
\begin{align}
 0={}&(d-n_1-2n_3-n_4)J_k(n_1,n_2,n_3,n_4,n_5)\notag\\
 &+n_1J_k(n_1+1,n_2-1,n_3,n_4,n_5)-n_1J_k(n_1+1,n_2,n_3-1,n_4,n_5)\notag\\
 &-n_4J_k(n_1,n_2,n_3-1,n_4+1,n_5)+n_4J_k(n_1,n_2,n_3,n_4+1,n_5-1)\,.
\end{align}
This means that we also obtain the same reduction for the most complicated integral:
\begin{equation}
 J_k(1,1,1,1,1)=\frac{2}{4-d}J_k(2,0,1,1,1)-\frac{2}{4-d}J_k(1,2,0,1,1)\,.
\end{equation}

In order to calculate the integrals appearing in the H-shaped diagrams, we are no longer able to give a general formula for the integrals with one index equal to zero in $d$ dimensions. Instead, we will calculate them explicitly in $d=3$. The first integral $J_k(1,0,2,1,0)$ still has a scaleless $q$ integration and would vanish in dimensional regularization, but for $d=3$ it is needed to cancel an infrared divergence in $J_k(1,0,2,0,1)$, so we have to keep it. The following two integrals $J_k(1,0,1,1,1)$ and $J_k(2,0,1,1,1)$ both have canceling infrared divergences in the massless case, but now those are separately removed through the mass term. Then we have
\begin{align}
 D_H=\frac{g^6}{T}\int_ke^{i\bm{k}\cdot\bm{r}}\biggl[&J_k(1,0,2,1,0)-J_k(1,0,2,0,1)-2J_k(1,0,1,1,1)+\frac{1}{2}J_k(1,1,0,1,1)\notag\\
 &+2\left(k^2+2m_D^2\right)J_k(2,0,1,1,1)-2\left(k^2+2m_D^2\right)J_k(2,1,0,1,1)\biggr]
\end{align}

We will now calculate these one by one:
\begin{align}
 &J_k(1,0,2,1,0)-J_k(1,0,2,0,1)=\int_p\frac{1}{\left((\bm{k}-\bm{p})^2+m_D^2\right)\left(\bm{p}^2+m_D^2\right)}\int_q\frac{\bm{q}^2-\bm{p}^2}{\left((\bm{q}-\bm{p})^2\right)^2\left(\bm{q}^2+m_D^2\right)}\notag\\
 &=\int_p\frac{1}{\left((\bm{k}-\bm{p})^2+m_D^2\right)\left(\bm{p}^2+m_D^2\right)}\int_0^\infty\frac{q^2dq}{4\pi^2}\int_{-1}^1dx\,\frac{q^2-p^2}{\left(q^2-2pqx+p^2\right)^2\left(q^2+m_D^2\right)}\notag\\
 &=\int_p\frac{1}{\left((\bm{k}-\bm{p})^2+m_D^2\right)\left(\bm{p}^2+m_D^2\right)}\int_0^\infty\frac{q^2dq}{2\pi^2}\frac{1}{\left(q^2-p^2\right)\left(q^2+m_D^2\right)}\notag\\
 &=\frac{m_D}{4\pi}\int_p\frac{1}{\left((\bm{k}-\bm{p})^2+m_D^2\right)\left(\bm{p}^2+m_D^2\right)^2}=\frac{m_D}{4\pi}\int_p\frac{1}{\left(\bm{p}^2+m_D^2\right)\left((\bm{k}-\bm{p})^2+m_D^2\right)^2}\notag\\
 &=\frac{m_D}{4\pi}\int_0^\infty\frac{p^2dp}{4\pi^2}\int_{-1}^1dx\,\frac{1}{\left(p^2+m_D^2\right)\left(k^2-2kpx+p^2+m_D^2\right)^2}\notag\\
 &=\frac{m_D}{4\pi}\int_0^\infty\frac{p^2dp}{2\pi^2}\frac{1}{\left(p^2+m_D^2\right)\left((k-p)^2+m_D^2\right)\left((k+p)^2+m_D^2\right)}\notag\\
 &=\frac{m_D}{4\pi}\frac{1}{8\pi m_D\left(k^2+4m_D^2\right)}=\frac{1}{32\pi^2\left(k^2+4m_D^2\right)}\,,
\end{align}
where we have taken the principal value for the $q=p$ pole in the $q$ integral. In the original expression for $D_H$, there was also a contribution where the roles of $p$ and $q$ were reversed, which we have eliminated through the redefinition $\bm{p}\leftrightarrow\bm{q}$. If this were kept without changes, then the pole would cancel between the two expressions, showing that the principal value is the right prescription to treat this artificial pole. We can also see that if we take $m_D\to0$ then the result agrees with the massless calculation for $d=3$.

The corresponding contribution to $D_H$ is then
\begin{equation}
 \frac{g^6}{T}\int_ke^{i\bm{k}\cdot\bm{r}}\left(J_k(1,0,2,1,0)-J_k(1,0,2,0,1)\right)=\frac{g^6}{32\pi^2T}\int_k\frac{e^{i\bm{k}\cdot\bm{r}}}{k^2+4m_D^2}=\frac{\alpha_\mathrm{s}^3e^{-2rm_D}}{2rT}\,.
\end{equation}

For the next integral, it is more convenient to include the $k$ integration from the beginning:
\begin{align}
 \int_ke^{i\bm{k}\cdot\bm{r}}J_k(1,0,1,1,1)&=\int_k\int_p\frac{e^{i\bm{k}\cdot\bm{r}}}{\left((\bm{k}-\bm{p})^2+m_D^2\right)\left(\bm{p}^2+m_D^2\right)}\int_q\frac{1}{(\bm{p}-\bm{q})^2\left(\bm{q}^2+m_D^2\right)}\notag\\
 &=\int_k\frac{e^{i\bm{k}\cdot\bm{r}}}{k^2+m_D^2}\int_p\frac{e^{i\bm{p}\cdot\bm{r}}}{p^2+m_D^2}\int_0^1dx\int_q\frac{1}{\left(q^2+x(1-x)p^2+xm_D^2\right)^2}\notag\\
 &=\frac{e^{-rm_D}}{4\pi r}\int_p\frac{e^{i\bm{p}\cdot\bm{r}}}{p^2+m_D^2}\int_0^1dx\,\frac{1}{8\pi\sqrt{x(1-x)p^2+xm_D^2}}\notag\\
 &=\frac{e^{-rm_D}}{4\pi r}\int_p\frac{e^{i\bm{p}\cdot\bm{r}}}{p^2+m_D^2}\frac{1}{4\pi p}\arctan\frac{p}{m_D}\notag\\
 &=\frac{e^{-rm_D}}{4\pi r}\int_0^\infty\frac{dp}{8\pi^3r}\frac{\sin(pr)}{p^2+m_D^2}\arctan\frac{p}{m_D}\notag\\
 &=\frac{e^{-rm_D}}{4\pi r}\int_{-\infty}^\infty\frac{dp}{16\pi^3r}\frac{-ie^{ipr}}{p^2+m_D^2}\arctan\frac{p}{m_D}\,.
\end{align}
We could replace the sine function by the exponential in the last step, because the cosine term gives an odd integrand and vanishes. The remaining $p$~integration can be put into the form of more standard integrals by deforming the contour in the complex plane. We can connect real $-\infty$ to $+\infty$ by a semicircle of infinite radius in the upper half-plane. However, the arctangent has a discontinuity along the imaginary axis starting from the pole at $im_D$, so we have to integrate around that.

The contributions from the circle segments at complex infinity vanish because of the exponential, so only the integrations along the imaginary axis and around the pole remain. For the first segment, we choose $p=i\kappa-\delta$ with $\kappa$ from $\infty$ to $m_D+\epsilon$; for the second segment we take $p=i\left(m_D+\epsilon e^{i\varphi}\right)$ with $\varphi$ from $\arctan(\delta/\epsilon)$ to $2\pi-\arctan(\delta/\epsilon)$; for the third segment we take $p=i\kappa+\delta$ with $\kappa$ from $m_D+\epsilon$ to $\infty$; for the infinitesimal parameters $\delta$ and $\epsilon$ we first take $\delta\to0$ and then $\epsilon\to0$ (an illustration of this contour can be found in Fig.~3 of~\cite{Nadkarni:1986cz}). Then we have
\begin{align}
 &\int_{-\infty}^\infty\frac{dp}{16\pi^3r}\frac{-ie^{ipr}}{p^2+m_D^2}\arctan\frac{p}{m_D}\notag\\
 &=\lim_{\delta,\epsilon\to0}\int_\infty^{m_D+\epsilon}\frac{d\kappa}{16\pi^3r}\frac{e^{-\kappa r-i\delta r}}{m_D^2-\kappa^2-2i\kappa\delta+\delta^2}\arctan\frac{i\kappa-\delta}{m_D}\notag\\
 &\phantom{=}-\lim_{\epsilon\to0}\int_0^{2\pi}\frac{\epsilon d\varphi}{16\pi^3r}\frac{ie^{i\varphi}e^{-rm_D\left(1+\epsilon e^{i\varphi}\right)}}{2im_D\epsilon\,e^{i\varphi}+\epsilon^2e^{2i\varphi}}\arctan\left(i+\frac{i\epsilon}{m_D}e^{i\varphi}\right)\notag\\
 &\phantom{=}+\lim_{\delta,\epsilon\to0}\int_{m_D+\epsilon}^\infty\frac{d\kappa}{16\pi^3r}\frac{e^{-\kappa r+i\delta r}}{m_D^2-\kappa^2+2i\kappa\delta+\delta^2}\arctan\frac{i\kappa+\delta}{m_D}\notag\\
 &=-\lim_{\epsilon\to0}\left[\int_{m_D+\epsilon}^\infty\frac{d\kappa}{16\pi^2r}\frac{e^{-\kappa r}}{\kappa^2-m_D^2}+\int_0^{2\pi}\frac{d\varphi}{16\pi^3r}\frac{e^{-m_Dr}}{4m_D}\ln\frac{\epsilon\,e^{i(\varphi-\pi)}}{2m_D}\right]\notag\\
 &=-\lim_{\epsilon\to0}\left[\int_{m_D+\epsilon}^\infty\frac{d\kappa}{32\pi^2rm_D}\left(\frac{e^{-\kappa r}}{\kappa-m_D}-\frac{e^{-\kappa r}}{\kappa+m_D}\right)+\frac{1}{16\pi^2r}\frac{e^{-r(m_D+\epsilon)}}{2m_D+\epsilon}\ln\frac{\epsilon}{2m_D}\right]\notag\\
 &=\int_{m_D}^\infty\frac{d\kappa}{32\pi^2rm_D}\left(\frac{e^{-\kappa r}}{\kappa+m_D}-r\ln(r(\kappa-m_D))e^{-\kappa r}\right)+\frac{e^{-rm_D}}{32\pi^2rm_D}\ln2rm_D\notag\\
 &=\frac{e^{-rm_D}}{32\pi^2rm_D}\left(\int_{2rm_D}^\infty dx\frac{e^{-x+2rm_D}}{x}-\int_{0}^\infty dx \,  e^{-x} \, \ln x  + \ln2rm_D\right)\notag\\
 &=\frac{e^{-rm_D}}{32\pi^2rm_D}\left(e^{2rm_D}\Gamma(0,2rm_D)+\gamma_E+\ln2rm_D\right)\,,
\end{align}
where $\Gamma(0,2rm_D)$ is the upper incomplete gamma function:
\begin{equation}
 \Gamma(s,x)=\int_x^\infty dt\,t^{s-1}e^{-t}\,.
\end{equation}
The resulting contribution to $D_H$ is
\begin{equation}
 -\frac{2g^6}{T}\int_ke^{i\bm{k}\cdot\bm{r}}J_k(1,0,1,1,1)=-\frac{\alpha_\mathrm{s}^3e^{-2rm_D}}{(rT)(rm_D)}\left(e^{2rm_D}\Gamma(0,2rm_D)+\gamma_E+\ln2rm_D\right)\,.
\end{equation}

The next integral we need is
\begin{align}
 J_k(1,1,0,1,1)&=\int_p\frac{1}{\left((\bm{k}-\bm{p})^2+m_D^2\right)\left(\bm{p}^2+m_D^2\right)}\int_q\frac{1}{\left((\bm{k}-\bm{q})^2+m_D^2\right)\left(\bm{q}^2+m_D^2\right)}\notag\\
 &=\left(\int_0^1dx\int_p\frac{1}{\left(p^2+x(1-x)k^2+m_D^2\right)^2}\right)^2\notag\\
 &=\left(\int_0^1dx\int_{-\infty}^\infty\frac{dp}{4\pi^2}\frac{p^2}{\left(p^2+x(1-x)k^2+m_D^2\right)^2}\right)^2\notag\\
 &=\left(\int_0^1dx\,\frac{1}{8\pi\sqrt{x(1-x)k^2+m_D^2}}\right)^2=\left(\frac{1}{4\pi k}\arctan\frac{k}{2m_D}\right)^2\,.
\end{align}
For the contribution to $D_H$, we use the same contour in the complex plane as in the previous calculation, except that the branch cut starts at $2im_D$ instead of $im_D$. We may also neglect the circle around this point, since the singularity is only logarithmic. Therefore we have
\begin{align}
 \frac{g^6}{2T}\int_ke^{i\bm{k}\cdot\bm{r}}J_k(1,1,0,1,1)&=\frac{g^6}{2T}\int_0^\infty\frac{k^2dk}{2\pi^2}\frac{\sin kr}{kr}\left(\frac{1}{4\pi k}\arctan\frac{k}{2m_D}\right)^2\notag\\
 &=\frac{g^6}{128\pi^4rT}\int_{-\infty}^\infty dk\frac{-ie^{ikr}}{k}\left(\arctan\frac{k}{2m_D}\right)^2\notag\\
 &=\frac{\alpha_\mathrm{s}^3}{2rT}\int_{2m_D}^\infty d\kappa\frac{e^{-\kappa r}}{\kappa}\ln\frac{\kappa+2m_D}{\kappa-2m_D}\notag\\
 &=\frac{\alpha_\mathrm{s}^3e^{-2rm_D}}{2rT}\int_0^\infty dx\frac{e^{-2rm_Dx}}{x+1}\ln\frac{x+2}{x}\,.
\end{align}

Finally, we have
\begin{align}
 J_k(2,0,1,1,1)&=\int_p\frac{1}{\left((\bm{k}-\bm{p})^2+m_D^2\right)^2\left(\bm{p}^2+m_D^2\right)}\int_q\frac{1}{(\bm{p}-\bm{q})^2\left(\bm{q}^2+m_D^2\right)}\notag\\
 &=\int_p\frac{1}{\left((\bm{k}-\bm{p})^2+m_D^2\right)^2\left(\bm{p}^2+m_D^2\right)}\frac{1}{4\pi p}\arctan\frac{p}{m_D}\notag\\
 &=\int_0^\infty\frac{p^2dp}{4\pi^2}\int_{-1}^1dx\,\frac{1}{\left(k^2-2kpx+p^2+m_D^2\right)^2\left(p^2+m_D^2\right)}\frac{1}{4\pi p}\arctan\frac{p}{m_D}\notag\\
 &=\int_0^\infty\frac{p^2dp}{2\pi^2}\frac{1}{\left(p^2+m_D^2\right)\left((k-p)^2+m_D^2\right)\left((k+p)^2+m_D^2\right)}\frac{1}{4\pi p}\arctan\frac{p}{m_D}\notag\\
 &=\frac{1}{32\pi^2}\left[\frac{1}{k^2\left(k^2+4m_D^2\right)}\ln\left(1+\frac{k^2}{4m_D^2}\right)+\frac{1}{km_D\left(k^2+4m_D^2\right)}\arctan\frac{k}{2m_D}\right].
\end{align}

In the $k$ integration over this last term, there would be an ultraviolet divergence, because the coefficient compensates the $1/(k^2+4m_D^2)$ denominator, hence this has to be canceled by the other integral with $n_1=2$:
\begin{align}
 J_k(2,1,0,1,1)&=\int_p\frac{1}{\left((\bm{k}-\bm{p})^2+m_D^2\right)^2\left(\bm{p}^2+m_D^2\right)}\int_q\frac{1}{\left((\bm{k}-\bm{q})^2+m_D^2\right)\left(\bm{q}^2+m_D^2\right)}\notag\\
 &=\left(\int_0^1dx\int_p\frac{2x}{\left(p^2+x(1-x)k^2+m_D^2\right)^3}\right)\left(\frac{1}{4\pi k}\arctan\frac{k}{2m_D}\right)\notag\\
 &=\left(\int_0^1dx\int_{-\infty}^\infty\frac{dp}{4\pi^2}\frac{2xp^2}{\left(p^2+x(1-x)k^2+m_D^2\right)^3}\right)\left(\frac{1}{4\pi k}\arctan\frac{k}{2m_D}\right)\notag\\
 &=\left(\int_0^1dx\,\frac{x}{16\pi\left(x(1-x)k^2+m_D^2\right)^{3/2}}\right)\left(\frac{1}{4\pi k}\arctan\frac{k}{2m_D}\right)\notag\\
 &=\left(\frac{1}{8\pi m_D}\frac{1}{k^2+4m_D^2}\right)\left(\frac{1}{4\pi k}\arctan\frac{k}{2m_D}\right)\,.
\end{align}

Together, they contribute to $D_H$ (using again the same contour with the branch cut starting at $2im_D$, but this time including the circle) as
\begin{align}
 &\frac{2g^6}{T}\int_ke^{i\bm{k}\cdot\bm{r}}\left(k^2+2m_D^2\right)\bigl[J_k(2,0,1,1,1)-J_k(2,1,0,1,1)\bigr]\notag\\
 &=\frac{2g^6}{T}\int_0^\infty\frac{dk}{64\pi^4}\frac{\sin kr}{kr}\frac{k^2+2m_D^2}{k^2+4m_D^2}\ln\left(1+\frac{k^2}{4m_D^2}\right)\notag\\
 &=\frac{\alpha_\mathrm{s}^3}{rT}\lim_{\epsilon\to0}\left[\int_{2m_D+\epsilon}^\infty\frac{d\kappa}{\kappa}\frac{2e^{-\kappa r}\left(\kappa^2-2m_D^2\right)}{\kappa^2-4m_D^2}+\int_0^{2\pi}\frac{i\epsilon e^{i\varphi}d\varphi}{2im_D\pi}\frac{2m_D^2e^{-2rm_D}}{4m_D\epsilon e^{i\varphi}}\ln\frac{\epsilon e^{i(\varphi-\pi)}}{m_D}\right]\notag\\
 &=\frac{\alpha_\mathrm{s}^3}{rT}\lim_{\epsilon\to0}\left[\int_{2m_D+\epsilon}^\infty d\kappa e^{-\kappa r}\left(\frac{1}{\kappa}+\frac{1}{2(\kappa-2m_D)}+\frac{1}{2(\kappa+2m_D)}\right)+\int_0^{2\pi}\frac{d\varphi}{4\pi}e^{-2rm_D}\ln\frac{\epsilon e^{i(\varphi-\pi)}}{m_D}\right]\notag\\
 &=\frac{\alpha_\mathrm{s}^3}{rT}\left[\Gamma(0,2rm_D)+\frac{e^{2rm_D}}{2}\Gamma(0,4rm_D)+\int_{2m_D}^\infty d\kappa\frac{re^{-\kappa r}}{2}\ln\frac{\kappa-2m_D}{m_D}\right]\notag\\
 &=\frac{\alpha_\mathrm{s}^3e^{-2rm_D}}{rT}\left[e^{2rm_D}\Gamma(0,2rm_D)+\frac{e^{4rm_D}}{2}\Gamma(0,4rm_D)-\frac{\gamma_E}{2}-\frac{1}{2}\ln rm_D\right]\,.
\end{align}

Combining all these results, we get the full expression for $D_H$ in the $1/r\sim m_D$ regime:
\begin{align}
 D_H=\frac{\alpha_\mathrm{s}^3e^{-2rm_D}}{2rT}&\biggl[1-\frac{2}{rm_D}\left(e^{2rm_D}E_1(2rm_D)+\gamma_E+\ln2rm_D\right)+\int_0^\infty dx\frac{e^{-2rm_Dx}}{x+1}\ln\frac{x+2}{x}\notag\\
 &+2e^{2rm_D} E_1(2rm_D)+e^{4rm_D}E_1(4rm_D)-\gamma_E-\ln rm_D\biggr]\,,
 \label{DHmass}
\end{align}
where we used that $\Gamma(0,x)=E_1(x)=\int_x^{\infty}d t e^{-t}/t$. 

To obtain $D_I$ for $1/r\sim m_D$ at $\mathcal{O}\left(g^4\right)$, we need the temporal gluon self-energy at one-loop order for momenta $k\sim m_D\ll T$. We have
\begin{equation}
 \Pi_{00}(k)=m_D^2-\delta Z_1k^2+\Pi_{00}^{(s)}(k)+\mathcal{O}\left(\alpha_s k^4/T^2\right)\,,
 \label{Pi00exp}
\end{equation}
where (after charge renormalization)
\begin{equation}
 \delta Z_1=\frac{\alpha_\mathrm{s}}{4\pi}\left[\frac{11}{3}N+\frac{2}{3}(1-4\ln2)n_f+2\beta_0\left(\gamma_E+\ln\frac{\mu}{4\pi T}\right)\right]\,,
\end{equation}
and $\Pi_{00}^{(s)}(k)$ is the static part (i.e., only involving zero modes) of the self-energy (see, e.g., Ref.~\cite{Brambilla:2010xn}: note that the static part of the gluon propagator in static gauge for the gauge parameter $\xi =0$ coincides with the static part of the gluon propagator in Coulomb gauge). The contribution to the self-energy coming from loop momenta of the order of the temperature scale appears as a power series in $k$, of which we have kept the first two terms: $m_D$ and $-\delta Z_1k^2$. In fact, the latter scales as $g^4$ and is already beyond the accuracy of our calculation in this regime; we have kept it in order to obtain the logarithm that fixes the scale of the running coupling at leading order (see the discussion in the main section, Secs.~\ref{secfreescreening} and~\ref{secfreescreeningresult}). Higher order terms are even more suppressed since $k\sim gT$ and can be neglected.

The contribution to $D_I$ from the quadratic term is given by
\begin{equation}
 \delta D_I=\frac{g^2}{T}\delta Z_1\int\frac{k^2e^{i\bm{k}\cdot\bm{r}}}{\left(k^2+m_D^2\right)^2}=\alpha_\mathrm{s}\delta Z_1e^{-rm_D}\left(\frac{1}{rT}-\frac{m_D}{2T}\right)\,.
 \label{DI1}
\end{equation}
The static one-loop self-energy gives the following contribution:
\begin{align}
 D_I^{(s)}&=g^4N\int_k\frac{e^{i\bm{k}\cdot\bm{r}}}{\left(\bm{k}^2+m_D^2\right)^2}\int_q\frac{4\left(\bm{k}^2\bm{q}^2-(\bm{k}\cdot\bm{q})^2\right)}{\left((\bm{k}+\bm{q})^2+m_D^2\right)\left(\bm{q}^2\right)^2}\notag\\*
 &=g^4N\int_k\frac{e^{i\bm{k}\cdot\bm{r}}}{\left(k^2+m_D^2\right)^2}\int_0^\infty\frac{dq}{\pi^2}\int_{-1}^1dx\frac{k^2\left(1-x^2\right)}{k^2+2kqx+q^2+m_D^2}\notag\\
 &=g^4N\int_k\frac{e^{i\bm{k}\cdot\bm{r}}}{\left(k^2+m_D^2\right)^2}\int_0^1\frac{dx}{\pi}\frac{\left(1-x^2\right)k^2}{\sqrt{\left(1-x^2\right)k^2+m_D^2}}\notag\\
 &=g^4N\int_k\frac{e^{i\bm{k}\cdot\bm{r}}}{\left(k^2+m_D^2\right)^2}\frac{1}{2\pi}\left(m_D+\frac{k^2-m_D^2}{k}\arctan\frac{k}{m_D}\right)\notag\\
 &=g^4N\int_0^\infty\frac{dk}{4\pi^3r}\frac{\sin kr}{\left(k^2+m_D^2\right)^2}\left(km_D+\left(k^2-m_D^2\right)\arctan\frac{k}{m_D}\right)\notag\\
 &=g^4N\int_{-\infty}^\infty\frac{dk}{8\pi^3r}\left[\frac{rm_De^{ikr}}{2\left(k^2+m_D^2\right)}+\frac{-i\left(k^2-m_D^2\right)e^{ikr}}{\left(k^2+m_D^2\right)^2}\arctan\frac{k}{m_D}\right]\notag\\
 &=\alpha_\mathrm{s}^2N\left[e^{-rm_D}+\lim_{\epsilon\to0}\left(-\int_{m_D+\epsilon}^\infty\frac{d\kappa\,e^{-\kappa r}}{r}\left(\frac{1}{(\kappa-m_D)^2}+\frac{1}{(\kappa+m_D)^2}\right)\right.\right.\notag\\
 &\phantom{=\alpha_\mathrm{s}^2N\Biggl[e^{-rm_D}+\lim_{\epsilon\to0}\Biggl(}+\left.\left.\left(\frac{1}{\epsilon r}+\ln\frac{\epsilon}{2m_D}+\frac{1}{2rm_D}\right)e^{-rm_D}\right)\right]\notag\\
 &=\alpha_\mathrm{s}^2N\left[2e^{-rm_D}+\lim_{\epsilon\to0}\left(\int_{m_D+\epsilon}^\infty d\kappa\,e^{-\kappa r}\left(\frac{1}{\kappa-m_D}+\frac{1}{\kappa+m_D}\right)+e^{-rm_D}\ln\frac{\epsilon}{2m_D}\right)\right]\notag\\
 &=\alpha_\mathrm{s}^2Ne^{-rm_D}\left[e^{2rm_D} E_1(2rm_D) + 2 - \gamma_E - \ln2rm_D\right]\,.
 \label{DI1s}
\end{align}
Using Eqs.~\eqref{DHmass}, \eqref{DI1}, and~\eqref{DI1s}, we reproduce the results for the singlet free energy and Polyakov loop correlators published in Refs.~\cite{Burnier:2009bk} and~\cite{Nadkarni:1986cz}, respectively.

\section{\texorpdfstring{Small $\bm{r}$ expansion of $\bm{F_S}$}{Small r expansion of F\_S}}
\label{Details}

A calculation of the contributions to $F_S$ from the scales $1/r$ and $\pi T$ was presented in Ref.~\cite{Burnier:2009bk} without relying on an expansion in $r\pi T$, which means that it is valid for any hierarchy between those two scales. The contribution from the zero mode has not been explicitly included in that calculation, therefore we add it here (the calculation is given at the end of this appendix).
The result of~\cite{Burnier:2009bk} reads
\begin{align}
 \left.\frac{2F_Q-F_S}{T}\right|_{1/r,T}={}&\frac{N^2-1}{2N}\frac{\alpha_\mathrm{s}}{rT}\left\{1+\frac{\alpha_\mathrm{s}}{4\pi}\left[\frac{11}{3}N+\frac{2}{3}\left(1-4\ln2\right)n_f+2\beta_0\left(\gamma_E+\ln\frac{\mu}{4\pi T}\right)\right]\right\}\notag\\
 &+\frac{N^2-1}{2N}\alpha_\mathrm{s}^2\left(\frac{2}{3}N+\frac{1}{3}n_f\right)r\pi T-\frac{N^2-1}{2}\alpha_\mathrm{s}^2\left(\frac{1}{2\varepsilon}-\frac{3}{2}+\gamma_E+\ln4\pi\mu^2r^2\right)\notag\\
 &+\frac{N^2-1}{2}\left[-\frac{\alpha_\mathrm{s}^2}{24r^2T^2}+\frac{\alpha_\mathrm{s}^2}{r\pi T}\int_1^\infty dx\,\left(-1+\frac{1}{x^2}-\frac{1}{2x^4}\right)\ln\left(1-e^{-4r\pi Tx}\right)\right]\notag\\
 &+\frac{\left(N^2-1\right)n_f}{4N}\frac{\alpha_\mathrm{s}^2}{r\pi T}\int_1^\infty dx\,\left(\frac{1}{x^2}-\frac{1}{x^4}\right)\ln\frac{1+e^{-2r\pi Tx}}{1-e^{-2r\pi Tx}}\,.
\end{align}
The $1/\epsilon$ pole corresponds to an infrared divergence in $F_Q$ when evaluated without the contribution from the scale $m_D$~\cite{Burnier:2009bk,Brambilla:2010xn}. We can expand the above expression in $r\pi T$ as a check of our calculation of $F_S$ at short distances.

The tricky part in the small $r$ expansion of this result lies in the $x$ integrations. If one expands straightforwardly in $r\pi T$, then the higher order terms lead to diverging $x$ integrals. In order to obtain an expansion of finite terms, we will integrate by parts until the integral in $x$ from $0$ to $1$ converges, then calculate the integral from $0$ to infinity exactly and subtract from that the integral from $0$ to $1$. This last part can then be expanded in $r\pi T$ without problems, because $x$ is no longer integrated to $\infty$.

We will only show this explicitly for the first term. The calculation for all other terms works in exactly the same fashion, only it requires more steps of integration by parts and thus becomes rather lengthy, therefore we will give just the results. We obtain
\begin{align}
 \int_1^\infty dx\,\ln\left(1-e^{-4r\pi Tx}\right)={}&-\ln\left(1-e^{-4r\pi T}\right)-\int_1^\infty dx\,\frac{4r\pi Tx}{e^{4r\pi Tx}-1}\notag\\
 ={}&-\ln 4r\pi T +\ln\frac{4r\pi T}{1-e^{-4r\pi T}}-\frac{\zeta(2)}{4r\pi T}+\int_0^1 dx\,\frac{4r\pi Tx}{e^{4r\pi Tx}-1}\notag\\
 ={}&-\ln 4r\pi T +2r\pi T-\frac{2}{3}(r\pi T)^2+\frac{4}{45}(r\pi T)^4+\dots-\frac{\pi}{24rT}\notag\\
 &+\int_0^1 dx\,\left(1-2r\pi Tx+\frac{4}{3}(r\pi T)^2x^2-\frac{16}{45}(r\pi T)^4x^4\right)+\dots\notag\\
 ={}&-\frac{\pi}{24rT}-\ln 4r\pi T +1+r\pi T-\frac{2}{9}(r\pi T)^2+\frac{4}{225}(r\pi T)^4+\dots\,,\\
 \int_1^\infty\frac{dx}{x^2}\ln\left(1-e^{-4r\pi Tx}\right)={}&\ln 4r\pi T +1-2r\pi T(1-\gamma_E-\ln2rT)-\frac{2}{3}(r\pi T)^2+\frac{4}{135}(r\pi T)^4\notag\\
 &+\dots\,,\\
 \int_1^\infty\frac{dx}{x^4}\ln\left(1-e^{-4r\pi Tx}\right)={}&\frac{1}{3}\ln 4r\pi T +\frac{1}{9}-r\pi T+\frac{2}{3}(r\pi T)^2-\frac{8\pi}{3}\zeta(3)r^3T^3+\frac{4}{45}(r\pi T)^4\notag\\
 &+\dots\,.
\end{align}

If we add those up, we get the full expansion for the integral appearing in the gluonic contribution:
\begin{align}
 \int_1^\infty& dx\,\left(-1+\frac{1}{x^2}-\frac{1}{2x^4}\right)\ln\bigl(1-e^{-4\pi rTx}\bigr)\notag\\
 ={}&\frac{\pi}{24rT}+\frac{11}{6}\ln 4r\pi T -\frac{1}{18}+2r\pi T\left(-\frac{5}{4}+\gamma_E+\ln 2rT \right)+\frac{4\pi}{3}\zeta(3)r^3T^3\notag\\
 &+\sum_{k=1}^\infty\left(\frac{1}{k(4k^2-1)}-\frac{1}{4k(2k-3)}\right)\frac{\zeta(1-2k)}{(2k-1)!}(4r\pi T)^{2k}\\
 ={}&\frac{\pi}{24rT}+\frac{11}{6}\ln 4r\pi T -\frac{1}{18}+2r\pi T\left(-\frac{5}{4}+\gamma_E+\ln 2rT \right)-\frac{7}{9}(r\pi T)^2\notag\\
 &+\frac{4\pi}{3}\zeta(3)r^3T^3-\frac{22}{675}(r\pi T)^4+\mathcal{O}\left((r\pi T)^6\right)\,,
\end{align}
where the intermediate expression gives the full series. We see that after the cubic term only even powers of $r\pi T$ appear.

For the fermionic integrals, we obtain
\begin{align}
 \int_1^\infty\frac{dx}{x^2}\ln\frac{1+e^{-2r\pi Tx}}{1-e^{-2\pi rTx}}&=-\ln r\pi T -1+2\ln(2)\,r\pi T-\frac{1}{3}(r\pi T)^2+\frac{7}{270}(r\pi T)^4+\dots\,,\\
 \int_1^\infty\frac{dx}{x^4}\ln\frac{1+e^{-2r\pi Tx}}{1-e^{-2r\pi Tx}}&=-\frac{1}{3}\ln r\pi T -\frac{1}{9}+\frac{1}{3}(r\pi T)^2-2\pi\zeta(3)r^3T^3+\frac{7}{90}(r\pi T)^4+\dots\,.
\end{align}
Accordingly, the combination appearing in the fermionic contribution of $F_S$ is given by
\begin{align}
 &\int_1^\infty dx\left(\frac{1}{x^2}-\frac{1}{x^4}\right)\ln\frac{1+e^{-2r\pi Tx}}{1-e^{-2r\pi Tx}}\notag\\
 &=-\frac{2}{3}\ln r\pi T-\frac{8}{9}+2\ln(2)\,r\pi T+2\pi\zeta(3)r^3T^3+\sum_{k=1}^\infty\frac{(2^{2k}-2)}{k(2k-1)(2k-3)}\frac{\zeta(1-2k)}{(2k-1)!}(2r\pi T)^{2k}\\
 &=-\frac{2}{3}\ln r\pi T-\frac{8}{9}+2\ln(2)\,r\pi T-\frac{2}{3}(r\pi T)^2+2\pi\zeta(3)r^3T^3-\frac{7}{135}(r\pi T)^4+\mathcal{O}\left((r\pi T)^6\right)\,.
\end{align}
The expansions all have the same structure: there is a logarithmic term, a few odd powers of $r\pi T$ at low orders, while for higher orders only even powers remain. The coefficients are rational numbers except for the terms where the power of $r\pi T$ is one less than the power of $x$ in the denominator of the integral.

If we insert all these expansions into the initial expression, we get
\begin{align}
 \left.\frac{2F_Q-F_S}{T}\right|_{1/r,T}={}&\frac{N^2-1}{2N}\frac{\alpha_\mathrm{s}}{rT}\left\{1+\frac{\alpha_\mathrm{s}}{4\pi}\left[\frac{31}{9}N-\frac{10}{9}n_f+2\beta_0(\gamma_E+\ln\mu r)\right]\right\}\notag\\
 &-\frac{N^2-1}{2N}\alpha_\mathrm{s}^2\left[N\left(\frac{1}{2\varepsilon}+1-\gamma_E+\ln\frac{\pi\mu^2}{T^2}\right)-n_f\ln2\right]\notag\\
 &-\frac{N^2-1}{18}\alpha_\mathrm{s}^2r\pi T+\frac{N^2-1}{2N}\left(\frac{4}{3}N+n_f\right)\zeta(3)\alpha_\mathrm{s}^2r^2T^2\notag\\
 &-\frac{N^2-1}{2N}\left(\frac{22}{675}N+\frac{7}{270}n_f\right)\alpha_\mathrm{s}^2(r\pi T)^3+\mathcal{O}\left((r\pi T)^5\right)\,.
\end{align}
Notice how the argument of the logarithm in the first line is now $\mu r$ instead of $\mu/(4\pi T)$. This is because in the unexpanded result there was no scale associated with the ultraviolet divergence, since we did not specify if $1/r$ or $\pi T$ was supposed to be larger. Now that we have expanded in $r\pi T$, we have set $1/r$ to be the largest scale and accordingly the logarithms associated with the ultraviolet divergence include that scale. Also the logarithm in the second line has changed its argument from $\mu r$ to $\mu/T$ for the same reason, because this logarithm is associated with the infrared divergence that gets cured by the contribution from the scale $m_D$, and the next higher scale is now $\pi T$. We see that the infrared divergence is the same as we got from the scale $\pi T$ contribution in the calculation of $D_I$.

We conclude this appendix with the calculation of the zero mode contribution. The $\mathcal{O}\left(\alpha_\mathrm{s}^2\right)$ zero mode contribution to $F_S$ coming from the gluon loop for $m_D=0$ is given by
\begin{align}
 2\left(N^2-1\right)g^4\int_k\frac{e^{i\k\cdot\r}}{\left(\bm{k}^2\right)^2}\int_q\frac{\bm{k}^2\bm{q}^2-(\k\cdot\q)^2}{\left(\bm{q}^2\right)^2(\q+\k)^2}\,.
\end{align}
The tadpole diagram is scaleless for $q_0=0$ and there is no zero mode in the fermion loop. We calculate this with the help of the following elementary integrals:
\begin{align}
 \int_k\frac{e^{i\bm{k}\cdot\bm{r}}}{\left(\bm{k}^2\right)^n}={}&\frac{\Gamma\left(\frac{d}{2}-n\right)}{2^{2n}\pi^{d/2}\Gamma(n)}r^{2n-d}\,,\label{Fourier}\\
 \int_q\frac{1}{\left(\bm{q}^2\right)^m\left((\bm{q}+\bm{k})^2\right)^n}={}&\frac{\Gamma\left(m+n-\frac{d}{2}\right)\Gamma\left(\frac{d}{2}-m\right)\Gamma\left(\frac{d}{2}-n\right)}{(4\pi)^{d/2}\Gamma(m)\Gamma(n)\Gamma(d-m-n)}k^{d-2m-2n}\,,\\
 \int_q\frac{q_i}{\left(\bm{q}^2\right)^m\left((\bm{q}+\bm{k})^2\right)^n}={}&-\frac{\Gamma\left(m+n-\frac{d}{2}\right)\Gamma\left(\frac{d}{2}-m+1\right)\Gamma\left(\frac{d}{2}-n\right)}{(4\pi)^{d/2}\Gamma(m)\Gamma(n)\Gamma(d-m-n+1)}k^{d-2m-2n}k_i\,,\\
 \int_q\frac{q_iq_j}{\left(\bm{q}^2\right)^m\left((\bm{q}+\bm{k})^2\right)^n}={}&\frac{\Gamma\left(m+n-1-\frac{d}{2}\right)\Gamma\left(\frac{d}{2}-m+1\right)\Gamma\left(\frac{d}{2}-n+1\right)}{2(4\pi)^{d/2}\Gamma(m)\Gamma(n)\Gamma(d-m-n+2)}k^{d-2m-2n+2}\delta_{ij}\notag\\
 &+\frac{\Gamma\left(m+n-\frac{d}{2}\right)\Gamma\left(\frac{d}{2}-m+2\right)\Gamma\left(\frac{d}{2}-n\right)}{(4\pi)^{d/2}\Gamma(m)\Gamma(n)\Gamma(d-m-n+2)}k^{d-2m-2n}k_ik_j\,,\\
 \int_q\frac{\bm{k}^2\bm{q}^2-(\bm{k}\cdot\bm{q})^2}{\left(\bm{q}^2\right)^m\left((\bm{q}+\bm{k})^2\right)^n}={}&\frac{(d-1)\Gamma\left(m+n-1-\frac{d}{2}\right)\Gamma\left(\frac{d}{2}-m+1\right)\Gamma\left(\frac{d}{2}-n+1\right)}{2(4\pi)^{d/2}\Gamma(m)\Gamma(n)\Gamma(d-m-n+2)}k^{d-2m-2n+4}\,,
\end{align}
where the third and fourth relation can be obtained from the previous one by taking the derivative with respect to $\bm{k}$, and the last relation is a combination of the second and the fourth.

The zero mode contribution is then given by
\begin{align}
 &2\left(N^2-1\right)g^4\int_k\frac{e^{i\k\cdot\r}}{\left(\bm{k}^2\right)^2}\int_q\frac{\bm{k}^2\bm{q}^2-(\k\cdot\q)^2}{\left(\bm{q}^2\right)^2(\q+\k)^2}\notag\\
 &=2\left(N^2-1\right)g^4\int_k\frac{e^{i\k\cdot\r}}{\left(\bm{k}^2\right)^{3-d/2}}\frac{(d-1)\Gamma\left(2-\frac{d}{2}\right)\Gamma\left(\frac{d}{2}-1\right)\Gamma\left(\frac{d}{2}\right)}{2(4\pi)^{d/2}\Gamma(d-1)}\notag\\
 &=2\left(N^2-1\right)g^4\frac{\Gamma(d-3)}{2^{6-d}\pi^{d/2}\Gamma\left(3-\frac{d}{2}\right)}\frac{(d-1)\Gamma\left(2-\frac{d}{2}\right)\Gamma\left(\frac{d}{2}-1\right)\Gamma\left(\frac{d}{2}\right)}{2(4\pi)^{d/2}\Gamma(d-1)}r^{6-2d}\notag\\
 &=\left(N^2-1\right)\alpha_\mathrm{s}^2\frac{(d-1)\Gamma(d-3)\Gamma\left(2-\frac{d}{2}\right)\Gamma\left(\frac{d}{2}-1\right)\Gamma\left(\frac{d}{2}\right)}{4\pi^{d-2}\Gamma\left(3-\frac{d}{2}\right)\Gamma(d-1)}r^{6-2d}\notag\\
 &=-\frac{N^2-1}{2}\alpha_\mathrm{s}^2\left(\frac{1}{2\epsilon}-\frac{3}{2}+\gamma_E+\ln4\pi+2\ln\mu r+\mathcal{O}(\epsilon)\right)\,.
\end{align}

\section{Cancellation of the magnetic scale contributions}
\label{scalemM}

In this appendix, we will show that all appearances of the magnetic scale in the Polyakov loop correlator cancel up to order $g^8$ in both hierarchies. First, the magnetic scale can only appear in spatial gluon propagators, which are not directly emitted from the Polyakov lines, so they have to emerge from temporal gluons. The only diagrams where this happens at the present order are one-loop $D_I$ and $D_H$. The momenta of the temporal gluons may be of order $1/r$ or $m_D$ (again, the temperature scale contributes at this order only with scaleless integrals or exponentially suppressed terms, depending on the hierarchy).

In the case where the momenta scale as $m_D$, and assuming the hierarchy $1/r\gg\pi T\gg m_D$, we have already seen that the leading term in the small $r$ expansion of $D_I$ is identical to $-2$ times the Polyakov loop, where the cancellation of the scale $m_M$ contributions has already been shown for orders $g^5$ and $g^6$ in~\cite{Berwein:2015ayt}. In the case of $D_H$, the magnetic scale contribution for temporal gluon momenta of order $m_D$ is beyond order $g^8$.

We now show the cancellation at order $g^8$ in the case when the temporal gluon momenta are of order $1/r$ for both hierarchies. We will show the calculation explicitly for $1/r\sim m_D$, the case for the other hierarchy follows straightforwardly by setting $m_D=0$. The leading contribution to $D_I$ with a one-loop self-energy of momentum $m_M$ is given by
\begin{align}
 D_I\Bigr|_{1/r,m_M}&=Ng^4\int_{k\sim1/r}\frac{e^{i\bm{k}\cdot\bm{r}}}{\left(\bm{k}^2+m_D^2\right)^2}\left(\frac{4k_ik_j}{\bm{k}^2+m_D^2}-\delta_{ij}\right)\int_{q\sim m_M}D_{ij}(q)\notag\\
 &=Ng^4\int_{k\sim1/r}e^{i\bm{k}\cdot\bm{r}}\nabla_{k,i}\left(-\frac{k_j}{\left(\bm{k}^2+m_D^2\right)^2}\right)\int_{q\sim m_M}D_{ij}(q)\notag\\
 &=\frac{Ng^4}{2}\int_{k\sim1/r}e^{i\bm{k}\cdot\bm{r}}\nabla_{k,i}\nabla_{k,j}\frac{1}{\bm{k}^2+m_D^2}\int_{q\sim m_M}D_{ij}(q)\notag\\
 &=-\frac{Ng^4}{2}r_ir_j\int_{k\sim1/r}\frac{e^{i\bm{k}\cdot\bm{r}}}{\bm{k}^2+m_D^2}\int_{q\sim m_M}D_{ij}(q)\notag\\
  &=-\frac{Ng^4}{2}\frac{e^{-rm_D}}{4\pi r}r_ir_j\int_{q\sim m_M}D_{ij}(q)\,,
\end{align}
where the second term in the first line comes from the tadpole.

The contribution from $D_H$ can be found in analogous fashion:
\begin{align}
 D_H\Bigr|_{1/r,m_M}&=\frac{g^6}{2T}\int_{k\sim1/r}\frac{2k_ie^{i\bm{k}\cdot\bm{r}}}{\left(\bm{k}^2+m_D^2\right)^2}\int_{p\sim1/r}\frac{2p_ie^{i\bm{p}\cdot\bm{r}}}{\left(\bm{p}^2+m_D^2\right)^2}\int_{q\sim m_M}D_{ij}(q)\notag\\
 &=\frac{g^6}{2T}\int_{k\sim1/r}e^{i\bm{k}\cdot\bm{r}}\nabla_{k,i}\frac{1}{\bm{k}^2+m_D^2}\int_{p\sim1/r}e^{i\bm{p}\cdot\bm{r}}\nabla_{p,j}\frac{1}{\bm{p}^2+m_D^2}\int_{q\sim m_M}D_{ij}(q)\notag\\
 &=-\frac{g^6}{2T}r_ir_j\int_{k\sim1/r}\frac{e^{i\bm{k}\cdot\bm{r}}}{\bm{k}^2+m_D^2}\int_{p\sim1/r}\frac{e^{i\bm{p}\cdot\bm{r}}}{\bm{p}^2+m_D^2}\int_{q\sim m_M}D_{ij}(q)\notag\\
 &=-\frac{g^6}{2T}\left(\frac{e^{-rm_D}}{4\pi r}\right)^2r_ir_j\int_{q\sim m_M}D_{ij}(q)\,.
\end{align}

These two contributions are of different order in $g$, but they contribute at the same order to the Polyakov loop correlator, because the one-loop $D_I$ has to be multiplied with its leading order result:
\begin{align}
 \exp\left[\frac{2F_Q-F_{Q\bar{Q}}}{T}\right]_{m_M}={}&\frac{N^2-1}{4N^2}D_I\Bigr|_{1/r}D_I\Bigr|_{1/r,m_M}-\frac{N^2-1}{4N}D_H\Bigr|_{1/r,m_M}+\mathcal{O}\left(g^9\right)\notag\\
 ={}&\left[-\frac{N^2-1}{4N^2}\frac{g^2e^{-rm_D}}{4\pi rT}\frac{Ng^4e^{-rm_D}}{8\pi r}+\frac{N^2-1}{4N}\frac{g^6}{2T}\left(\frac{e^{-rm_D}}{4\pi r}\right)^2\right]\notag\\
 &\times r_ir_j\int_{q\sim m_M}D_{ij}(q)+\mathcal{O}\left(g^9\right)\notag\\
 ={}&\mathcal{O}\left(g^9\right)\,.
\end{align}
Note that the leading order of $D_I$ is either of order $g^2$ or $g^3$, depending on the hierarchy, but the product with the scale $m_M$ contribution from $D_I$ is of order $g^8$ in both cases (the integral over the spatial gluon propagator is of dimension one, hence proportional to $m_M \sim g^2T$). This cancellation is independent of the actual form of the resummed magnetic scale gluon propagator, and it is valid in general $d$ dimensions [where one just has to replace the Yukawa potential $e^{-rm_D}/4\pi r$ with the $d$-dimensional integral over $e^{i\bm{k}\cdot\bm{r}}/\left(\bm{k^2}+m_D^2\right)$ in each case].

\section{Relation to other forms of resummation}
\label{app:pisarski}

The results of this paper relate to a calculation published in~\cite{Pisarski:2015xea}, which we will discuss here. The authors of~\cite{Pisarski:2015xea} performed a partial resummation of the perturbative series for the Polyakov loop correlator and the singlet free energy correlator (which they call Wilson loop, but since they neglect any contributions involving the spatial Wilson lines, both functions are identical), and they find an unexpected behavior at short distances. While the calculation itself appears to be correct, some of their conclusions may not be.

The resummation includes all diagrams where gluons of momentum $\sim1/r$ without any loop insertions are exchanged between the two Polyakov lines; any other contribution is neglected. As such it is well defined, but gauge dependent. They choose static gauge (SG) $\partial_0A_0=0$ and we believe this to be the source of their unexpected results. Performing the same kind of resummation in Coulomb gauge (CG) leads to a different result.

We may use the exponentiated expression of Eq.~\eqref{F1exp}. In the corresponding discussion, we have already argued that all diagrams where gluons can be separated into a left and a right part by a line cutting the two Polyakov loops such that no gluon crosses this line do not contribute to the exponent. In other words, for diagrams made of unresummed gluon propagators a necessary condition to appear in the exponent is that the gluons cross. However, for such diagrams in Coulomb gauge the delta function in the propagator makes all diagrams with more than one gluon vanish. Hence the result of this resummation in Coulomb gauge is simply the exponential of $D_I$.

Comparing this result with the one in static gauge from~\cite{Pisarski:2015xea} for SU(2), we have
\begin{align}
 \mathcal{W}_\mathrm{SG}&\approx(1+z)\cosh(z)+(2+z)\sinh(z)=1+3z+\frac{3}{2}z^2+\dots\,,\\
 \mathcal{W}_\mathrm{CG}&\approx\exp(3z)=1+3z+\frac{9}{2}z^2+\dots\,,
\end{align}
where $z=g^2/16\pi rT$ and we have expanded for small $z$. We see that the first order term is the same, but the second order is not. This confirms our previous statement that this resummation is gauge dependent. However, since the Wilson loop is gauge invariant (if the spatial Wilson lines are included), the difference between both gauges must be contained in terms that were neglected in this resummation. A gauge invariant expression could be obtained from a resummation of all terms of order $z^n$, however, in static gauge not all such terms come from ladder diagrams without loop insertions. We will show this at $\mathcal{O}\left(z^2\right)$.

There are two sources for the discrepancy between both gauges, the first comes from the singular part in the static gauge gluon self-energy. At one-loop order this is given by (see, e.g., Ref.~\cite{Brambilla:2010xn})
\begin{equation}
 \Pi_{00}(0,k\gg\pi T)_{\rm sing}=-\frac{Ng^2|\k|^3}{192T}\,.
\end{equation}
If we include this contribution in the one-gluon exchange, we get in SU(2)
\begin{equation}
 \frac{3g^2}{4T}\int_k\frac{e^{i\k\cdot\r}}{\k^2+\Pi_{\rm sing}}=\frac{3g^2}{16\pi rT}+\frac{g^4}{(16\pi)^2r^2T^2}+\dots=3\left(z+\frac{z^2}{3}+\dots\right)\,.
\end{equation}
Therefore instead of $z$ one should insert $\widetilde{z}=z+z^2/3+\dots$ into the resummed expression for $\mathcal{W}_\mathrm{SG}$ in order not to neglect any contribution of order $z^n$ from the singular part of the self-energy.

\begin{figure}[t]
 \centering
 \includegraphics[width=0.8\linewidth]{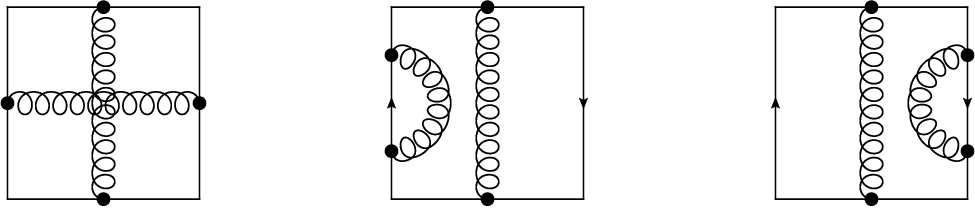}
 \caption{Additional diagrams in the Wilson loop that contribute with terms of order $z^2$.}
 \label{strings}
\end{figure}

The second source of the discrepancy comes from the neglected contributions of the spatial Wilson lines. There are three diagrams with one gluon between the two Polyakov lines and one gluon connected to the spatial Wilson lines (cf.\ Fig.~\ref{strings}). The first diagram has a color factor $-(N^2-1)/4N^2$ and the other two have $(N^2-1)^2/4N^2$, and it is straightforward to show that the sum of the three diagrams is equivalent to the first diagram with a coefficient $-\left(N^2-1\right)/4$.

The spatial gluon propagator for large momenta has a term of order $1/T^2$:
\begin{equation}
 D_{ij}(k_0\neq0,\k)=\frac{k_ik_j}{k_0^2\,\k^2}+\mathcal{O}\left(\frac{1}{\k^2}\right)\,.
\end{equation}
With this, the crossed diagram gives a contribution of order $z^2$ (again with $N=2$):
\begin{align}
 \delta\mathcal{W}&=-\frac{3}{4}\left(\frac{g^2}{T}\int_k\frac{e^{i\k\cdot\r}}{\k^2}\right)\left((ig)^2\int_0^1ds_1r_i\int_1^0ds_2r_j{\sum_K}'\hspace{-18pt}\int\,e^{i\k\cdot\r(s_1-s_2)}\frac{k_ik_j}{k_0^2\,\k^2}\right)\notag\\
 &=-\frac{3}{4}\left(\frac{g^2}{4\pi rT}\right)\left(-\frac{g^2}{24\pi rT}\right)=2z^2\,.
\end{align}

Coulomb gauge has neither a singular part in the one-loop self-energy nor a term of order $1/T^2$ in the spatial gluon propagator, hence the tree-level one-gluon exchange diagram already contains all terms of order $z^n$. If we put all contributions in static gauge together, we indeed get the same result as in Coulomb gauge for the SU(2) Wilson loop:
\begin{equation}
 \mathcal{W}_\mathrm{SG}=1+3\left(z+\frac{z^2}{3}\right)+\frac{3}{2}z^2+2z^2+\mathcal{O}\left(z\alpha_\mathrm{s},z^3\right)=1+3z+\frac{9}{2}z^2+\mathcal{O}\left(z\alpha_\mathrm{s},z^3\right)\,.
\end{equation}

There are, in fact, different versions of static gauge, which differ in the gauge fixing of the spatial gluons; here we used the one of Ref.~\cite{Brambilla:2010xn}. A different version of static gauge might give different expressions for $\Pi_{\rm sing}$ and $D_{ij}$, but also in this case the two contributions described above will be necessary to get the full result for $\mathcal{W}$ at $\mathcal{O}\left(z^2\right)$.

The other part of~\cite{Pisarski:2015xea} deals with the large $N$ limit. The result they obtain in this case for the Wilson loop is given by a Bessel function:
\begin{equation}
 \mathcal{W}_\mathrm{SG}=I_0\left(2\sqrt{z}\right)=1+z+\frac{z^2}{4}  +\dots\,,
\end{equation}
where now $z=g^2N/8\pi rT$. For Coulomb gauge in the planar limit, the resummation works in the same way as before and we have
\begin{equation}
 \mathcal{W}_\mathrm{CG}=\exp{z}=1+z+\frac{z^2}{2}+\dots\,.
\end{equation}
Taking the planar limit for the other two results we get $z^2/12$ from the singular part of the self-energy and $z^2/6$ from the diagrams involving the spatial Wilson lines. We see also here that if we add these two contributions to the tree-level one-gluon exchange result in static gauge, then both gauges agree up to $\mathcal{O}\left(z^2\right)$.

So far, we have only considered small $z$ expansions. In~\cite{Pisarski:2015xea} there is also a discussion on the large $z$ limit, which corresponds to $rT\ll\alpha_\mathrm{s}$ or $rT\ll\alpha_\mathrm{s}N$. We disagree with their conclusions. In order to take the limit $z\to\infty$ one really has to include all terms of order $z^n$ in the resummation, and, as we just saw, this has not been done in~\cite{Pisarski:2015xea}. There will also be higher powers of $z$ from multiple gluon exchanges between the spatial Wilson lines and higher powers in the expansion of the propagators in the singular self-energy. Since those terms were not included in the resummations, there is no reason to trust the results for large $z$. The authors have commented on a strange behavior of the Wilson loop for large $z$ and interpreted it as a side effect of the planar limit, while in our view it is due to an incomplete resummation and gauge dependence. In Coulomb gauge, there are no contributions of order $z^n$ from gluon exchanges between the spatial Wilson lines and there are also no singular terms in the self-energy up to one-loop order. We do not know if at a higher loop order a singular term may appear in the self-energy, but assuming that it does not, then the resummed result of the Wilson loop is also valid in the large $z$ limit and shows exactly the Coulombic behavior that is expected.

Apart from the Wilson loop, Ref.~\cite{Pisarski:2015xea} also discusses the Polyakov loop correlator. There the picture is similar, the leading term in the small $z$ expansion of their resummed result reproduces the known expression, but the next order term is missing the contribution from the singular part of the self-energy [cf.\ the $1/(rT)^3$ term in Eq.~(57) of Ref.~\cite{Brambilla:2010xn}]. Then the large $z$ limit does not reproduce the right behavior, because the resummation is incomplete.

Assuming that the Coulomb gauge does not have singular contributions from the self-energy at higher orders, we may take the $z\to\infty$ limit in Eq.~\eqref{PLCDiag} without problems. The contribution from the adjoint free energy becomes exponentially suppressed, and the Polyakov loop correlator is given by the exponential of the singlet free energy alone. 

\bibliography{ref}
\end{document}